\documentclass{article}
\IfFileExists{arxiv.sty}{
  \usepackage{arxiv}
}{
  \usepackage[margin=1in]{geometry}
}

\usepackage[utf8]{inputenc}
\usepackage[T1]{fontenc}
\usepackage{booktabs}
\usepackage{amsmath}
\usepackage{amssymb}
\usepackage[authoryear]{natbib}
\usepackage{microtype}
\usepackage{nicefrac}
\usepackage{url}
\usepackage{xcolor}
\usepackage{caption}
\usepackage{subcaption}
\usepackage{rotating}
\usepackage{listings}
\usepackage{appendix}
\usepackage{placeins}
\usepackage[hidelinks]{hyperref}
\usepackage{csquotes}
\usepackage{lscape}
\usepackage{soul}

\IfFileExists{siunitx.sty}{\usepackage{siunitx}}{}

\newcommand{\bibliostyle}{abbrvnat}

\graphicspath{{pics/}}
\newcommand{\currentappendixletter}{A}
\newcommand{\appendixtitle}[1]{\section*{Appendix \currentappendixletter. #1}}
\newcommand{\setappendixletter}[1]{%
  \gdef\currentappendixletter{#1}%
  \setcounter{section}{0}%
  \renewcommand{\thesection}{\currentappendixletter.\arabic{section}}%
  \renewcommand{\theHsection}{\currentappendixletter.\arabic{section}}%
  \renewcommand{\theHsubsection}{\currentappendixletter.\arabic{section}.\arabic{subsection}}%
}
\newcommand{\amsappendixA}{\setappendixletter{A}}
\newcommand{\amsappendixB}{\setappendixletter{B}}

\date{}

\title{Towards operational data-driven regional weather forecasting: extreme-event skill and near-real-time verification with HRRRCast}

\author{
  Daniel Abdi\thanks{These authors contributed equally.}\thanks{CIRES = Cooperative Institute for Research in Environmental Sciences; CIRA = Cooperative Institute for Research in the Atmosphere; NOAA-GSL = NOAA Global Systems Laboratory; NOAA-OMD = NOAA Office of Model Development.} \\
  CIRES/GSL\\
  \texttt{daniel.abdi@noaa.gov}
  \And
  Isidora Jankov\footnotemark[1] \\
  NOAA-GSL
  \And
  Vanderlei Vargas \\
  CIRA/GSL
  \And
  Samuel Trahan \\
  CIRES/GSL
  \And
  Paul Madden \\
  CIRES/GSL
  \And
  Emily Carpenter \\
  CIRES/GSL
  \And
  Travis Wilson \\
  NOAA-GSL
  \And
  Jeff Hamilton \\
  NOAA-GSL
  \And
  Molly Smith \\
  CIRES/GSL
  \And
  Lindsay Blank \\
  CIRA/GSL
  \And
  Linlin Cui \\
  Lynker at NOAA-OMD\\
  \And
  Jun Wang \\
  NOAA-OMD
  \And
  Matthew Mahalik \\
  NOAA-GSL
}

\begin{document}

\maketitle

\begin{abstract}
Improving convective-scale skill in data-driven regional weather models requires better representing rare
high-impact events, training at native resolution, broadening variable output, and demonstrating 
performance in real-time settings. Here, we evaluate these challenges using HRRRCast, a data-driven
convection-allowing emulator trained on the High-Resolution Rapid Refresh (HRRR) analysis data.
HRRRCast combines native 3-km training, moderate sigmoid-weighted loss reweighting for reflectivity and precipitation, and broader system
improvements including expansion of the training archive from 3 to 9 years (2016--2025),
increased vertical resolution, expanded outputs, and architectural upgrades. The weighted loss improves
high-threshold skill while preserving common-event behavior. Native-resolution
training avoids aliasing from 6-km subsampling and substantially improves fractions skill score at
convective scales, with gains that persist across pooling windows. HRRRCast has also been running in
near-real-time mode for the past year. In seven months of independent observation-based
verification, HRRRCast ensemble means improve several near-surface metrics relative to
operational HRRR and maintain strong precipitation and reflectivity performance. Single-member reflectivity skill is
strongest at 20--30 dBZ; at 40 dBZ, skill is near parity at 6-km pooling and better at pooling
radii $\geq$30 km. Precipitation skill remains competitive through 1 in/h and extends to about 3 in/h in the best recent configuration. A focused comparison with HREF further shows useful low-cost
probabilistic guidance for 3-h precipitation. At storm scale,
HRRRCast exhibits a displacement--structure trade-off, with generally better storm placement than
storm morphology.
\end{abstract}

\section*{Significance Statement}
This study examines how data-driven regional weather forecasting can be made more operationally useful.
Using HRRRCast as a testbed, we show that improved treatment of rare
high-impact events, native 3-km training, and a much longer training archive improve storm-related forecast skill,
hourly precipitation performance, and several near-surface measures, while also providing useful
low-cost ensemble guidance. The results show that the field can move beyond
proof-of-concept demonstrations toward near-real-time practical complementarity with traditional
weather models. They also identify an important remaining challenge: improving the fine-scale
structure of predicted storms.

\section{Introduction}
\label{sec:introduction}

Convection-allowing models (CAMs) have become essential tools for predicting high-impact weather
phenomena such as severe thunderstorms, heavy precipitation, and flash flooding. Operating at
km-scales, CAMs explicitly resolve many convective processes without relying on convective
parameterization schemes, thereby producing more realistic forecasts of mesoscale weather features.
The High-Resolution Rapid Refresh (HRRR) model, operational since 2014 and running at 3-km
resolution over the contiguous United States (CONUS), exemplifies this capability and provides
crucial guidance for operational forecasters \citep{dowell2022}.

Despite their success, physics-based CAMs like HRRR remain computationally expensive. A single
18-hour HRRR forecast requires substantial high-performance computing resources, limiting the
frequency of model runs, the ensemble size, and the ability to explore alternative configurations
for research. These computational constraints motivate the development of fast, data-driven
emulators that can complement traditional numerical weather prediction (NWP) systems.

In \citet{abdi39}, we introduced HRRRCast, an ensemble data-driven emulator trained on HRRR analyses
data to produce rapid convection-allowing forecasts over CONUS. HRRRCast demonstrated that using a
simple architecture, one could generate ensemble forecasts with competitive skill at
light-to-moderate reflectivity thresholds (20--30 dBZ), while significantly reducing computational
cost relative to HRRR. The model was trained on sub-sampled 6-km data spanning three years and
incorporated future GFS boundary conditions to improve long-lead skill. While this initial version
represented a significant step toward operational machine learning-based regional forecasting,
several limitations remained: (i) training on sub-sampled 6-km data introduced aliasing that
corrupted fine-scale spectral structure in the training targets; (ii) performance at extreme precipitation and
reflectivity thresholds lagged behind HRRR; and (iii) the limited training dataset (3 years)
constrained the model's ability to capture the full climatological range of CONUS weather regimes.

In this paper, we use HRRRCast to examine several ingredients needed for more operationally
useful regional AI weather forecasting. The key contributions are:
\begin{enumerate}
\item \textbf{Native-resolution training:} We demonstrate what is gained by training on the native
3-km HRRR grid rather than on sub-sampled 6-km fields.
\item \textbf{Extreme-event optimization:} We demonstrate that a moderate sigmoid-weighted loss for
reflectivity and hourly precipitation improves high-threshold skill while preserving common-event
behavior.
\item \textbf{Improvements for operational performance:} We expand the
training archive to nearly a decade, broaden multivariate outputs and vertical levels, and
introduce architectural upgrades.
\item \textbf{Operational/observational verification:} We demonstrate that the resulting system is
broadly comparable to operational HRRR in independent observation-based verification, including
lead-time die-off diagnostics and threshold-dependent behavior.
\item \textbf{Storm-scale error characterization:} We demonstrate and interpret a consistent
displacement--structure contrast, including lead-time-dependent fragmentation and object-scale
morphology biases.
\end{enumerate}

HRRRCast has been running in near-real-time for nearly one year. The observation-based
verification with the Model Analysis Tool Suite (MATS) \citep{turner2020} therefore focuses on
the most recent seven months, when the native 3-km HRRRCast system was running. Over this period,
HRRRCast single-member forecasts are broadly comparable to operational HRRR across multiple
verification metrics, while its ensembles provide additional probabilistic guidance at minimal
computational cost. Because rapid ensemble guidance is a key intended use case, we also include a
focused comparison with HREF for probabilistic precipitation forecasting. These results show
improved HRRRCast skill for relative humidity, dewpoint, and 10-m winds. Hourly precipitation
skill is broadly competitive through 1.00 in and improves further in the best recent
configuration. Reflectivity skill also improves at 20 and 30 dBZ; at 40 dBZ, behavior is scale
dependent, with near parity at 6-km pooling and HRRRCast advantages at larger pooling radii.
Fractions skill score (FSS) analyses further show that native 3-km training yields 
substantial gains at convection-permitting scales that persist across pooling windows. Together, 
these results position HRRRCast as a viable complement to physics-based CAMs for rapid ensemble 
generation, real-time decision support, and exploratory sensitivity studies.

Beyond aggregate skill gains, we examine residual storm-scale errors in AI regional weather
models. This analysis reveals a consistent contrast between storm placement and storm morphology,
with HRRRCast generally showing better displacement behavior than structural realism at longer
leads. Using subhourly lead-time-conditioned inference as a diagnostic, we further show that
temporally smooth evolution can coexist with persistent storm-structure bias. Section~\ref{sec:storm_scale} then
evaluates these behaviors and discusses possible mechanisms.

The remainder of this paper is organized as follows. Section~\ref{sec:methodology} describes the
updated methodology, including the training loss design and architectural enhancements.
Section~\ref{sec:dataset} summarizes the expanded training dataset and preprocessing pipeline.
Section~\ref{sec:results} presents the main near-real-time verification results, including
observation-based diagnostics and a focused ensemble comparison against HREF; supplemental RMSE
verification against HRRR analyses is deferred to Appendix~\ref{append:analysis_rmse}.
Section~\ref{sec:ablation_studies} presents attribution and ablation studies, including
native-resolution and loss-reweighting comparisons. Section~\ref{sec:storm_scale} focuses on
storm-scale behavior and interpretation, including displacement, structure, and
subhourly-cadence limitations. Section~\ref{sec:conclusions} presents conclusions and
Section~\ref{sec:future_directions} outlines directions for future work.

\section{Methodology}
\label{sec:methodology}
This section describes the key methodological improvements in HRRRCast relative to the
original model presented in \citet{abdi39}. The updates include: (i) training on
the native 3-km HRRR grid rather than sub-sampled 6-km data, (ii) an expanded training dataset
spanning nearly a decade of HRRR and GFS analyses, (iii) a sigmoid-weighted balanced loss function
to improve extreme-event skill, and (iv) architectural improvements required for 3-km training.

\subsection{Native 3-km grid training}
\label{sec:native_3km_training}
The original HRRRCast was trained on HRRR data sub-sampled to 6-km resolution to manage memory and
computational constraints. While this approach enabled rapid prototyping, spatial coarsening
introduces aliasing artifacts that corrupt fine-scale spectral structure in the training targets.
By training directly on the native 3-km HRRR grid, the model learns physically consistent fine-scale
structure characteristic of convection-allowing scales. 

To accommodate the four-fold increase in grid points relative to 6-km training, we employ
memory-efficient engineering strategies including mixed-precision training (\texttt{float16}
storage with selective \texttt{float32} computation), gradient checkpointing, and optimized data
loading pipelines. These techniques enable end-to-end 3-km model training on eight H100 GPUs
without sacrificing model capacity or batch size.

\subsection{Expanded training dataset and multivariate outputs}
\label{sec:expanded_dataset_multivariate}
Relative to the original model, the training corpus is expanded from three years to nearly a decade,
substantially broadening coverage of sampled weather regimes across seasons and
event types. In parallel, the multivariate learning setup is expanded to include finer vertical
structure and a larger prognostic variable set, with additional diagnostic fields derived for
hazardous-weather applications.

Conceptually, this update shifts HRRRCast from a narrower proof-of-concept configuration toward a
more complete CAM-emulator training target. Detailed data curation, variable definitions, vertical
level design, and accumulated-precipitation target construction are described in Section~\ref{sec:dataset}.

\subsection{Sigmoid-weighted loss for extreme events}
\label{sec:sigmoid_loss}
A fundamental limitation of standard mean squared error (MSE) training is the tendency to optimize
the conditional mean of the target distribution, which systematically underweights rare,
high-intensity events \citep{shi2017deep,ravuri2021skillful}. In our training distribution,
more than 90\% of samples are low-threshold events, so uniform MSE primarily optimizes common
cases and under-emphasizes extremes. For precipitation and reflectivity forecasting, this class
imbalance mainly degrades high-threshold skill.

To address this, we implement a sigmoid-weighted balanced MSE loss applied selectively to composite
reflectivity (REFC) and accumulated precipitation (APCP). The weighting function applies a smooth
multiplicative factor that increases from approximately $1\times$ for common events to $3\times$
for rare, high-intensity events (defined as roughly $4\sigma$ to $10\sigma$ in normalized anomaly
space). For REFC, $\sigma \approx 5$ dBZ in the training normalization, so the ramp begins in the
lower convective-reflectivity range (reflectivity values around 20 dBZ, consistent with
Fig.~\ref{fig:weighted_mse}); APCP uses the same $4\sigma$--$10\sigma$ criterion in its own
transformed, variable-specific normalized space. This design provides controlled emphasis on
extreme events while avoiding the unbounded weight growth that can destabilize training or produce unrealistic forecast amplitudes.

The weighting is applied only to REFC and APCP because these variables exhibit the most pronounced
heavy-tailed distributions and are most critical for operational severe weather applications. All
other forecast variables (temperature, winds, geopotential height, etc.) continue to use standard
MSE. The rationale for this variable selection, comparisons with alternative weighting
formulations, domain-size context, and the tuning experiments used to select the final schedule
are provided in Appendix~\ref{append:sigmoid_loss}.

\subsection{Architectural improvements}
\label{sec:architectural_improvements}
Several architectural enhancements complement the training and data improvements. We introduce
dedicated preprocessing branches that process HRRR, GFS, and (for diffusion training)
noised-HRRR inputs through separate residual stems before fusion, enabling more flexible input
handling. These modifications, combined with memory-efficient engineering (gradient
checkpointing, spatially grouped convolutions), enable stable training at native 3-km resolution.
We also replace squeeze-and-excitation (SE) blocks with Convolutional Block Attention Modules
(CBAM; \citealp{woo2018cbam}), which provide both channel-wise and spatial attention.
Complete architectural specifications are provided in Appendix~A.

\section{Dataset}
\label{sec:dataset}
This section summarizes the data sources and preprocessing choices used for training and
verification, with emphasis on aspects most relevant for interpreting Section~\ref{sec:results} results. Complete
details on data sources, variable selection, preprocessing, normalization, version harmonization,
and quality control procedures are provided in Appendix~A.

\subsection{Data sources}
\label{sec:data_sources}
The primary training target is the HRRR model, which has been operational over CONUS since 2014 at
3-km horizontal resolution with hourly cycling \citep{dowell2022}. Over the 2016--2025 training
period, HRRR underwent several version updates that introduced changes to model physics, data
assimilation procedures, and output variable specifications. 

GFS data provide large-scale atmospheric context and future boundary conditions. We interpolate GFS data
to the HRRR grid and include them as additional input channels, enabling the model to perform 
physically consistent downscaling implicitly from coarse global guidance to convection-allowing scales.

\subsection{Training/testing splits and verification periods} 
\label{sec:training_testing_splits}
The production 3-km training dataset is split chronologically into a training period from
2016-10-01T00 through 2025-06-05T23 and a testing period from 2025-06-06T00 through
2025-10-05T23. Additionally, the near-real-time MATS verification archive considered in 
this study spans 2025-10-05T23 through 2026-10-09T23, and the focused seven-month subset 
discussed in Section~\ref{sec:mats_nrt_verification} is 2026-03-20T00 through 2026-10-09T23.

\subsection{Training variables and vertical resolution}
\label{sec:training_variables_vertical_resolution}
The model predicts a comprehensive prognostic variable set consisting of 3D fields (temperature,
geopotential height, u/v wind components, specific humidity, and vertical velocity) on 20 pressure
levels spanning 1000--200 hPa, compared with 12 levels in the original model, to better resolve the
vertical thermodynamic and kinematic structure relevant to convection; 2D surface and
near-surface fields (2-m temperature, 2-m dewpoint, 10-m u/v winds, mean sea level pressure,
and surface pressure); accumulated hourly precipitation; and composite reflectivity. Relative to the
original model, we add 18 surface/integrated variables, including hourly accumulated precipitation
(APCP), and treat them as prognostic variables in the same autoregressive framework (i.e., they
are both network inputs and prediction targets at subsequent steps). In addition to these
prognostic fields, we compute a broad set of surface and upper-air diagnostic variables relevant
to hazardous-weather forecasting, bringing HRRRCast closer to the breadth of output variables
available from HRRR.

\subsection{Accumulated precipitation targets}
\label{sec:apcp_targets}
Accumulated variables introduce several training challenges. In the HRRR analyses dataset,
APCP (and even instantaneous precipitation rate) is zeroed out, so precipitation targets must
be taken from HRRR forecasts rather than analyses. If reliable analyzed instantaneous
precipitation-rate fields were available, they would provide a more direct supervision target.
In future work, it may be preferable to use MRMS hourly precipitation as the training target to
reduce known HRRR precipitation biases and to improve consistency with other variables trained
against HRRR analyses. We also found that the vertical velocity field (VVEL) in HRRR analyses 
contains spurious values; therefore, we use the 1-h forecast VVEL field instead of the analyzed field.

Another challenge is selecting the accumulation period for precipitation targets. HRRR provides
hourly accumulated precipitation, whereas HRRRCast is trained to predict three lead times (1-h,
3-h, and 6-h). One alternative is to predict each corresponding accumulation directly. However,
this effectively creates three strongly correlated precipitation targets with different sparsity/climatology. 
We therefore predict only the last-hour accumulation as the prognostic precipitation target. 
This choice is aligned with hourly autoregressive stepping, keeps precipitation as a single 
variable in training, and still allows 3-h and 6-h accumulations to be reconstructed consistently 
by summing hourly forecasts.

\subsection{Preprocessing and normalization}
\label{sec:preprocessing_normalization}
All input and target fields are normalized using climatological statistics computed from the training
dataset. For variables with highly skewed distributions (e.g., precipitation), we apply 
variance-stabilizing transformations prior to normalization to improve
numerical conditioning and training stability. Native 3-km training requires careful attention to
storage and I/O efficiency; we store preprocessed data in \texttt{float16} format and employ
mixed-precision training with selective \texttt{float32} computation in loss evaluation and
gradient-sensitive operations. These engineering considerations are detailed in Appendix~A.

\section{Results}
\label{sec:results}

Results in this section focus on near-real-time observation-based verification against
MRMS through MATS, followed by a focused operational comparison against HREF.

\subsection{Near-real-time verification against observations using MATS}
\label{sec:mats_nrt_verification}

\begin{figure*}[tbp]
\centering
\begin{subfigure}[t]{0.32\textwidth}
\centering
\includegraphics[width=\textwidth]{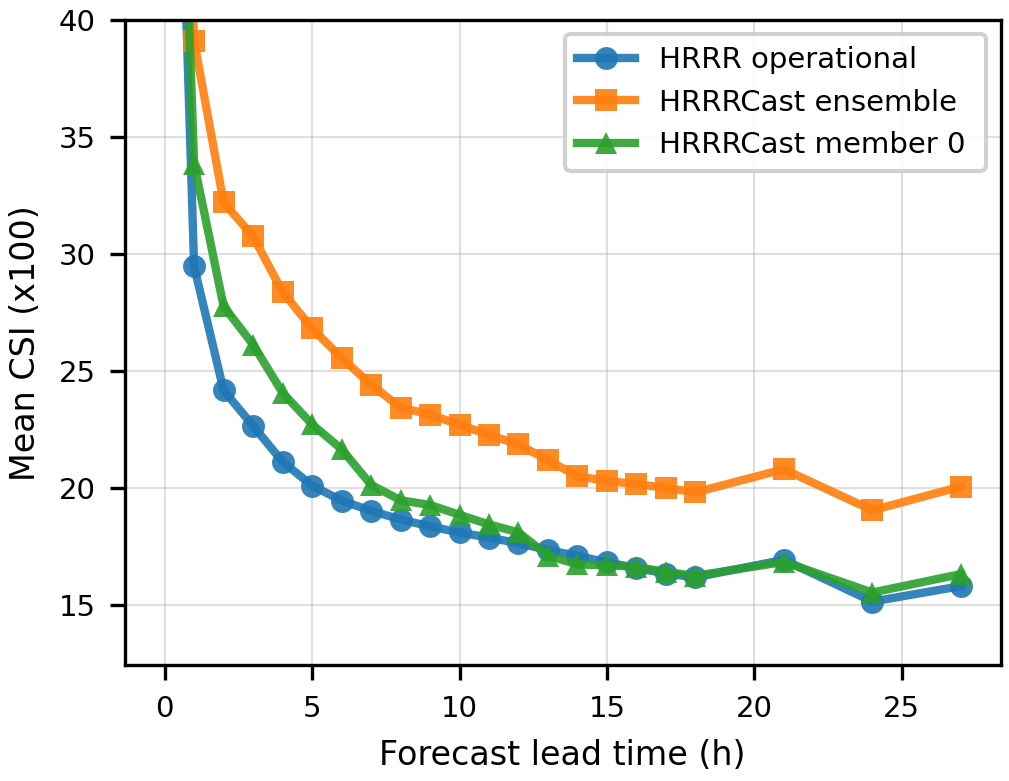}
\caption{CSI at 20 dBZ}
\end{subfigure}\hfill
\begin{subfigure}[t]{0.32\textwidth}
\centering
\includegraphics[width=\textwidth]{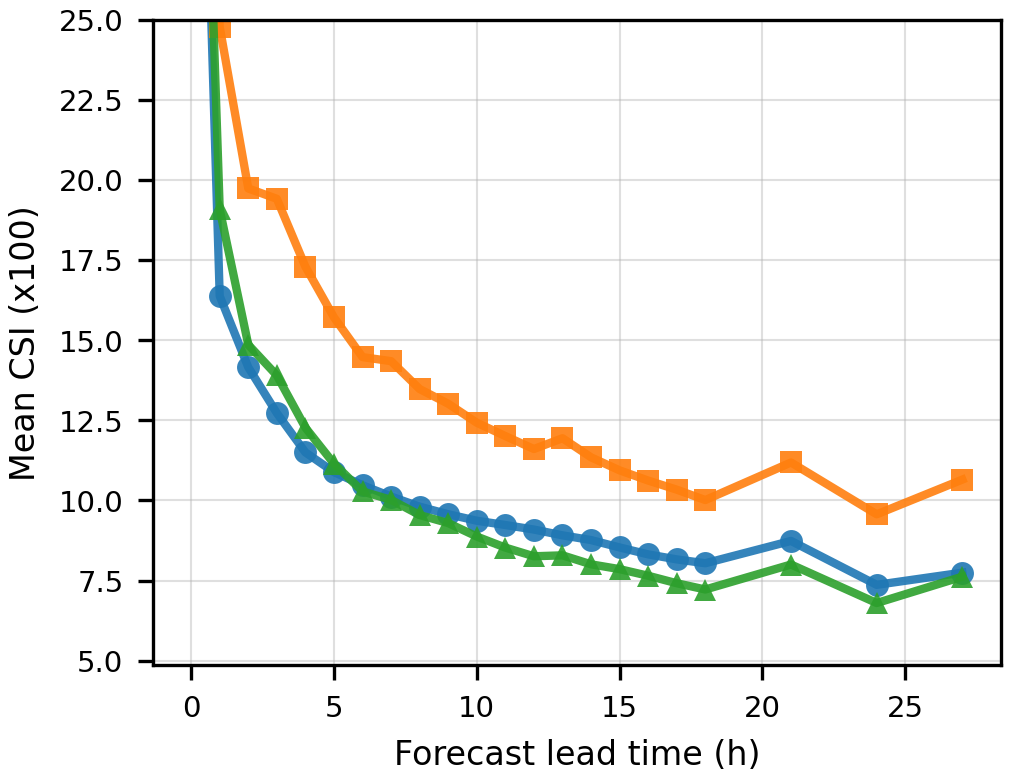}
\caption{CSI at 30 dBZ}
\end{subfigure}\hfill
\begin{subfigure}[t]{0.32\textwidth}
\centering
\includegraphics[width=\textwidth]{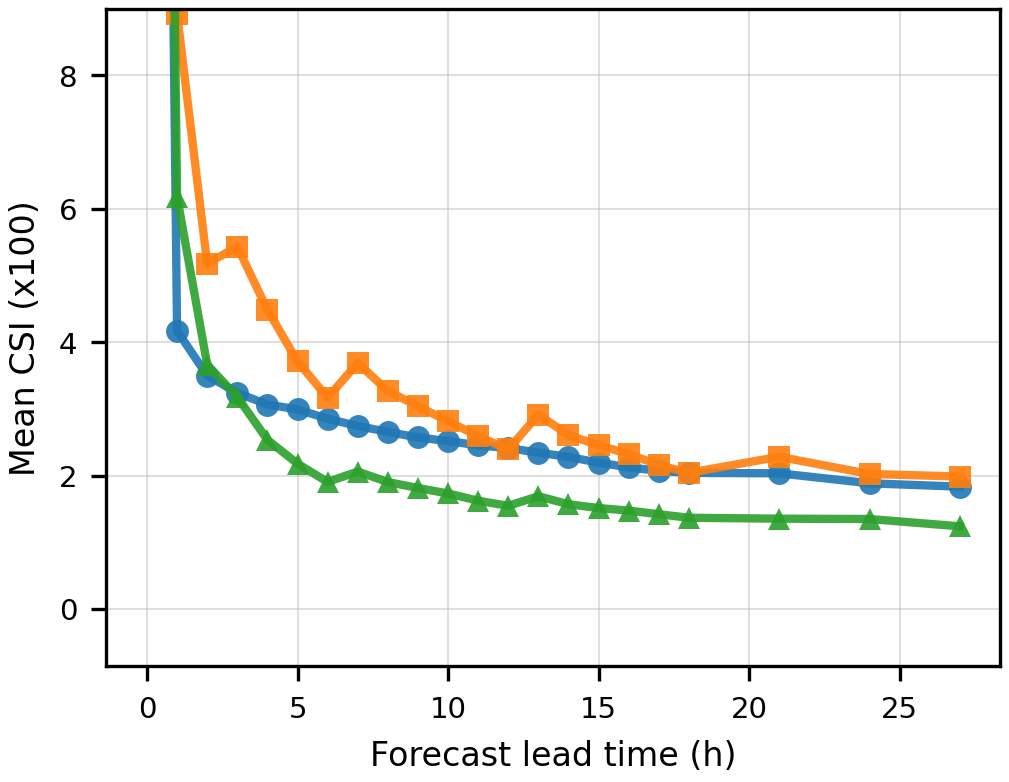}
\caption{CSI at 40 dBZ}
\end{subfigure}
\begin{subfigure}[t]{0.32\textwidth}
\centering
\includegraphics[width=\textwidth]{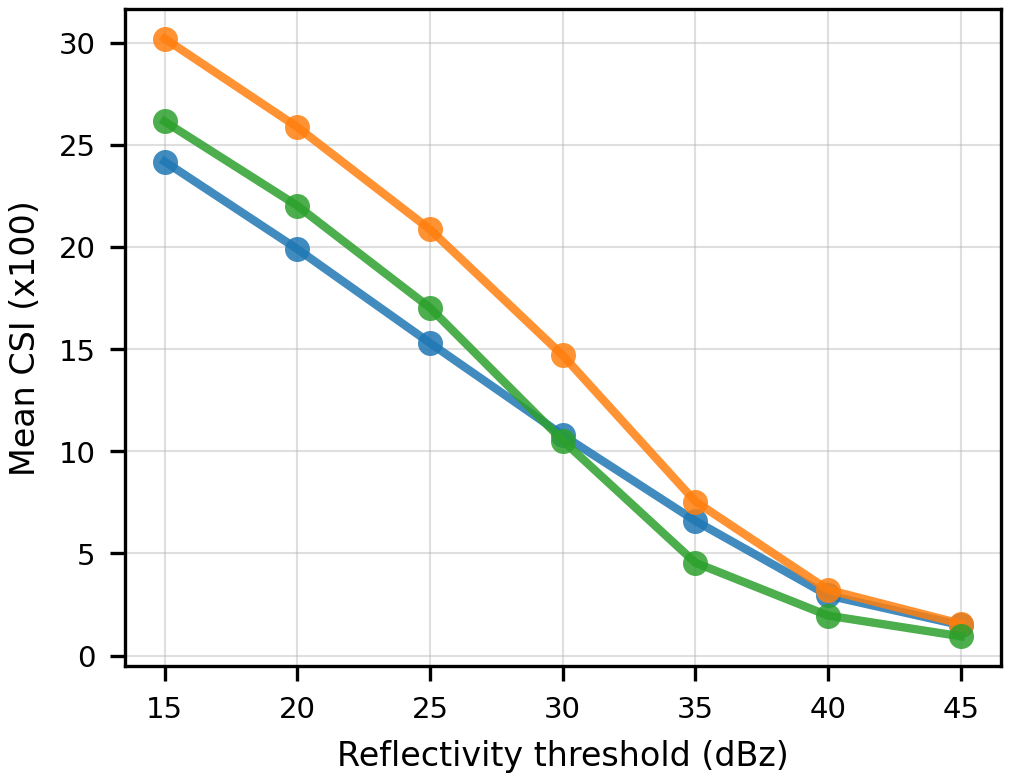}
\caption{6h forecast}
\end{subfigure}
\begin{subfigure}[t]{0.32\textwidth}
\centering
\includegraphics[width=\textwidth]{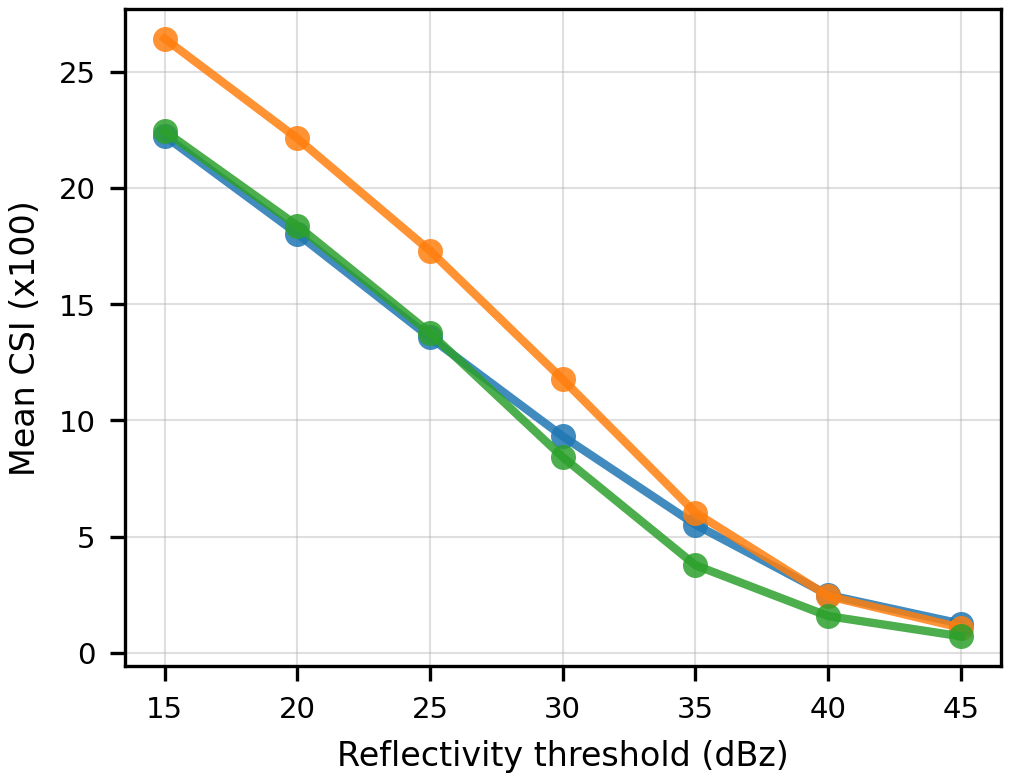}
\caption{12h forecast}
\end{subfigure}
\begin{subfigure}[t]{0.32\textwidth}
\centering
\includegraphics[width=\textwidth]{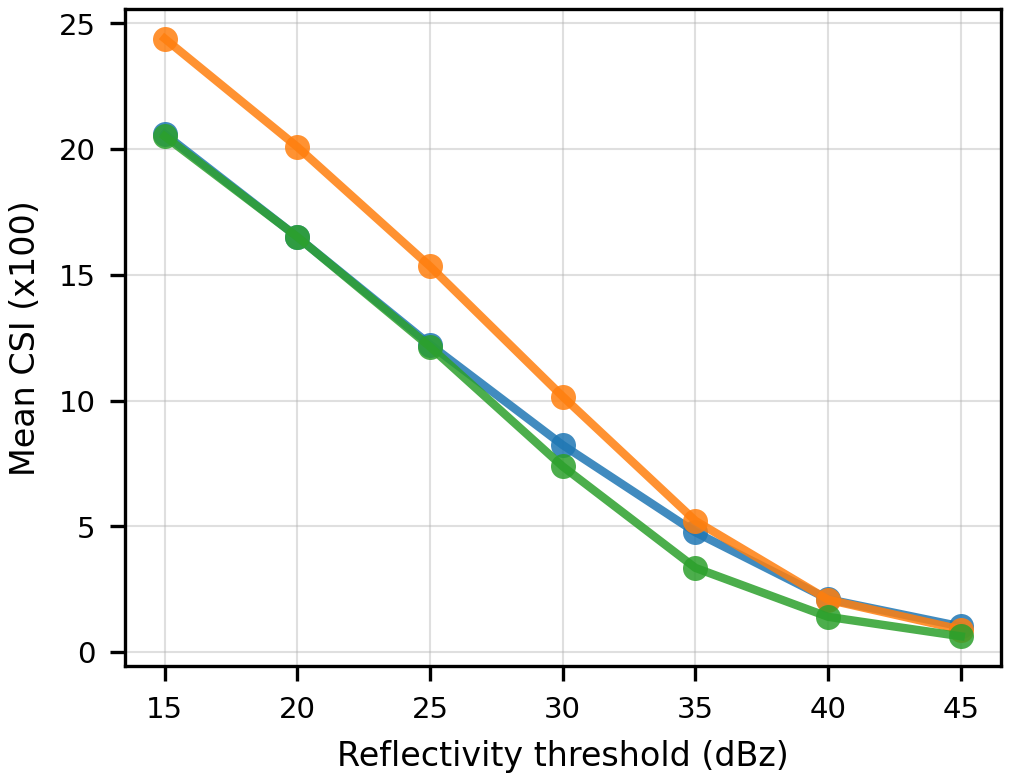}
\caption{18h forecast}
\end{subfigure}
\caption{Composite-reflectivity CSI over the seven-month evaluation period.
Top row: lead-time die-off at 20, 30, and 40~dBZ; bottom row: threshold-dependent
CSI at 6-, 12-, and 18-h leads. The HRRRCast ensemble generally outperforms
operational HRRR, but the two converge at 40--45~dBZ as CSI approaches zero;
HRRRCast member 0 remains slightly lower at these highest thresholds.}
\label{fig:mats_figs_reflectivity}
\end{figure*}


A key distinction from retrospective verification is that the near-real-time archive spans
three native 3-km HRRRCast configurations, with the strongest version operating only during
the final two-month period. Aggregate statistics should therefore be interpreted with
that model evolution in mind. Solid lines denote the full approximately seven-month archive; where
shown, dashed lines denote the final two-month subset. The near-real-time ensemble
used 9 members for most of the archive and 18 members after July 1, 2026. Member 0 uses
unshifted GFS forcing, while the other members use phase-shifted GFS boundary-condition inputs;
see Section~\ref{sec:near_real_time_config} for details.

\begin{figure*}[tbp]
\centering
\begin{subfigure}[t]{0.32\textwidth}
\centering
\includegraphics[width=\textwidth]{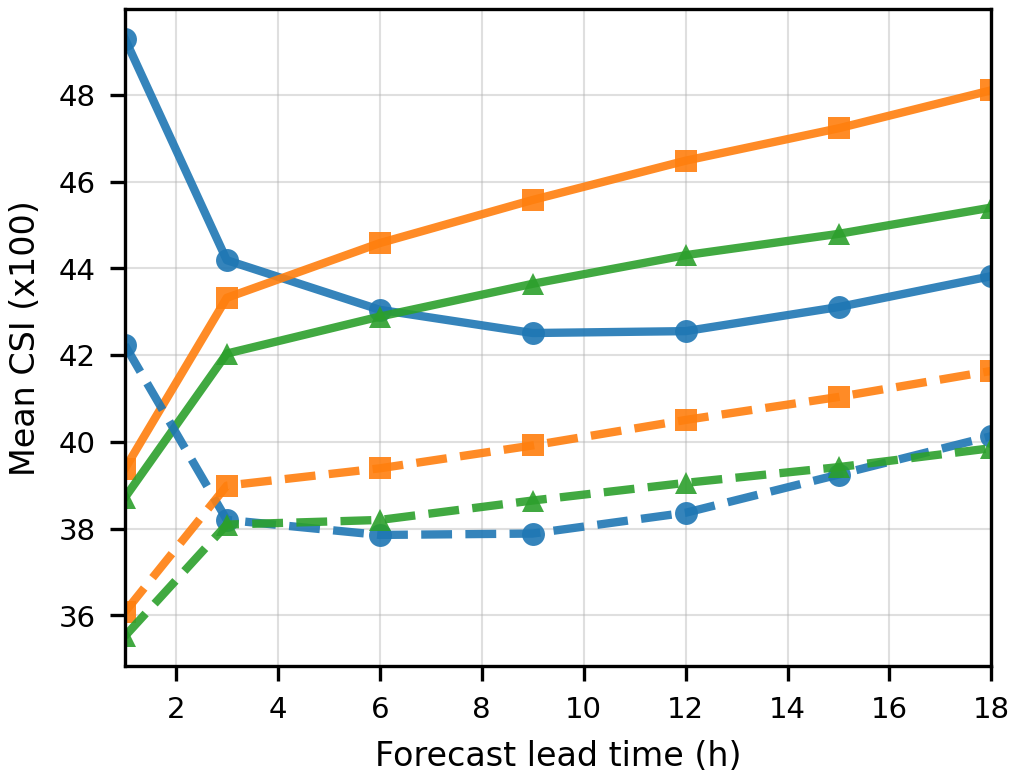}
\caption{CSI at 0.01 in}
\end{subfigure}\hfill
\begin{subfigure}[t]{0.32\textwidth}
\centering
\includegraphics[width=\textwidth]{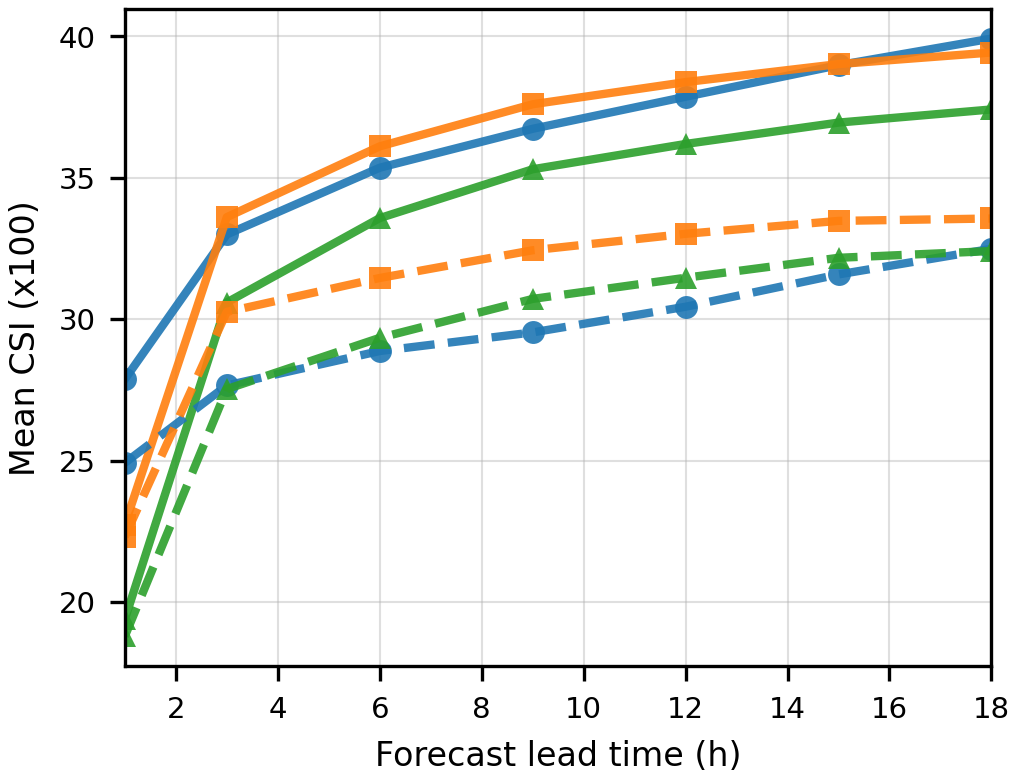}
\caption{CSI at 0.10 in}
\end{subfigure}\hfill
\begin{subfigure}[t]{0.32\textwidth}
\centering
\includegraphics[width=\textwidth]{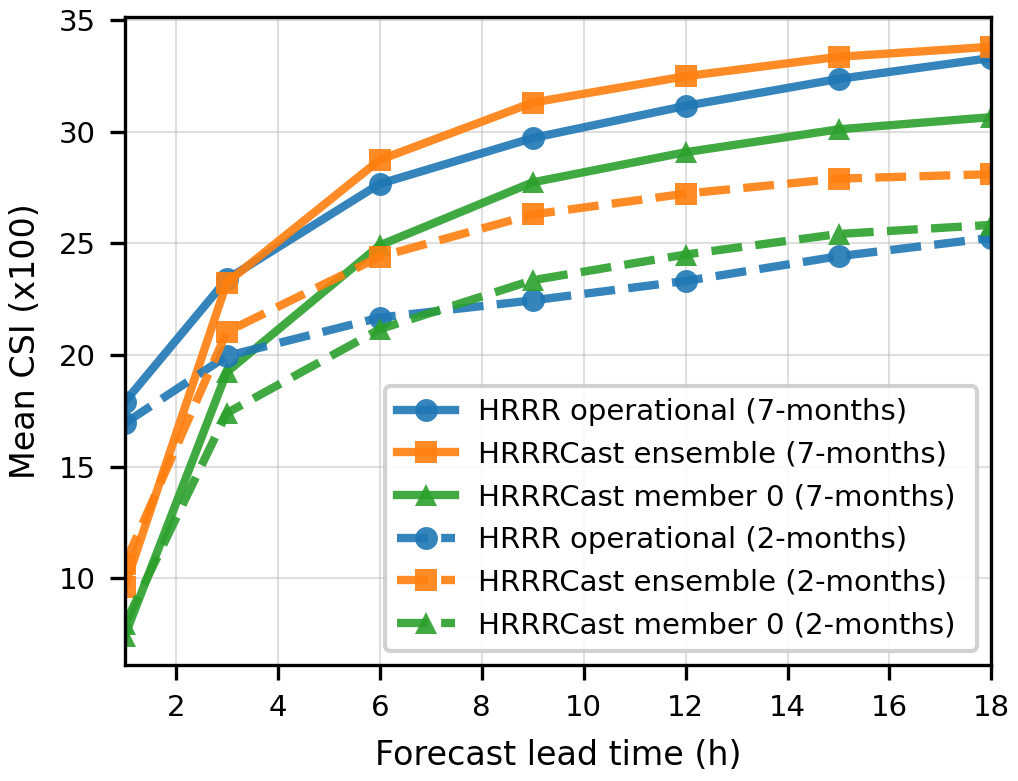}
\caption{CSI at 0.25 in}
\end{subfigure}\hfill
\begin{subfigure}[t]{0.32\textwidth}
\centering
\includegraphics[width=\textwidth]{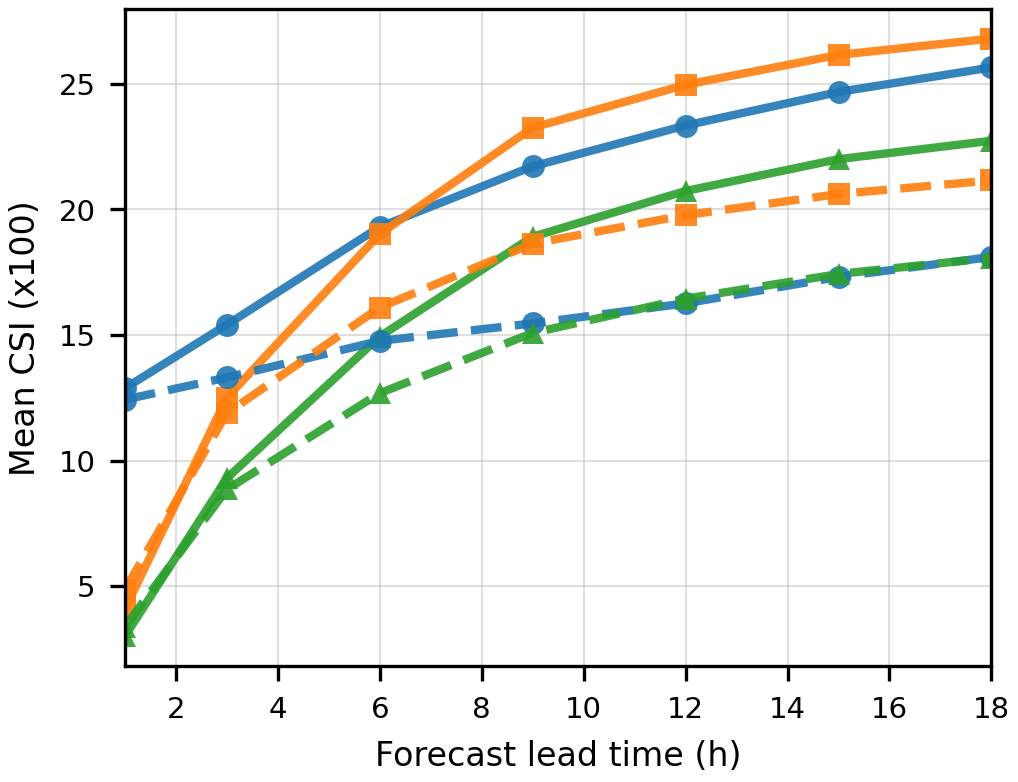}
\caption{CSI at 0.50 in}
\end{subfigure}\hfill
\begin{subfigure}[t]{0.32\textwidth}
\centering
\includegraphics[width=\textwidth]{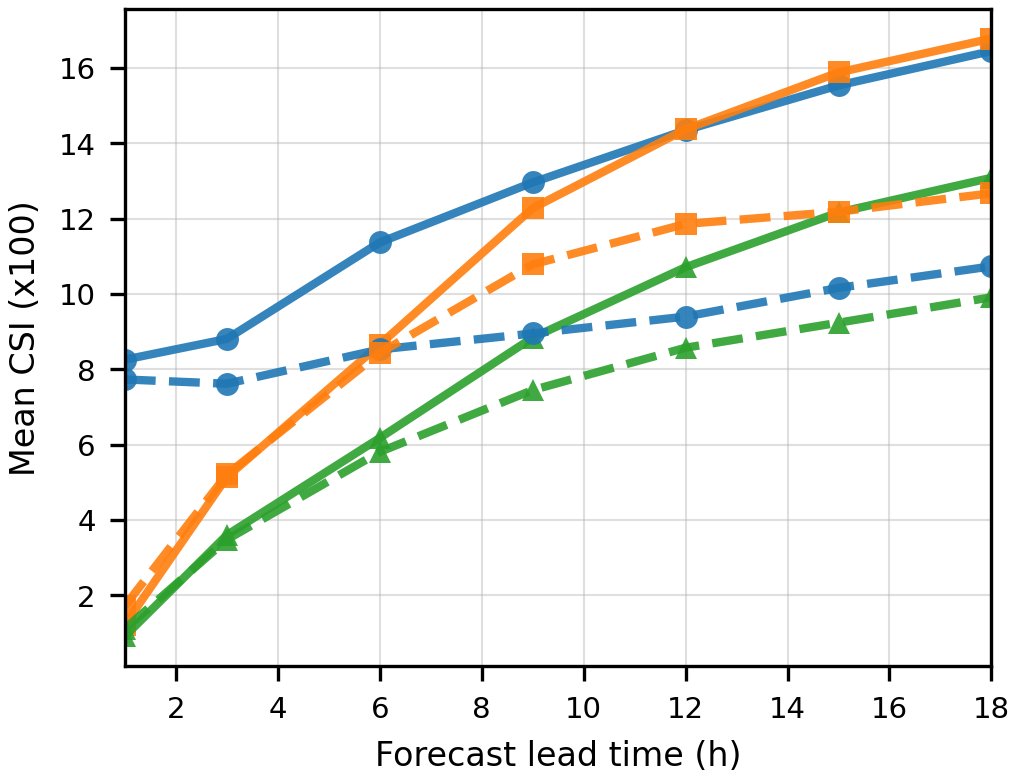}
\caption{CSI at 1.00 in}
\end{subfigure}
\begin{subfigure}[t]{0.32\textwidth}
\centering
\includegraphics[width=\textwidth]{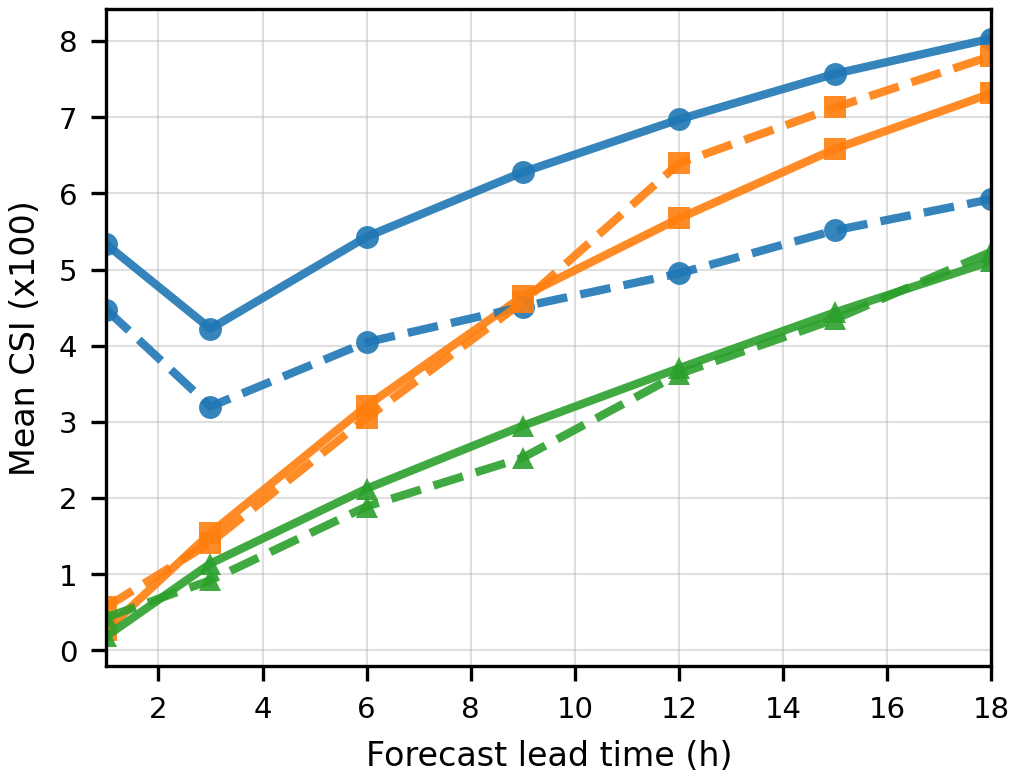}
\caption{CSI at 2.00 in}
\end{subfigure}

\caption{Lead-time die-off of hourly-precipitation CSI across thresholds.
Each panel shows CSI versus forecast lead for a different precipitation threshold.
The final two-month subset (dotted lines) shows the strongest HRRRCast gains.}
\label{fig:mats_figs_precip_dieoff}
\end{figure*}

\begin{figure*}[btp]
\centering
\textbf{Full seven-month archive}\\[0.2em]

\begin{subfigure}[t]{0.32\textwidth}
\centering
\includegraphics[width=\textwidth]{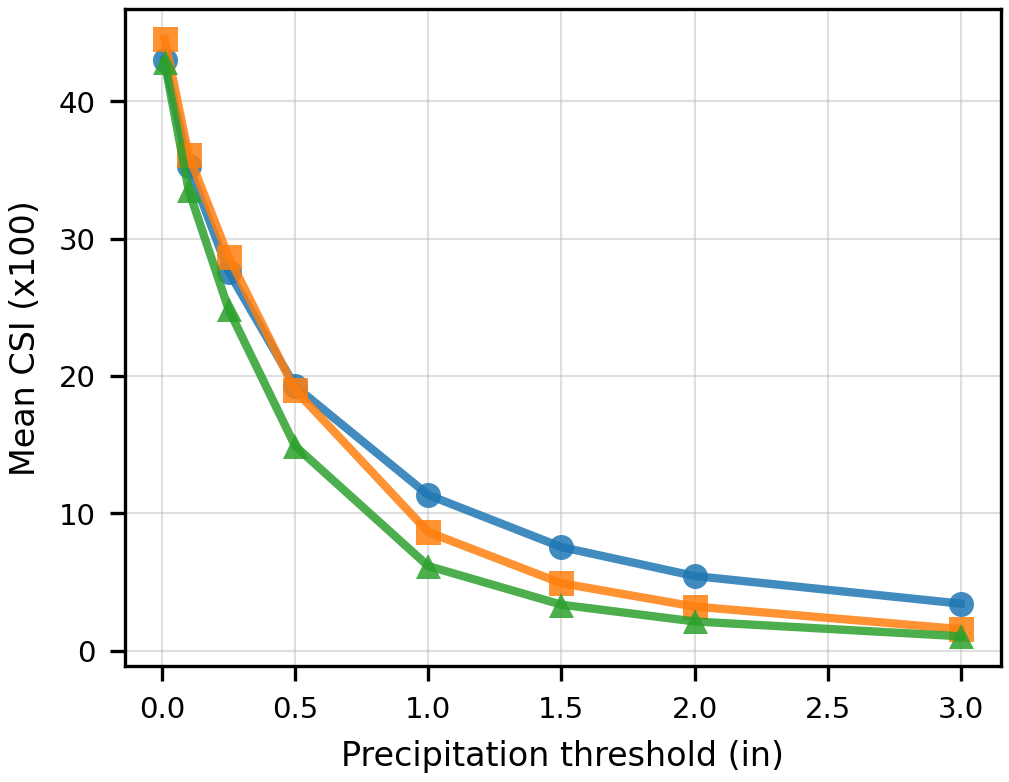}
\caption{6h forecast}
\end{subfigure}
\begin{subfigure}[t]{0.32\textwidth}
\centering
\includegraphics[width=\textwidth]{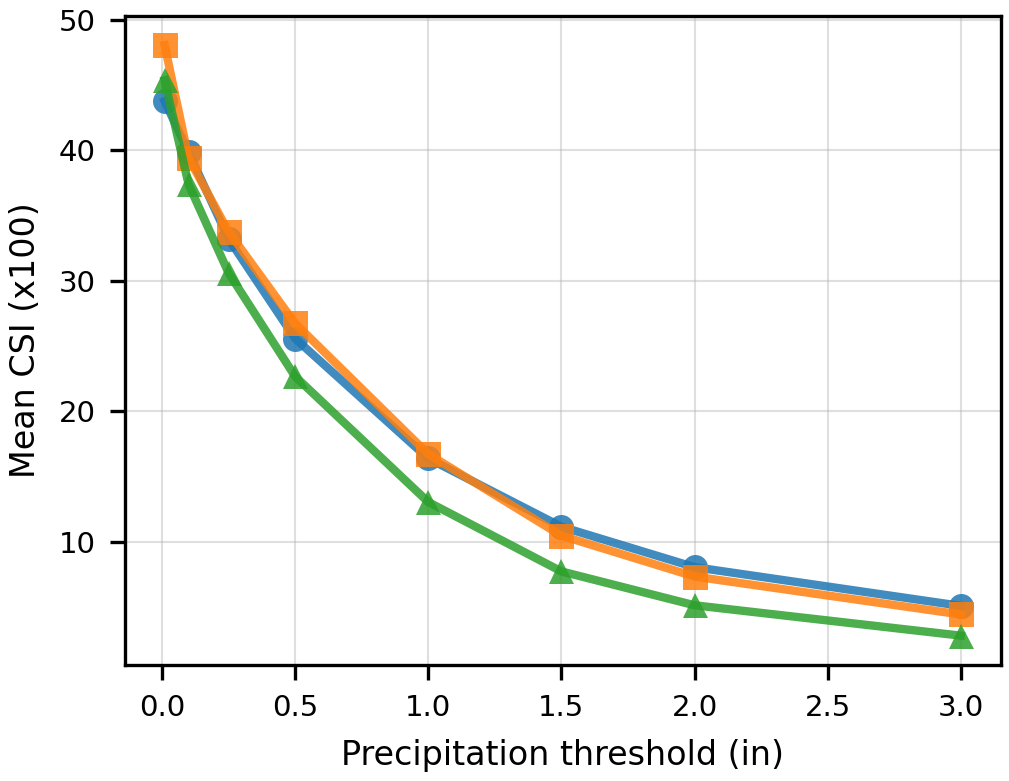}
\caption{12h forecast}
\end{subfigure}
\begin{subfigure}[t]{0.32\textwidth}
\centering
\includegraphics[width=\textwidth]{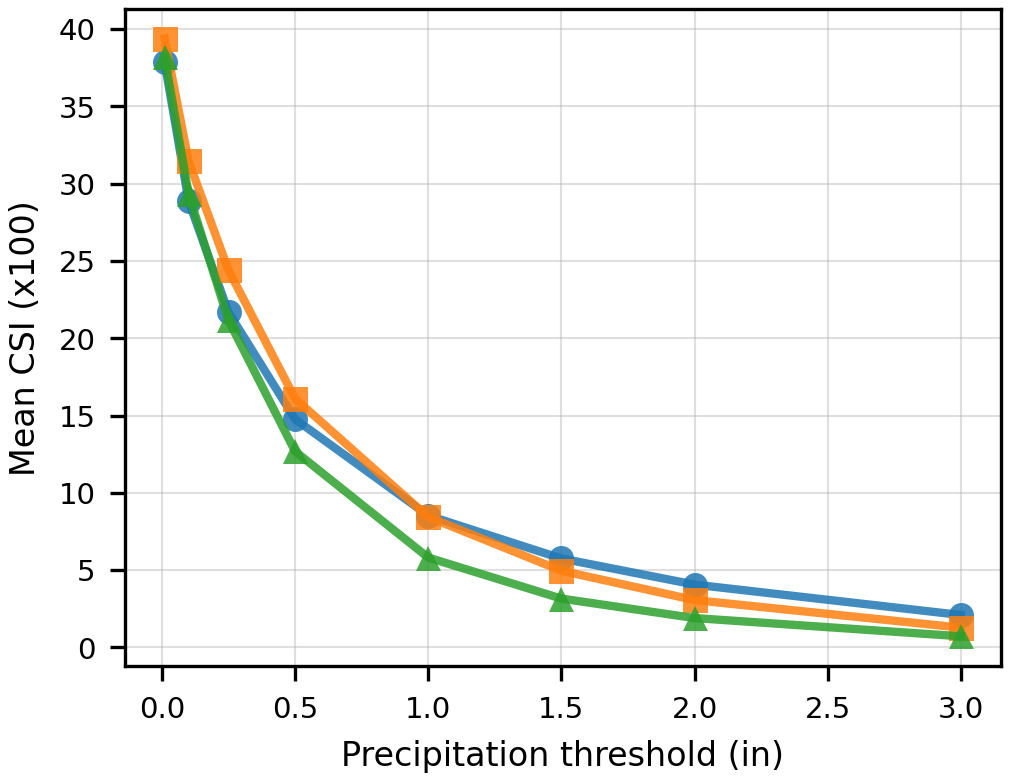}
\caption{18h forecast}
\end{subfigure}

\vspace{0.2em}

\textbf{Final two-month subset}\\[0.2em]

\begin{subfigure}[t]{0.32\textwidth}
\centering
\includegraphics[width=\textwidth]{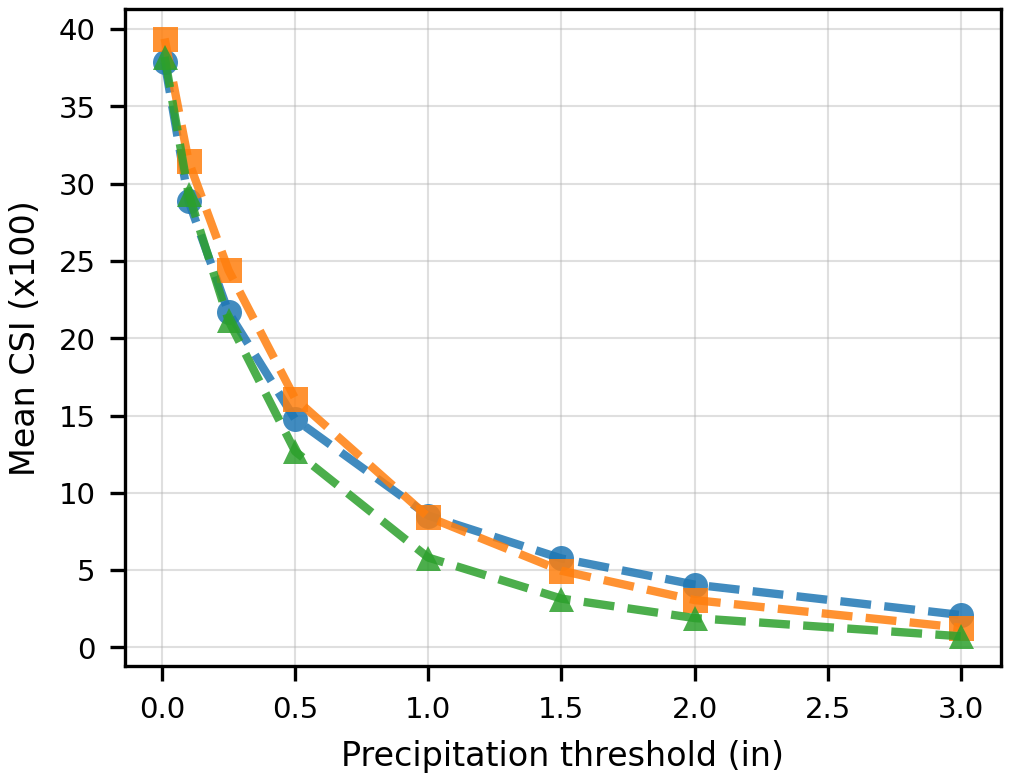}
\caption{6h forecast}
\end{subfigure}
\begin{subfigure}[t]{0.32\textwidth}
\centering
\includegraphics[width=\textwidth]{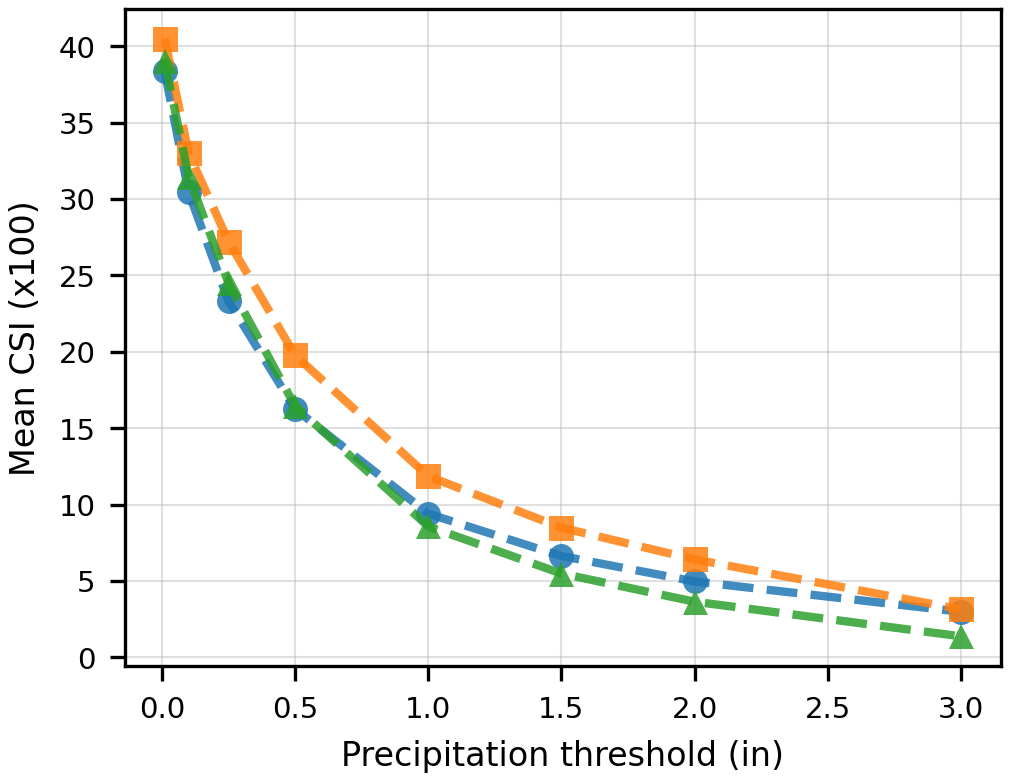}
\caption{12h forecast}
\end{subfigure}
\begin{subfigure}[t]{0.32\textwidth}
\centering
\includegraphics[width=\textwidth]{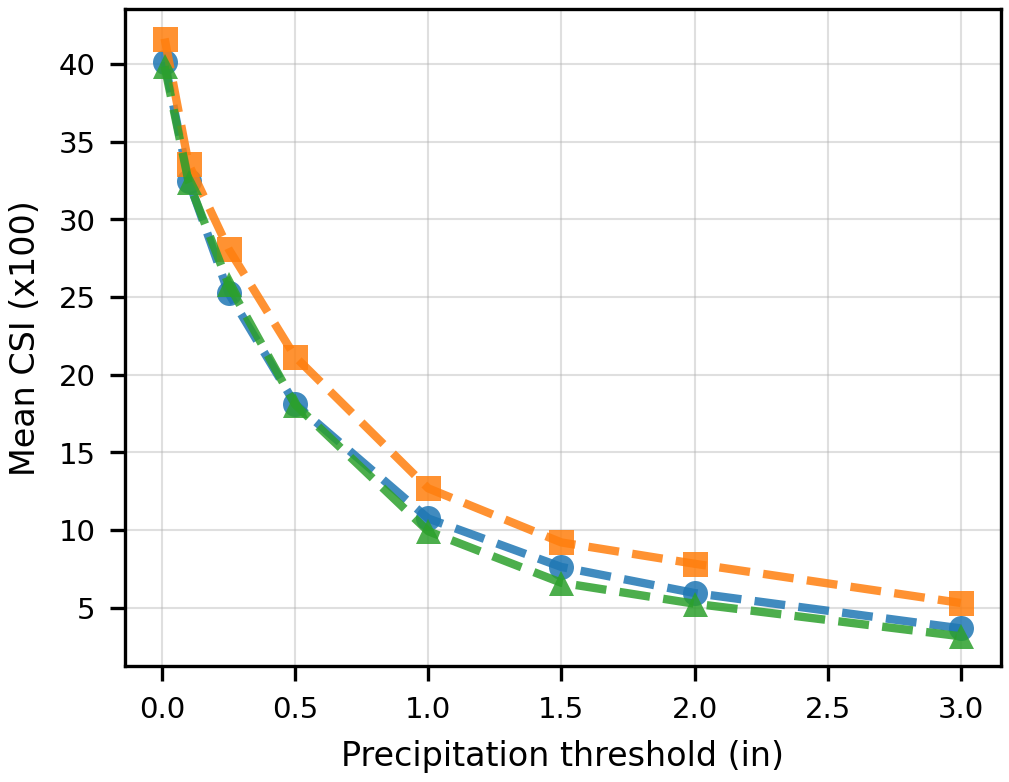}
\caption{18h forecast}
\end{subfigure}

\caption{Threshold-dependent hourly-precipitation CSI for 6-, 12-, and 18-h forecasts.
The top row shows the full seven-month archive; the second row shows the mature
recent subset. In the latest configuration, HRRRCast shows a clear improvement in
precipitation skill up to about 3 in/h, and the performance advantage becomes
more pronounced at 12- and 18-h lead times.}
\label{fig:mats_figs_precip_threshold}
\end{figure*}

Composite-reflectivity CSI in the lead-time die-off summaries depends on threshold
(Fig.~\ref{fig:mats_figs_reflectivity}a--c). At all reflectivity thresholds, the HRRRCast
ensemble mean (orange) remains consistently higher than operational HRRR across most lead times,
with the expected decline in CSI as forecast range increases. For deterministic single-member
comparisons, HRRRCast member 0 is higher than HRRR at 20 dBZ, approximately equal at 30 dBZ,
and slightly lower at 40 dBZ. This threshold dependence is consistent with greater difficulty at
the most intense convective cores. The fixed-lead threshold curves
(Figs.~\ref{fig:mats_figs_reflectivity}d--f) further show that the single-member
crossover threshold shifts lower with lead time. The ensemble mean, by contrast, 
retains a clear advantage over HRRR through 40~dBZ across all three lead times, with 
the three curves converging as skill approaches zero at 45~dBZ, suggesting that ensemble 
pooling extends meaningful CSI improvements into the intense-convection regime even as the 
deterministic advantage narrows. It is worth noting that the CSI diagnostics here use strict 
grid-to-grid matching rather than neighborhood-based verification and are therefore more 
sensitive to small displacement errors.

Hourly precipitation CSI in the lead-time die-off summaries likewise reveals
threshold-dependent behavior (Figs.~\ref{fig:mats_figs_precip_dieoff} and
\ref{fig:mats_figs_precip_threshold}). At 0.01 and 0.10 in, the HRRRCast ensemble (orange)
is generally better than HRRR operational across most leads in both the full seven-month
archive and the mature subset, and HRRRCast member 0 also becomes competitive after the
first few forecast hours. At 0.25 and 0.50 in, the ensemble maintains a clear advantage
over HRRR through much of the forecast in both archives, with the mature subset showing
that the best native 3-km HRRRCast configuration strengthens that advantage further.
At 1.00 in, the distinction between the two archives is especially important: in the full
seven-month aggregate, the ensemble is near parity to modestly better than HRRR, whereas in
the mature subset the HRRRCast ensemble is clearly better than HRRR across all plotted leads.
HRRRCast member 0 remains below the ensemble mean at this threshold, but in the mature
subset it is still competitive even at 1.00 in, especially at longer leads.

The additional fixed-lead threshold summaries (6-, 12-, and 18-h forecasts) reinforce this picture.
Across these panels, the mature-subset curves show clearer HRRRCast improvement than
the full seven-month aggregate, with the ensemble generally above HRRR across much of the
threshold range and with the strongest separation at 12 and 18 h. The full-period curves
are more mixed at some thresholds, especially at 6 h, but still indicate broadly
competitive precipitation CSI and useful discrimination into the more extreme tail.

\begin{figure*}[h]
\centering
\begin{subfigure}[t]{0.32\textwidth}
\centering
\includegraphics[width=\textwidth]{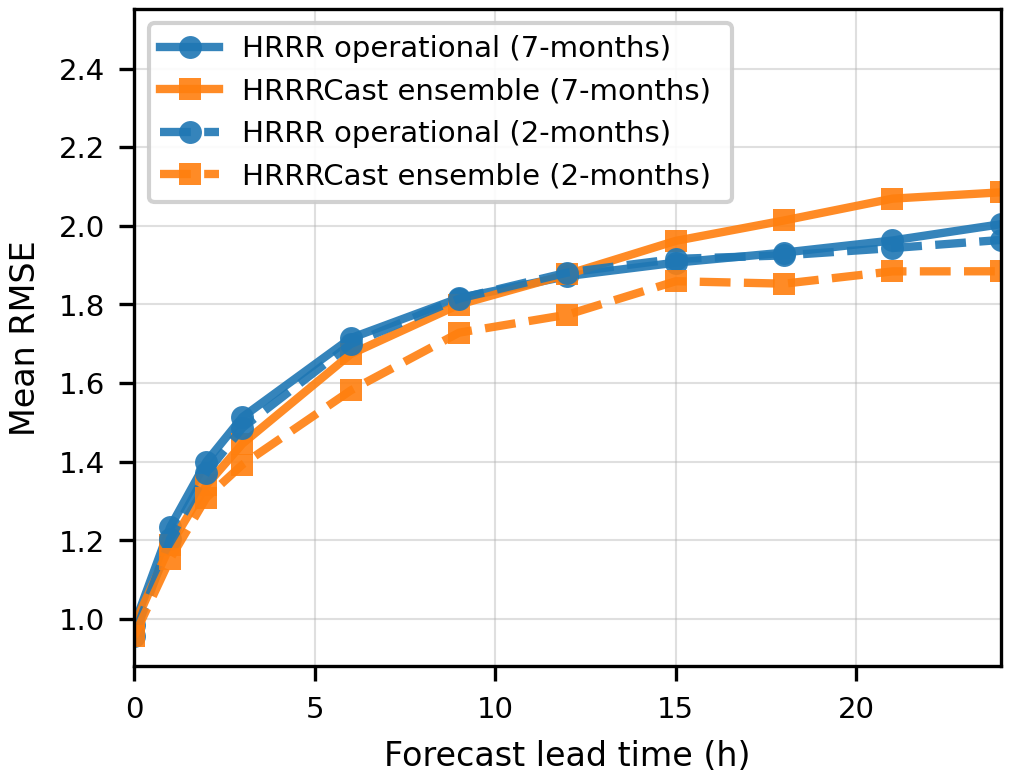}
\caption{2-m temperature RMSE (°C)}
\end{subfigure}\hfill
\begin{subfigure}[t]{0.32\textwidth}
\centering
\includegraphics[width=\textwidth]{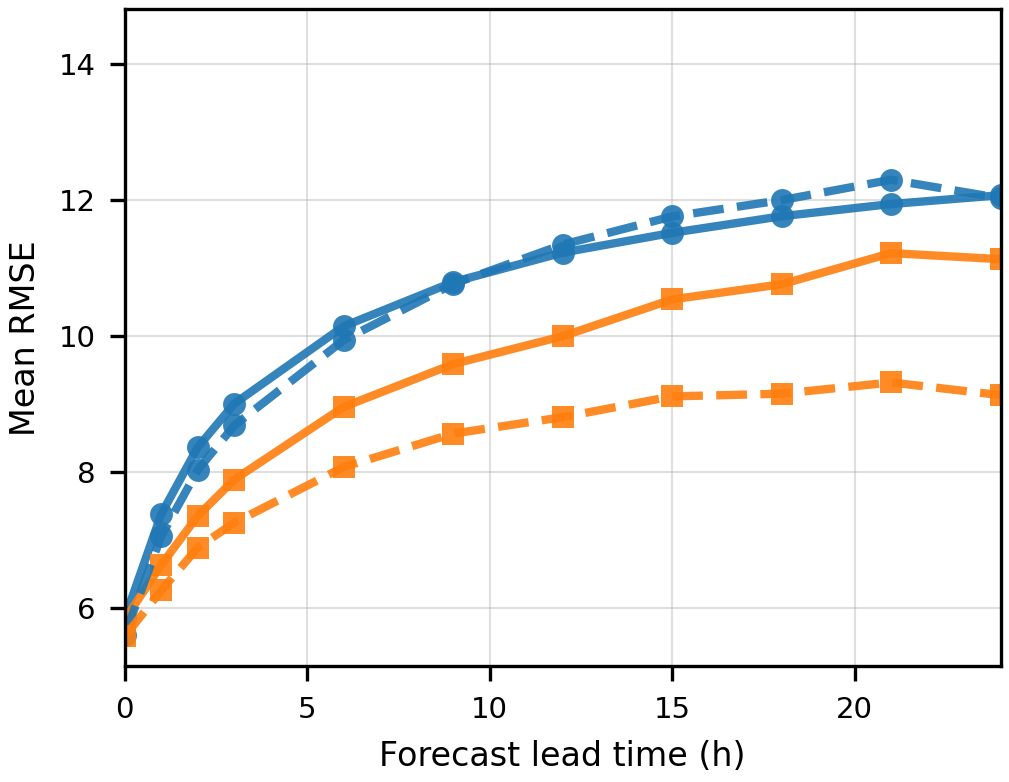}
\caption{2-m relative humidity RMSE (\%)}
\end{subfigure}\hfill
\begin{subfigure}[t]{0.32\textwidth}
\centering
\includegraphics[width=\textwidth]{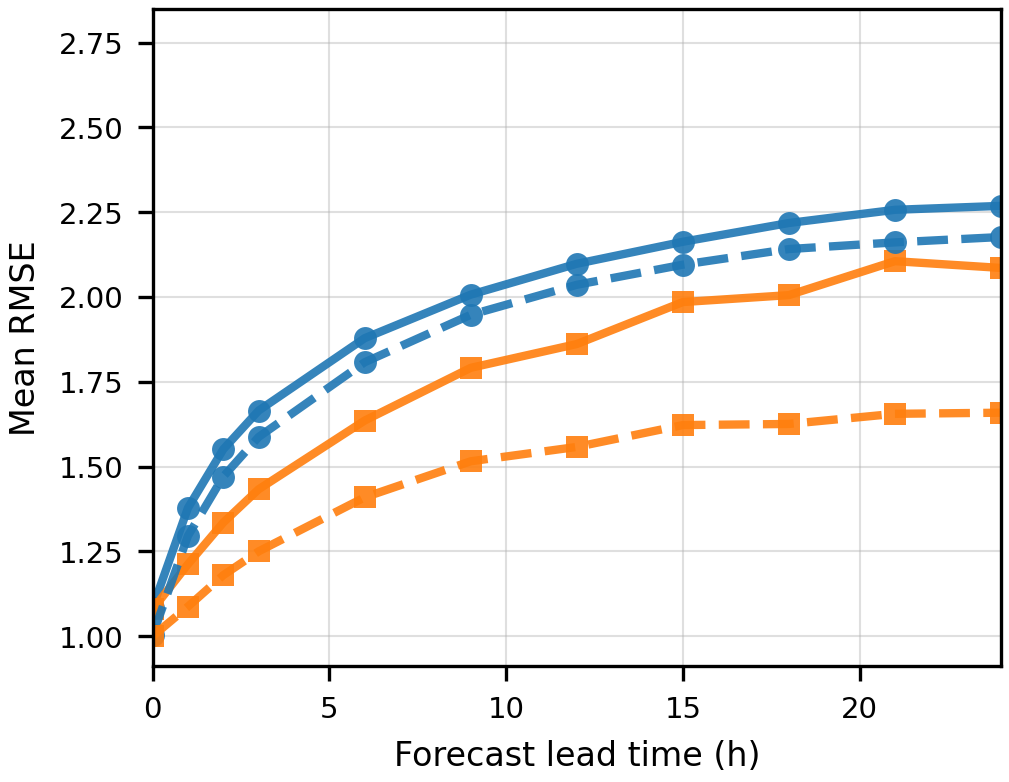}
\caption{2-m dewpoint RMSE (°C)}
\end{subfigure}

\vspace{0.6em}

\begin{subfigure}[t]{0.32\textwidth}
\centering
\includegraphics[width=\textwidth]{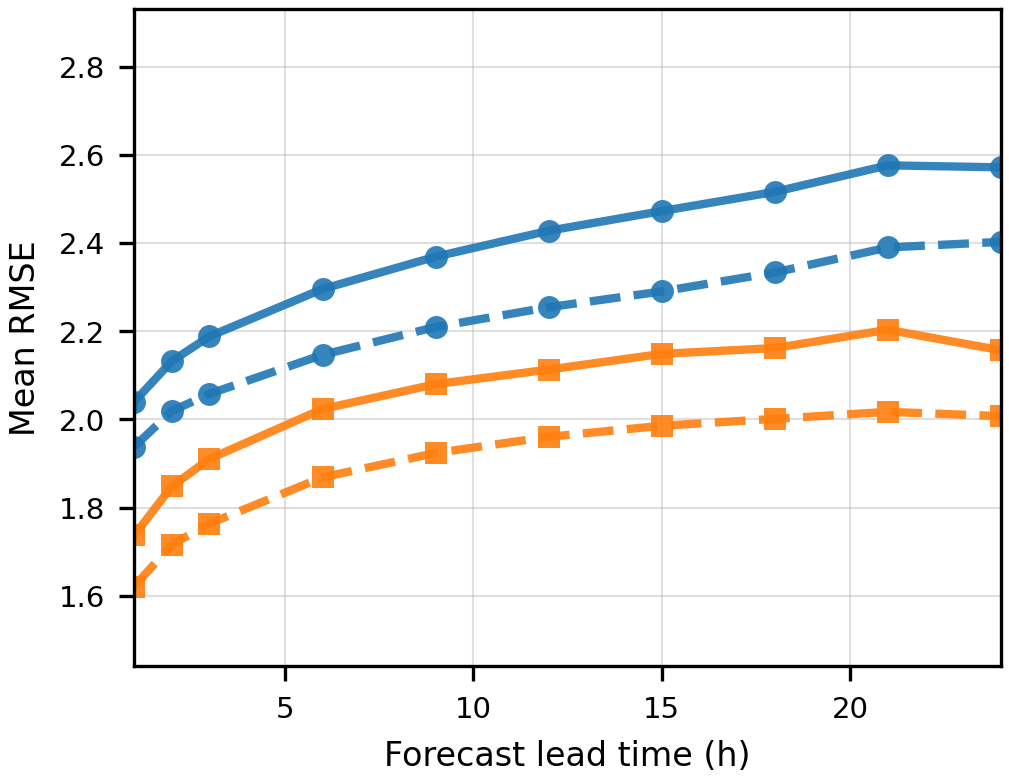}
\caption{10-m wind speed RMSE (m s$^{-1}$)}
\end{subfigure}\hfill
\begin{subfigure}[t]{0.32\textwidth}
\centering
\includegraphics[width=\textwidth]{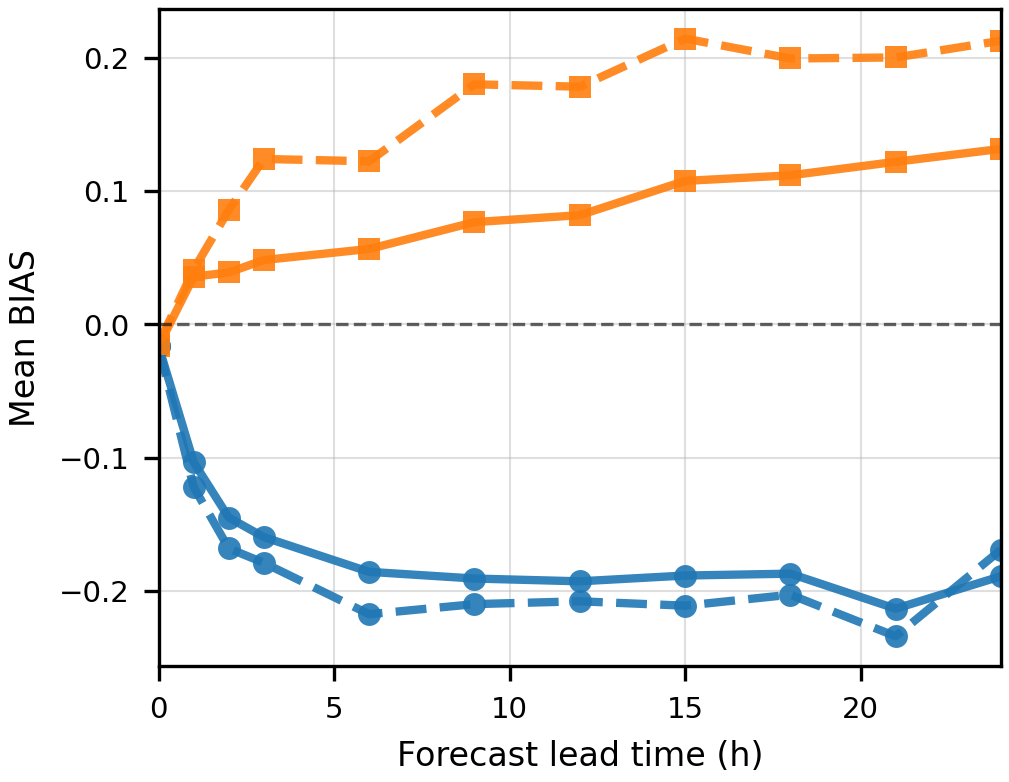}
\caption{10-m zonal wind bias (m s$^{-1}$)}
\end{subfigure}\hfill
\begin{subfigure}[t]{0.32\textwidth}
\centering
\includegraphics[width=\textwidth]{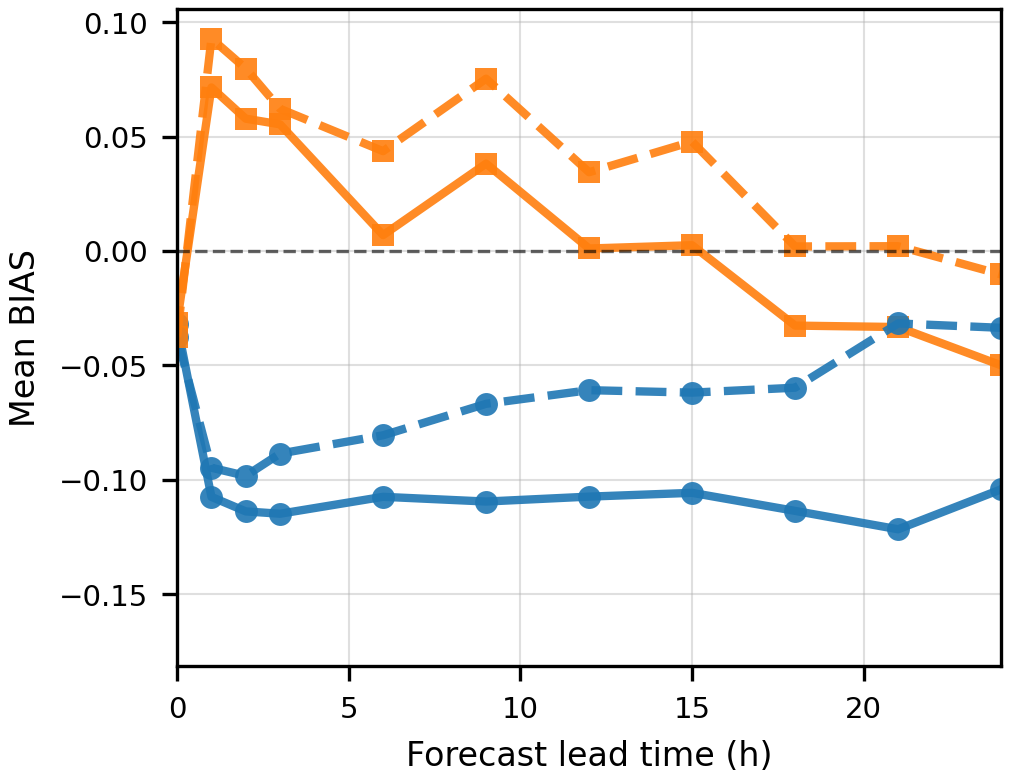}
\caption{10-m meridional wind bias (m s$^{-1}$)}
\end{subfigure}
\caption{Lead-time die-off verification for near-surface variables. Solid lines denote the full 
approximately seven-month evaluation window, and dashed lines denote the final two-month subset.}
\label{fig:mats_figs_surface}
\end{figure*}

For near-surface thermodynamic and wind RMSE (Fig.~\ref{fig:mats_figs_surface}a--d),
HRRRCast shows robust skill gains for relative humidity, dewpoint, and 10-m wind speed across most
lead times. Relative humidity RMSE is consistently lower for HRRRCast, and the two-month
dashed subset indicates that the mature 3-km configuration further widens that advantage. Dewpoint RMSE
shows a similar lead-time evolution, with HRRRCast maintaining lower error from the shortest leads
through the day-1 range. Wind-speed RMSE improvements are likewise systematic, including the longer
lead times where aggregate errors are largest.
For 2-m temperature RMSE, the full-period aggregate is more mixed: HRRRCast is near parity to
slightly better at several shorter and intermediate leads, while HRRR operational remains slightly
lower at parts of the curve. Over the final month, however, HRRRCast has lower RMSE across all four
near-surface variables, including 2-m temperature, suggesting that the mixed full-period
temperature result is influenced by earlier near-real-time configurations rather than by the
mature model alone.

Vector wind bias diagnostics (Fig.~\ref{fig:mats_figs_surface}e--f) indicate
that HRRRCast substantially reduces the strong negative zonal-wind bias present in HRRR
operational across most lead times in both the full-period and final-month summaries. For
meridional wind, both systems stay closer to zero than in the zonal component, but HRRRCast
generally remains less negatively biased through most of the forecast.

Together, these MATS results show that HRRRCast maintains clear reflectivity CSI gains at all
thresholds, provides broadly competitive to improved precipitation CSI, and improves the
background near-surface state (humidity, dewpoint, temperature, wind speed),
especially in the mature final-month configuration.

\begin{figure*}[tbp]
\centering
\setlength{\tabcolsep}{2pt}
\renewcommand{\arraystretch}{1.1}
\begin{tabular}{@{}ccc@{}}
\includegraphics[height=0.138\textwidth]{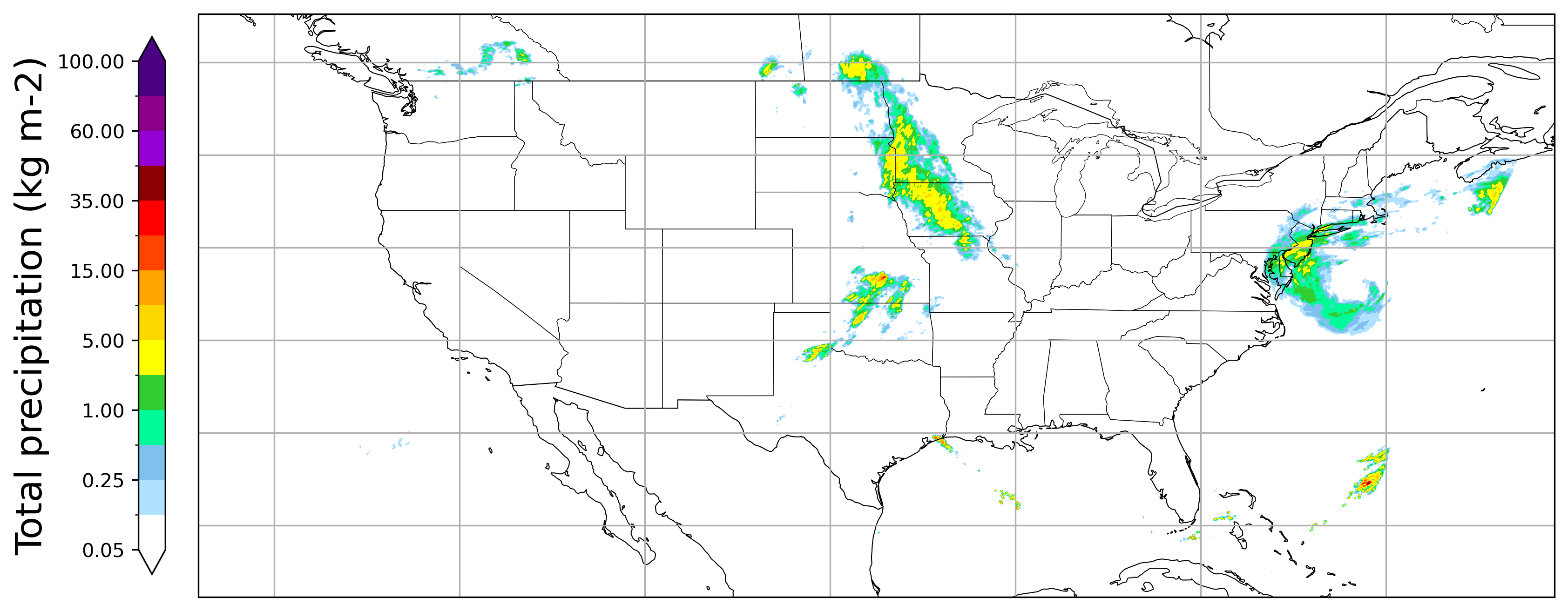} &
\includegraphics[height=0.138\textwidth]{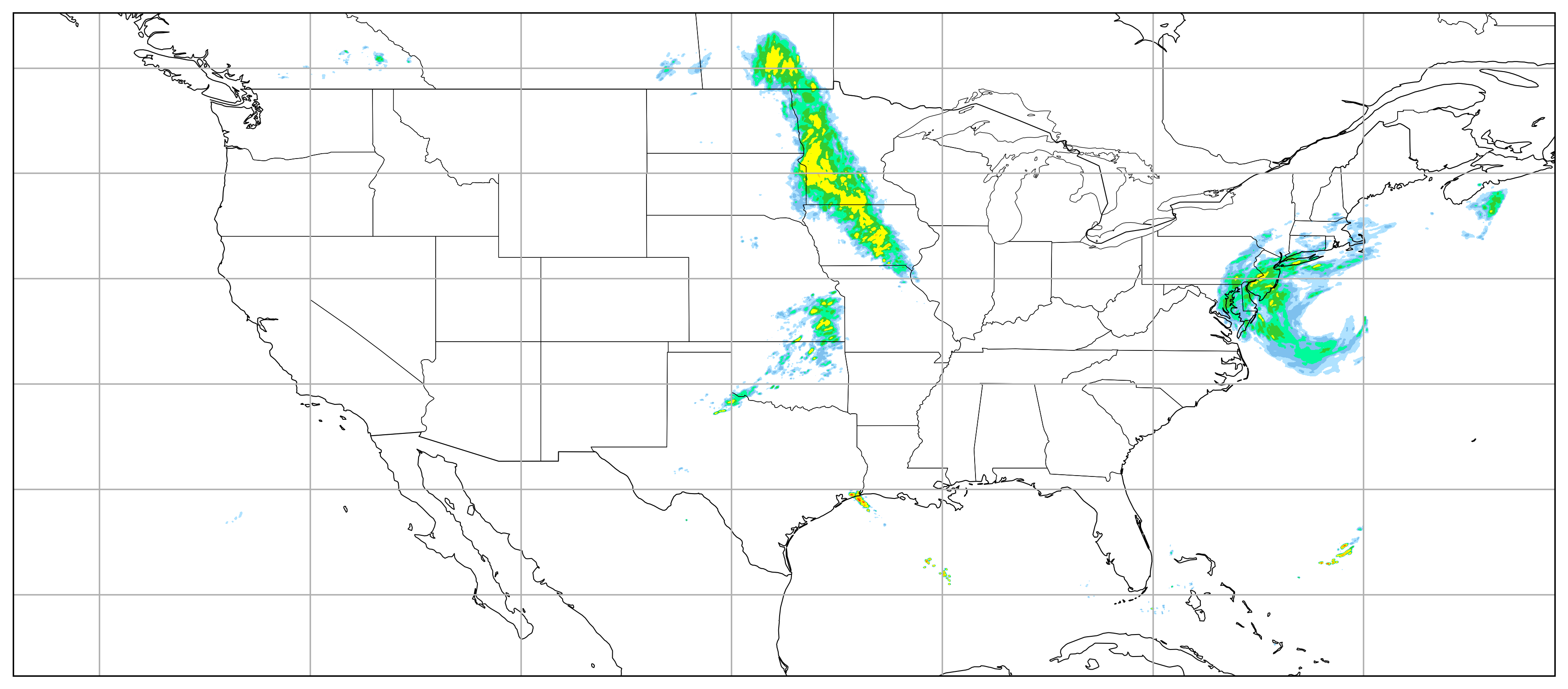} &
\includegraphics[height=0.138\textwidth]{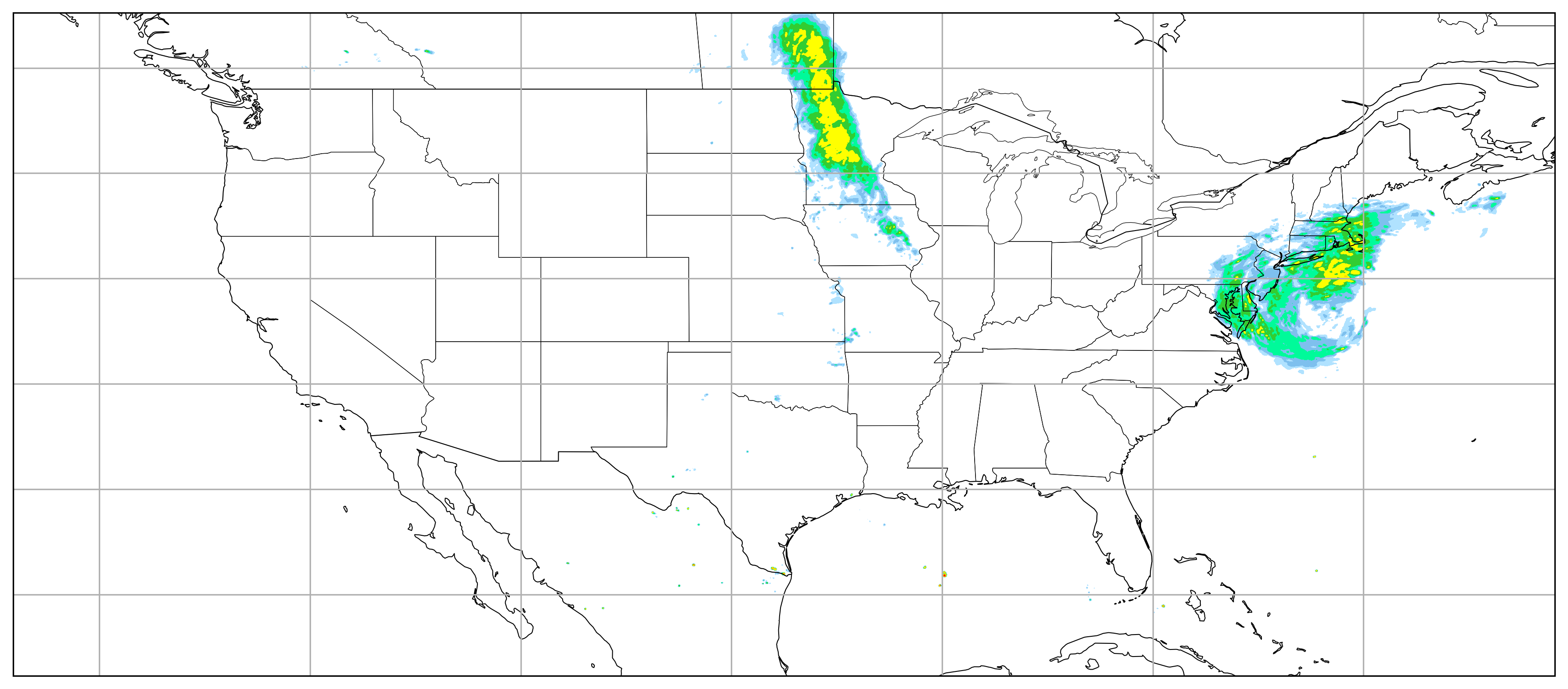} \\
\includegraphics[height=0.138\textwidth]{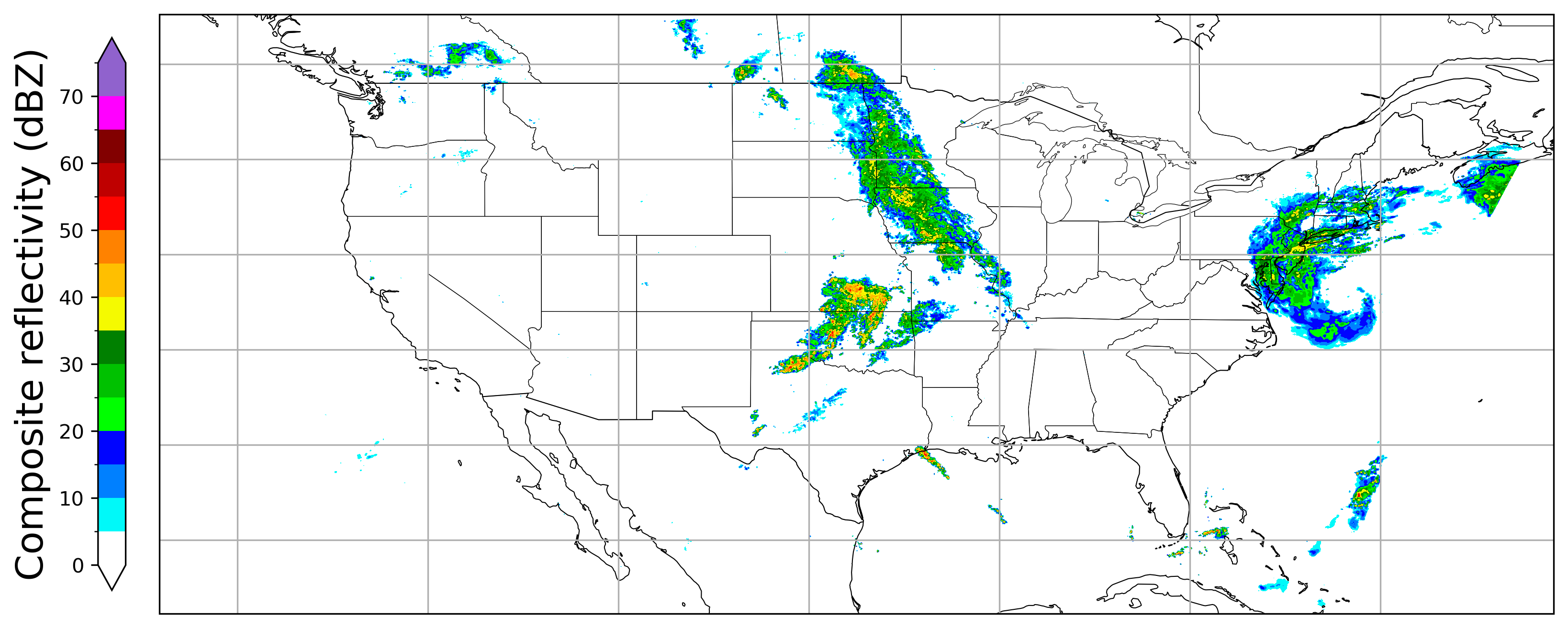} &
\includegraphics[height=0.138\textwidth]{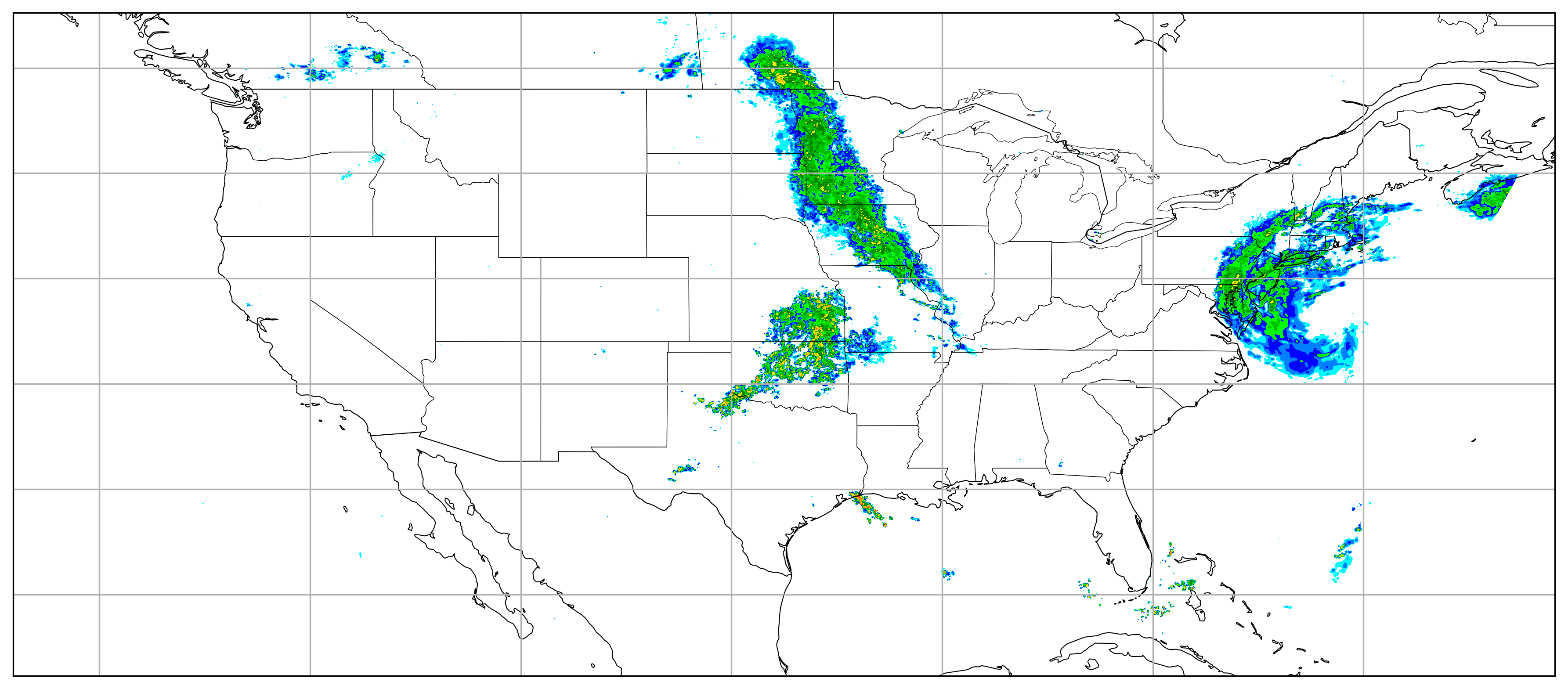} &
\includegraphics[height=0.138\textwidth]{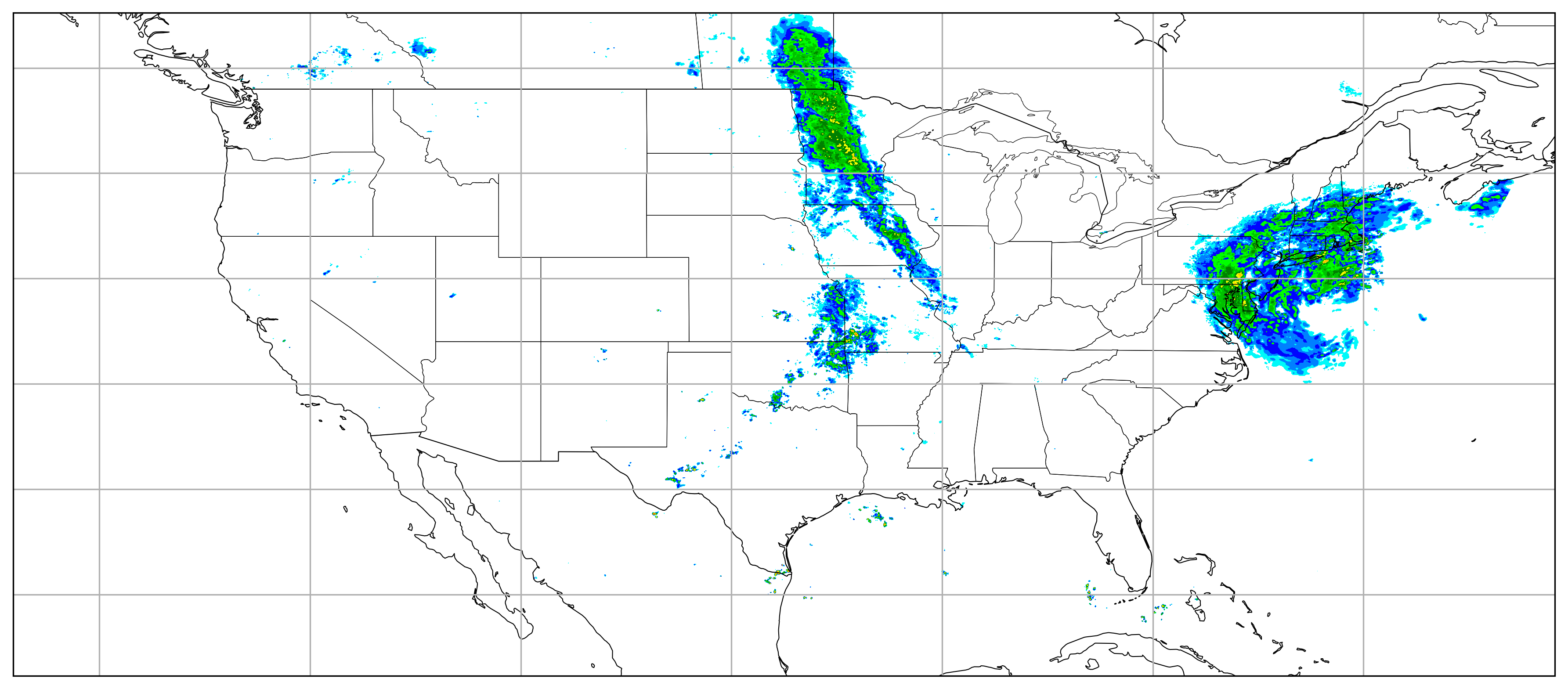} \\
\includegraphics[height=0.138\textwidth]{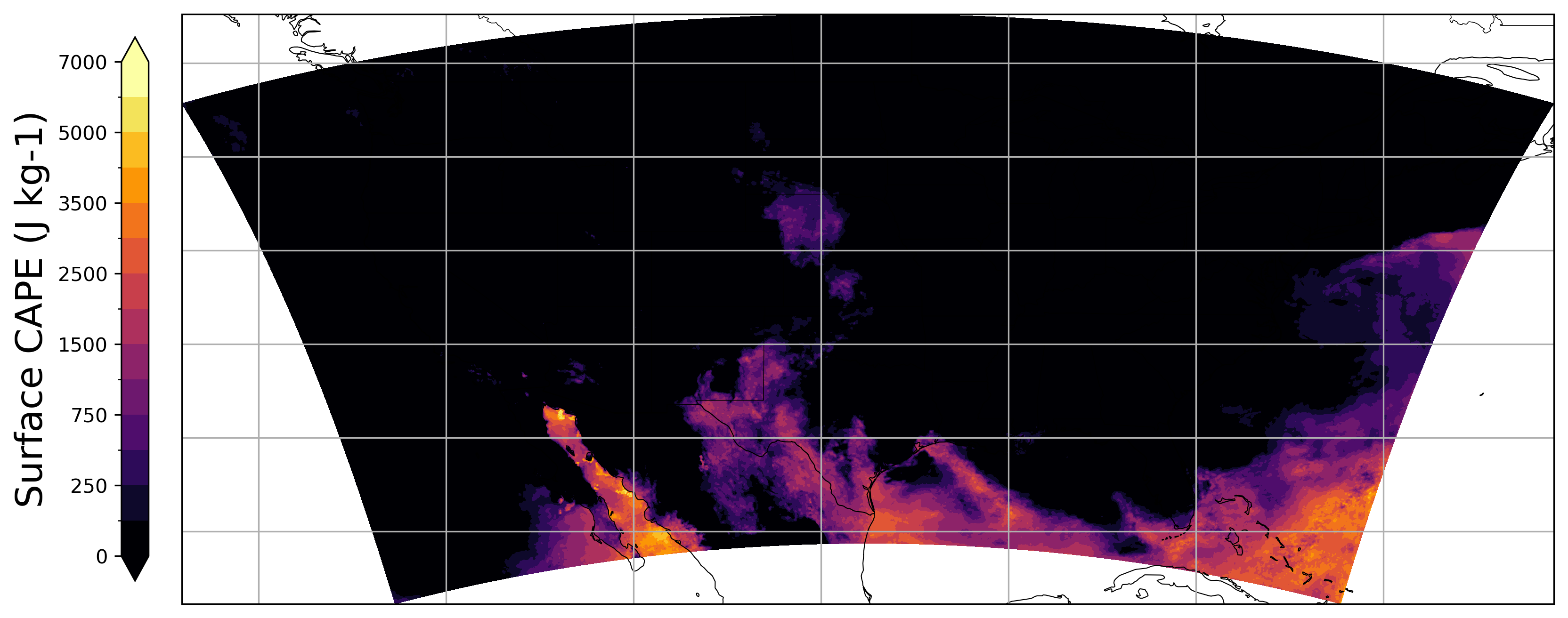} &
\includegraphics[height=0.138\textwidth]{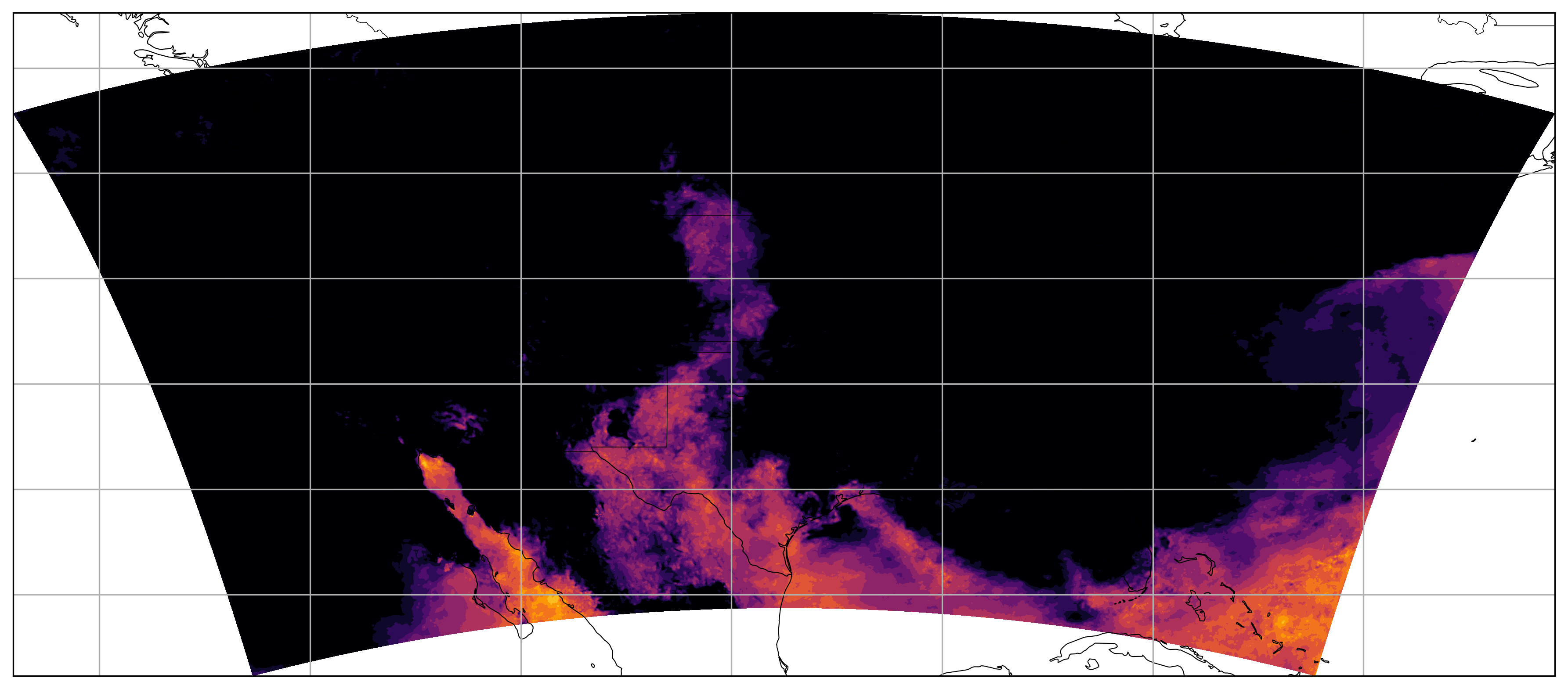} &
\includegraphics[height=0.138\textwidth]{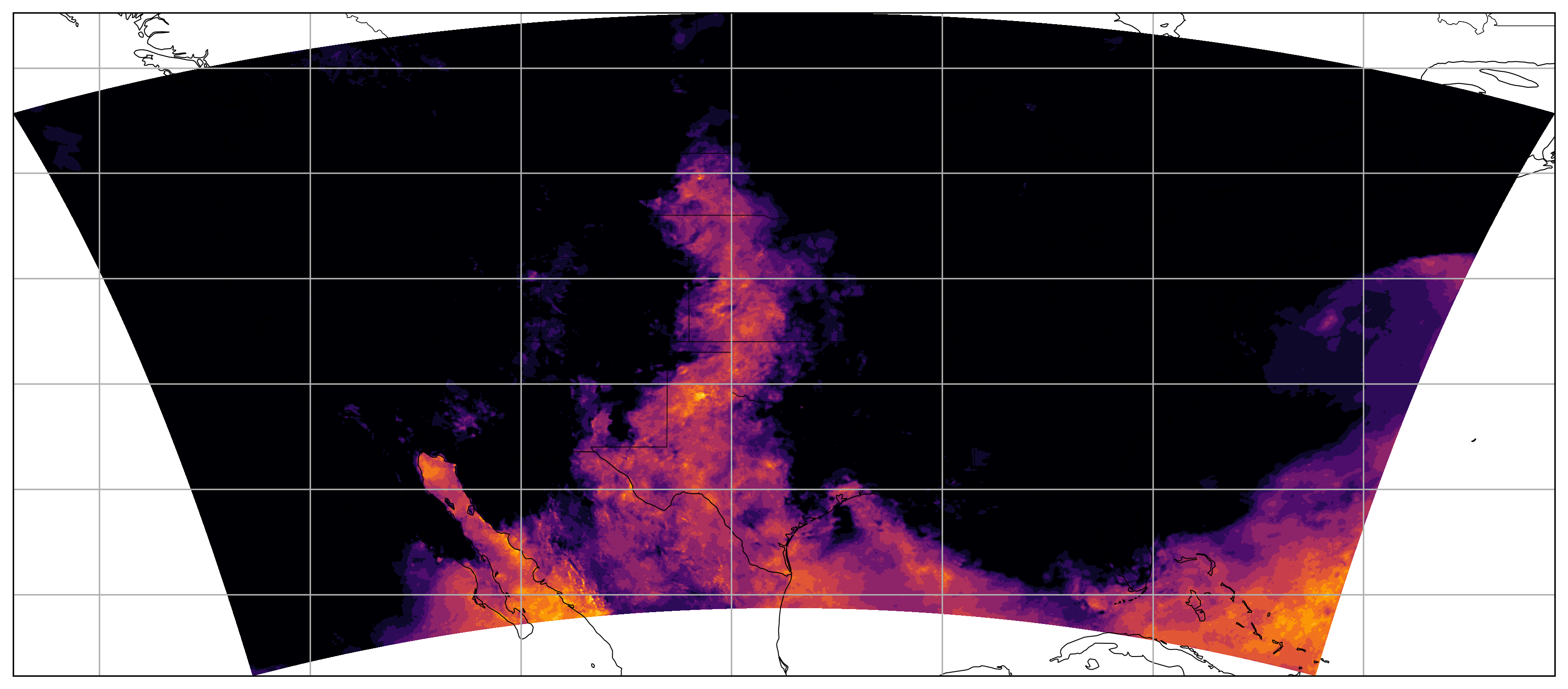} \\
\includegraphics[height=0.138\textwidth]{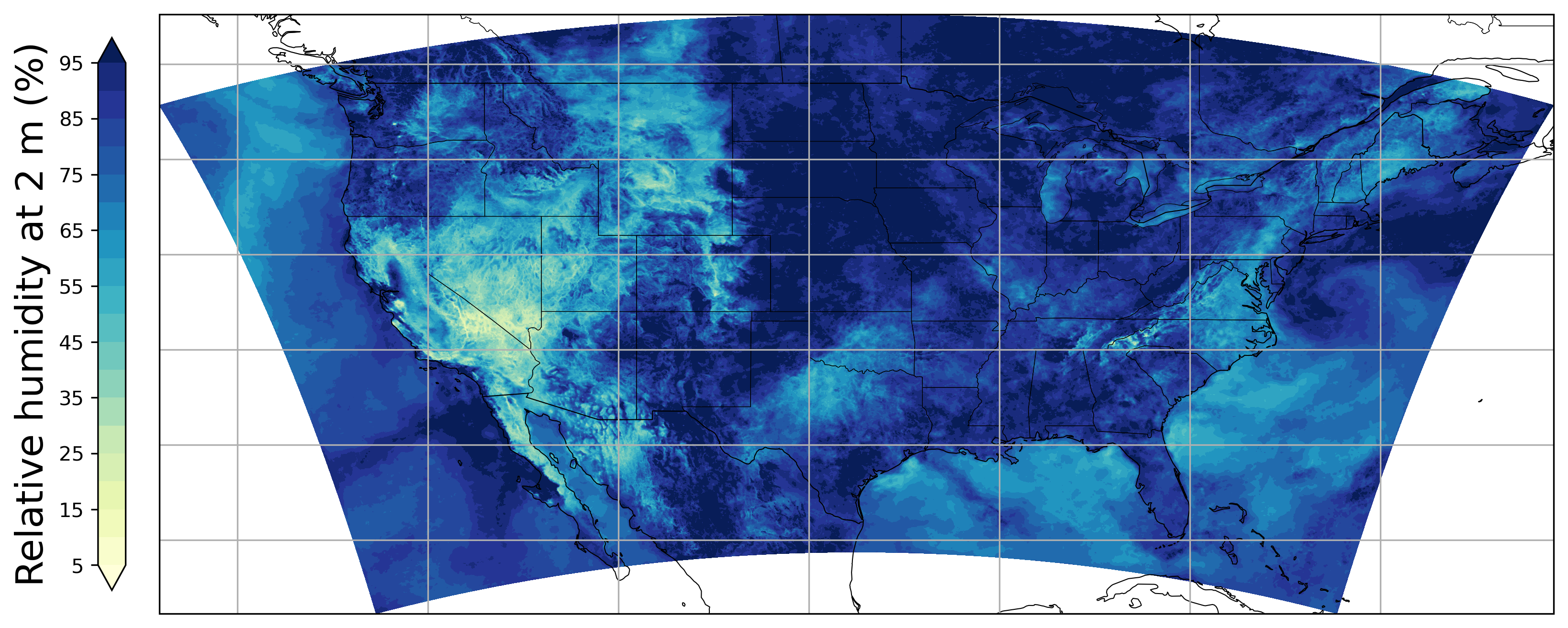} &
\includegraphics[height=0.138\textwidth]{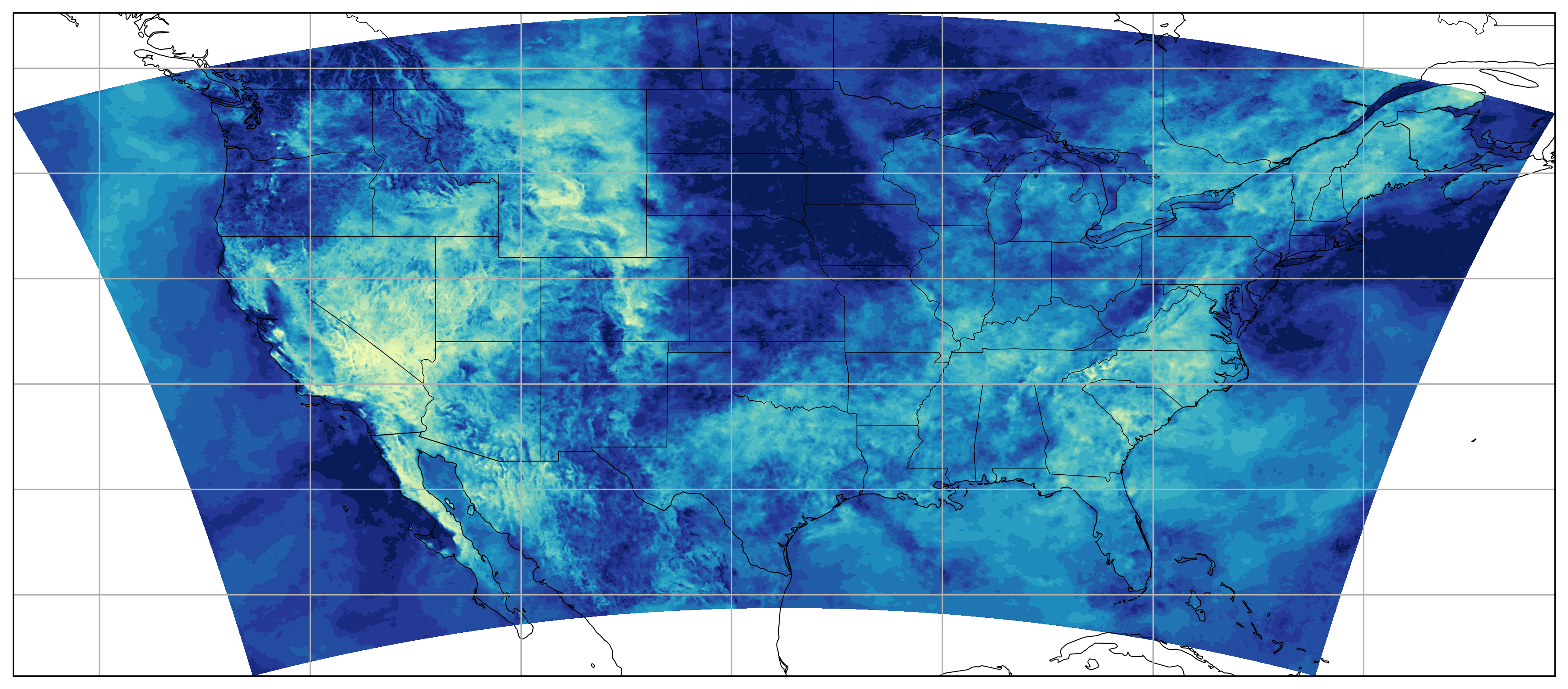} &
\includegraphics[height=0.138\textwidth]{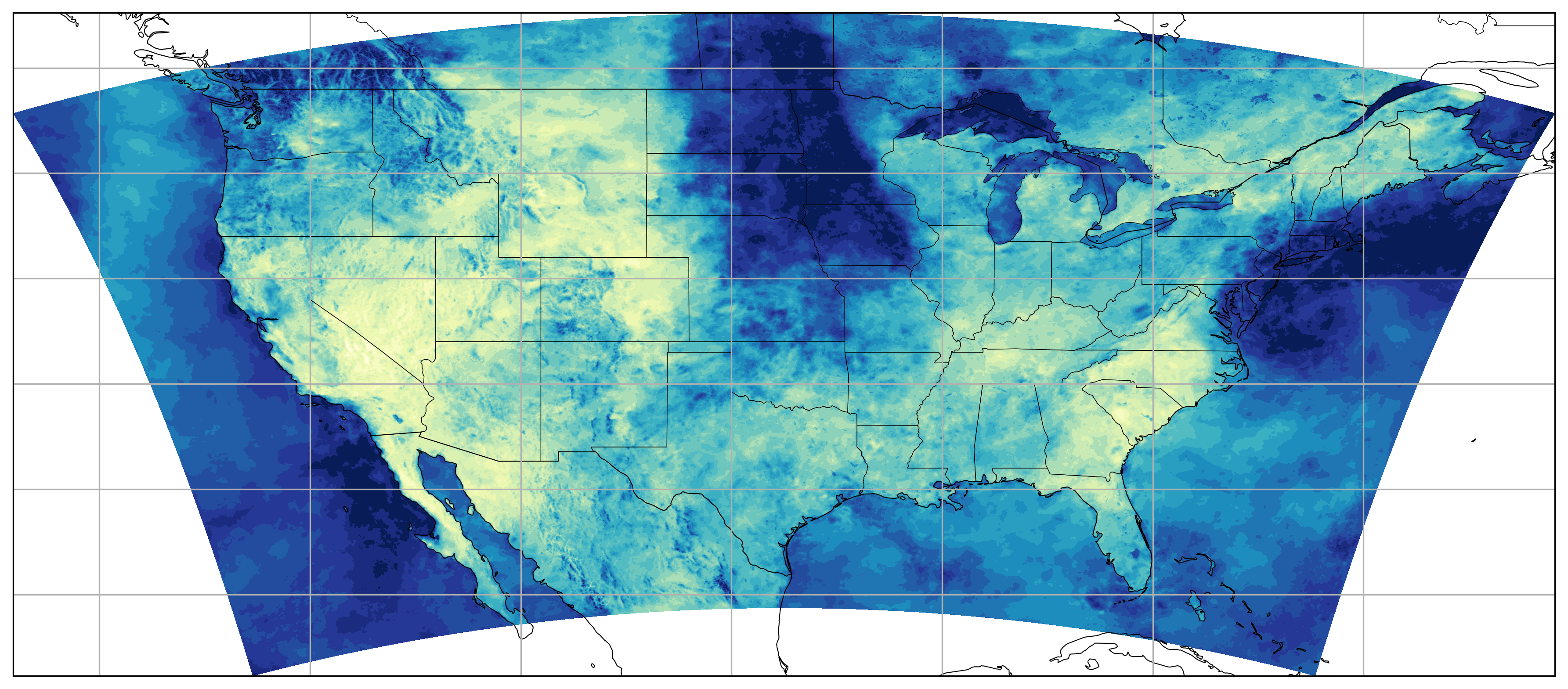} \\
\includegraphics[height=0.138\textwidth]{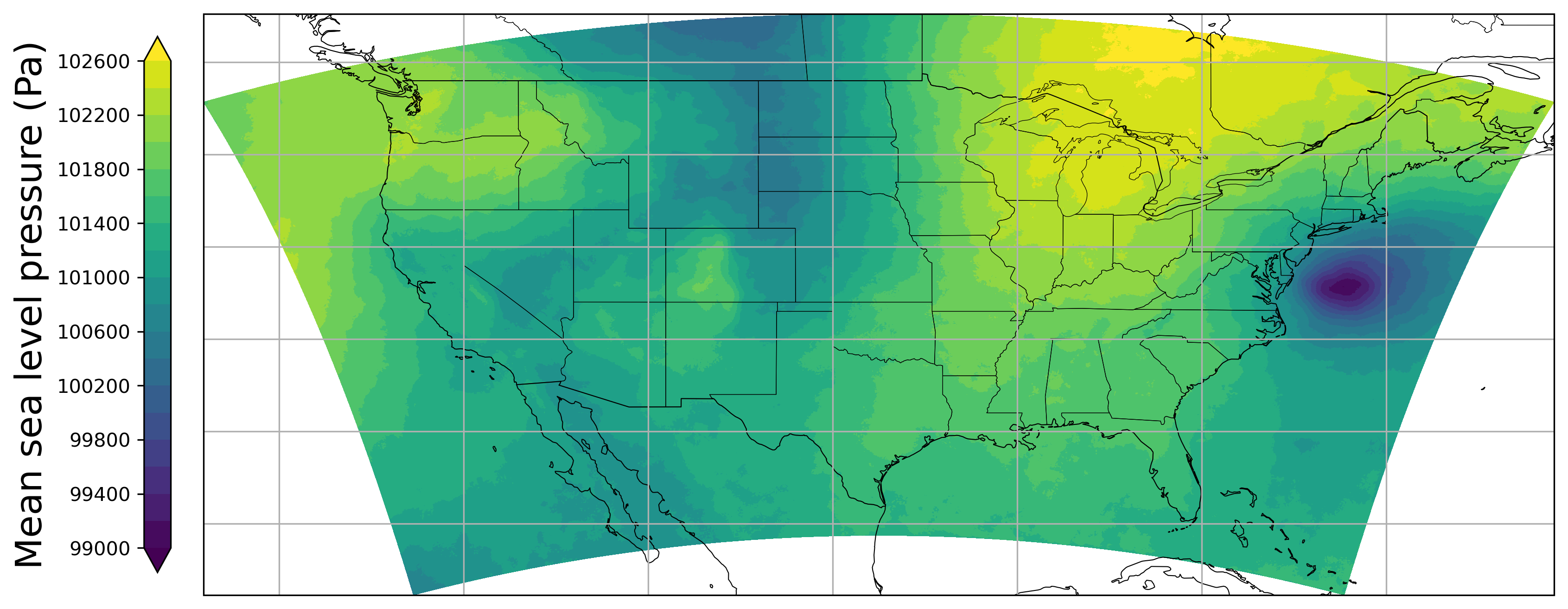} &
\includegraphics[height=0.138\textwidth]{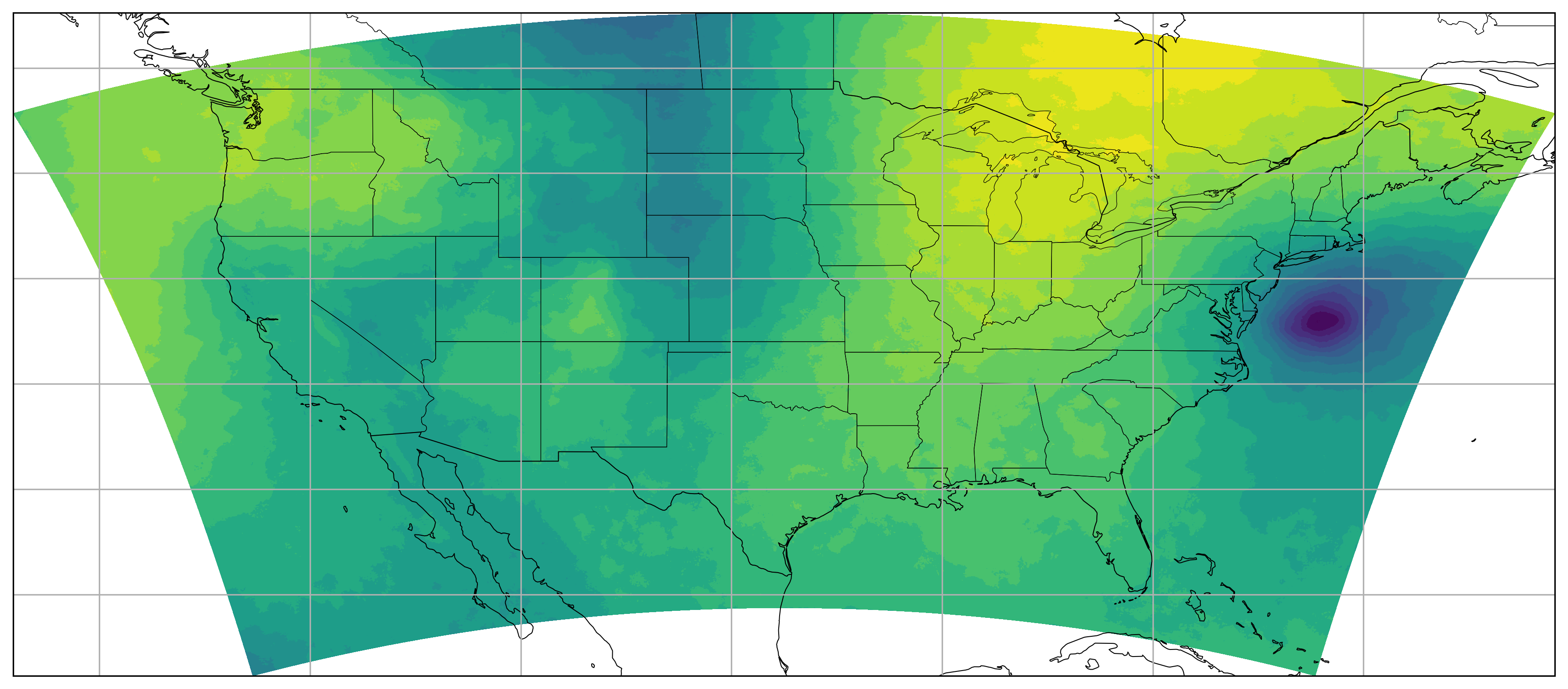} &
\includegraphics[height=0.138\textwidth]{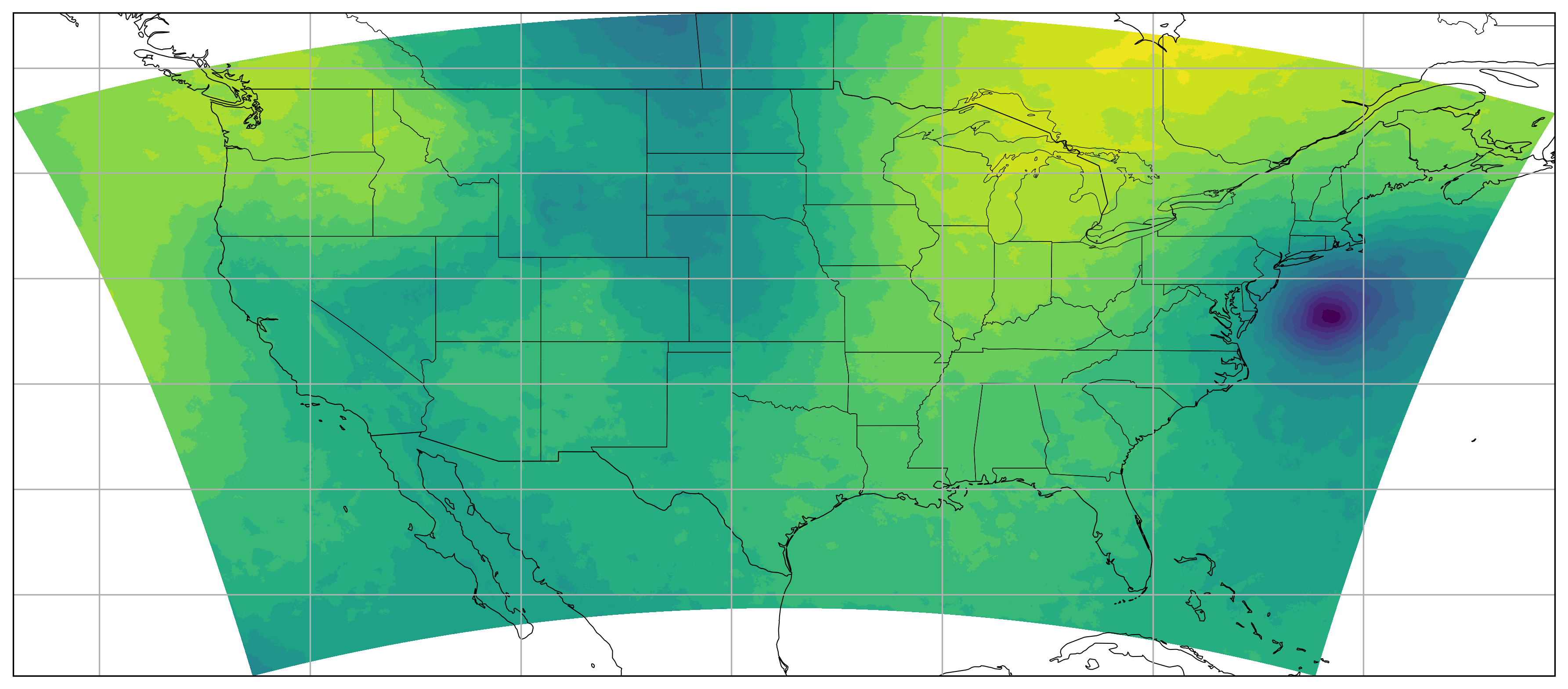} \\
\includegraphics[height=0.138\textwidth]{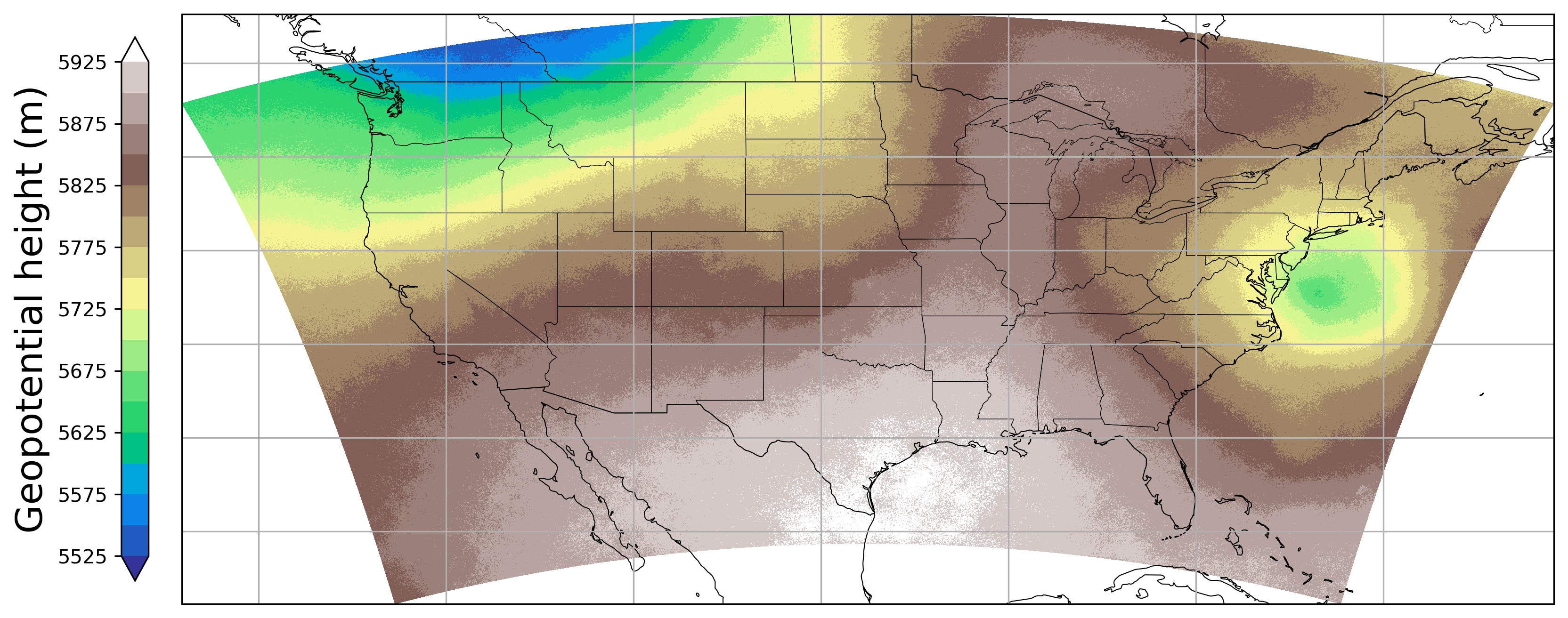} &
\includegraphics[height=0.138\textwidth]{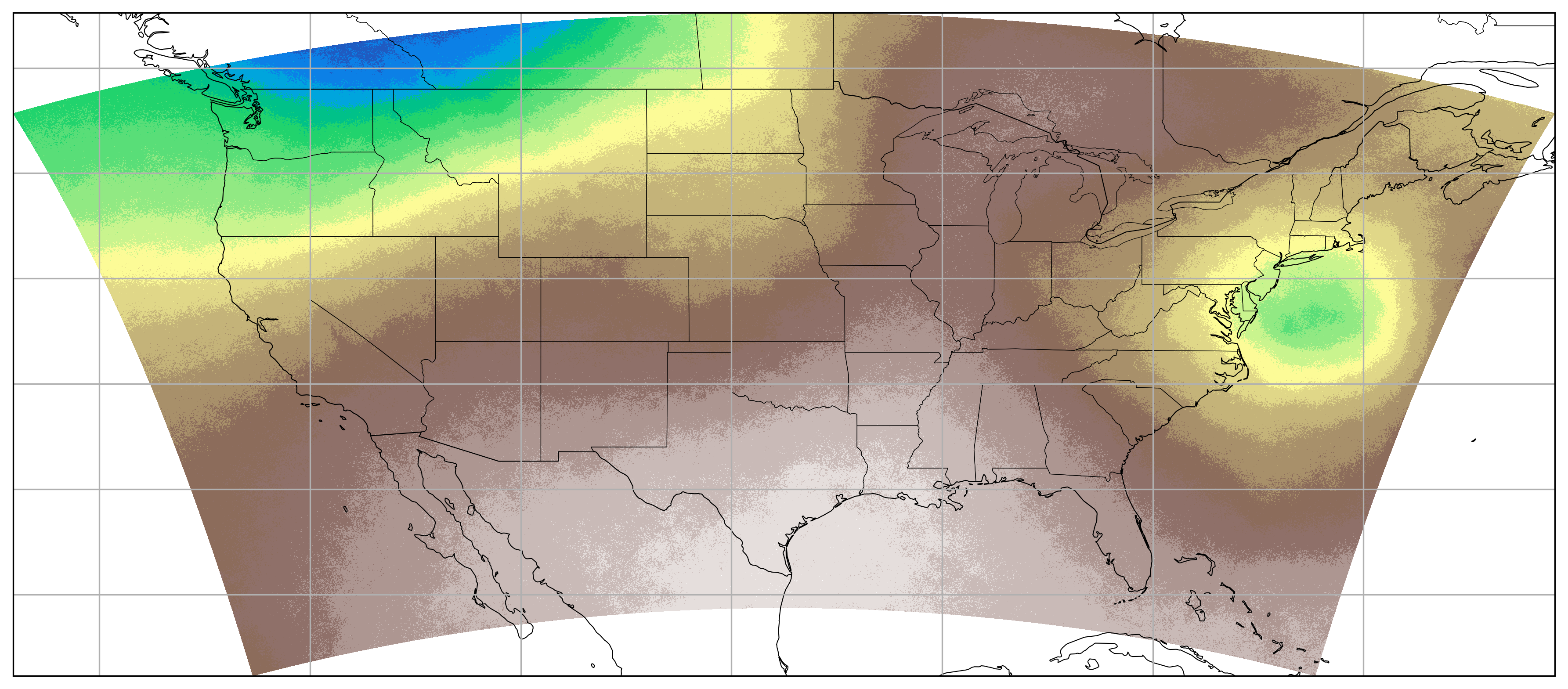} &
\includegraphics[height=0.138\textwidth]{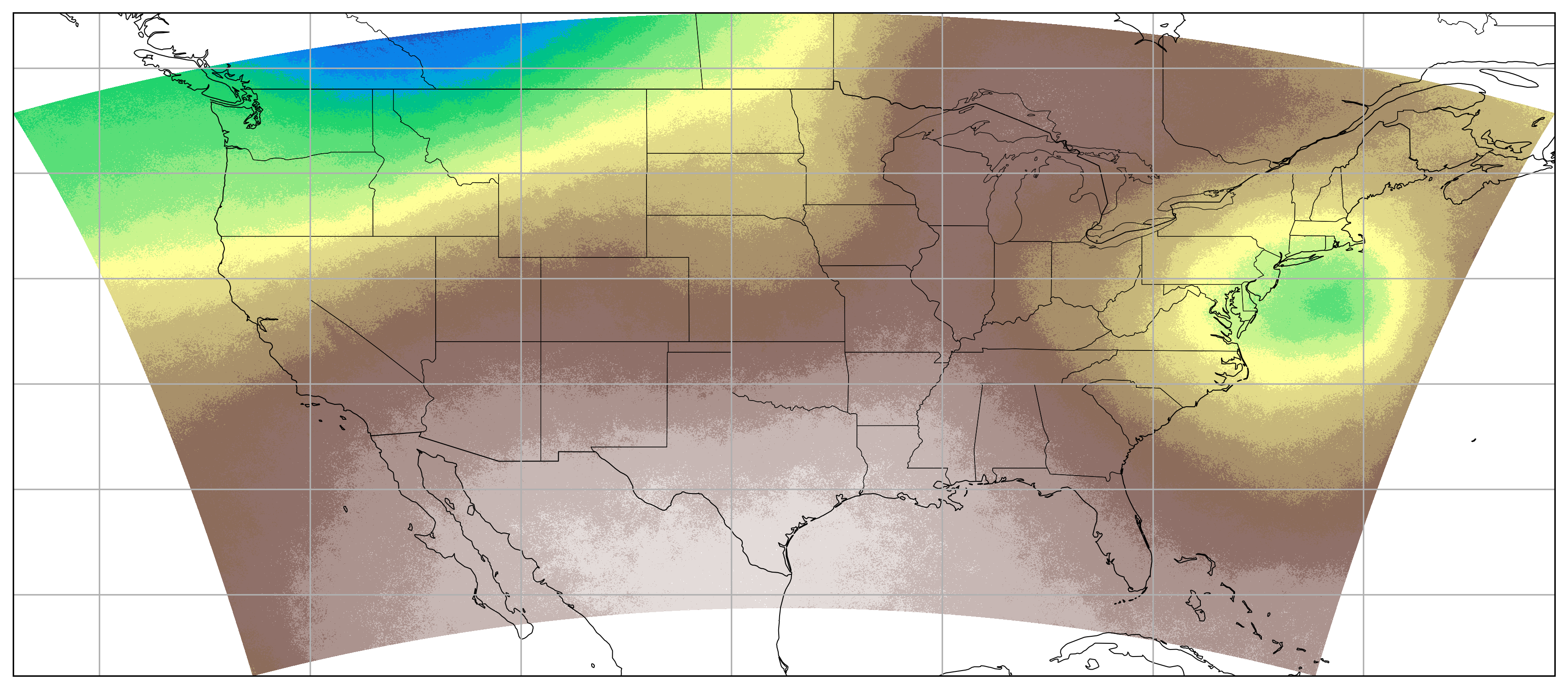} \\
\includegraphics[height=0.138\textwidth]{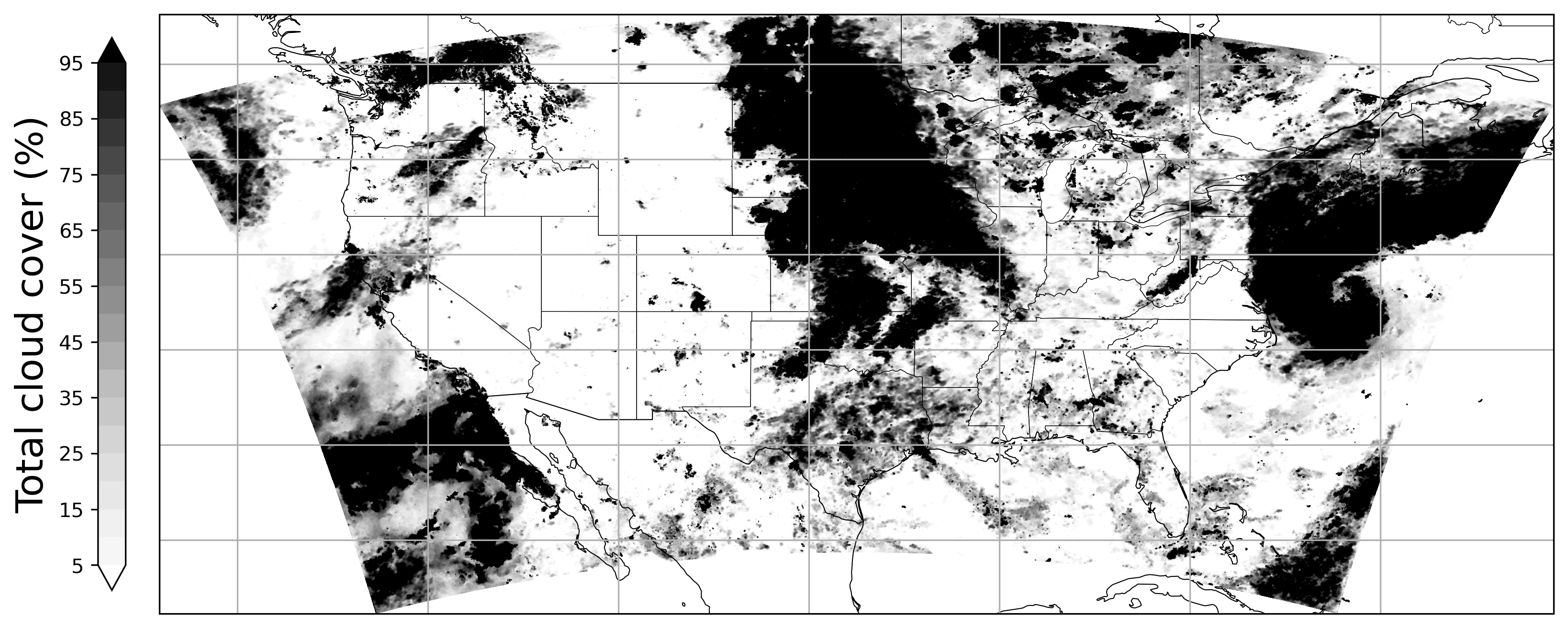} &
\includegraphics[height=0.138\textwidth]{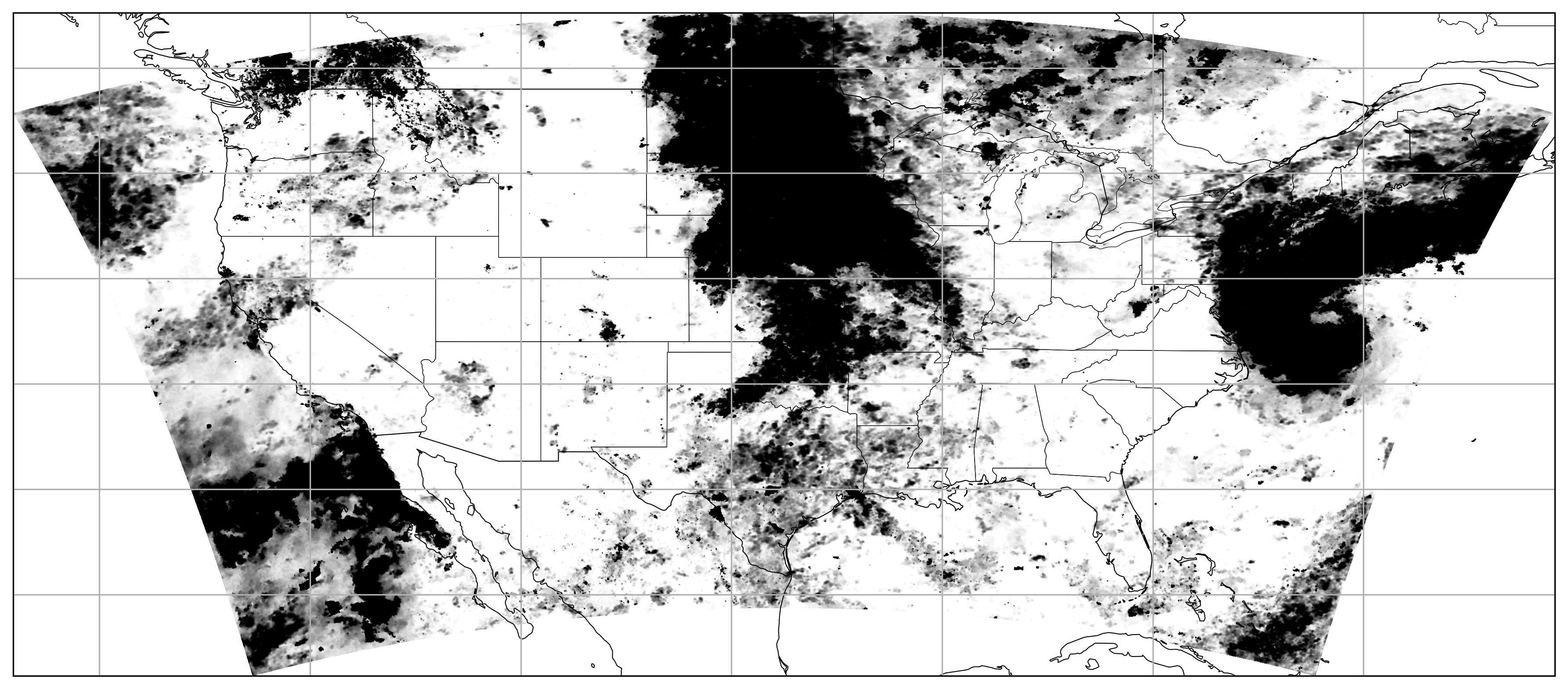} &
\includegraphics[height=0.138\textwidth]{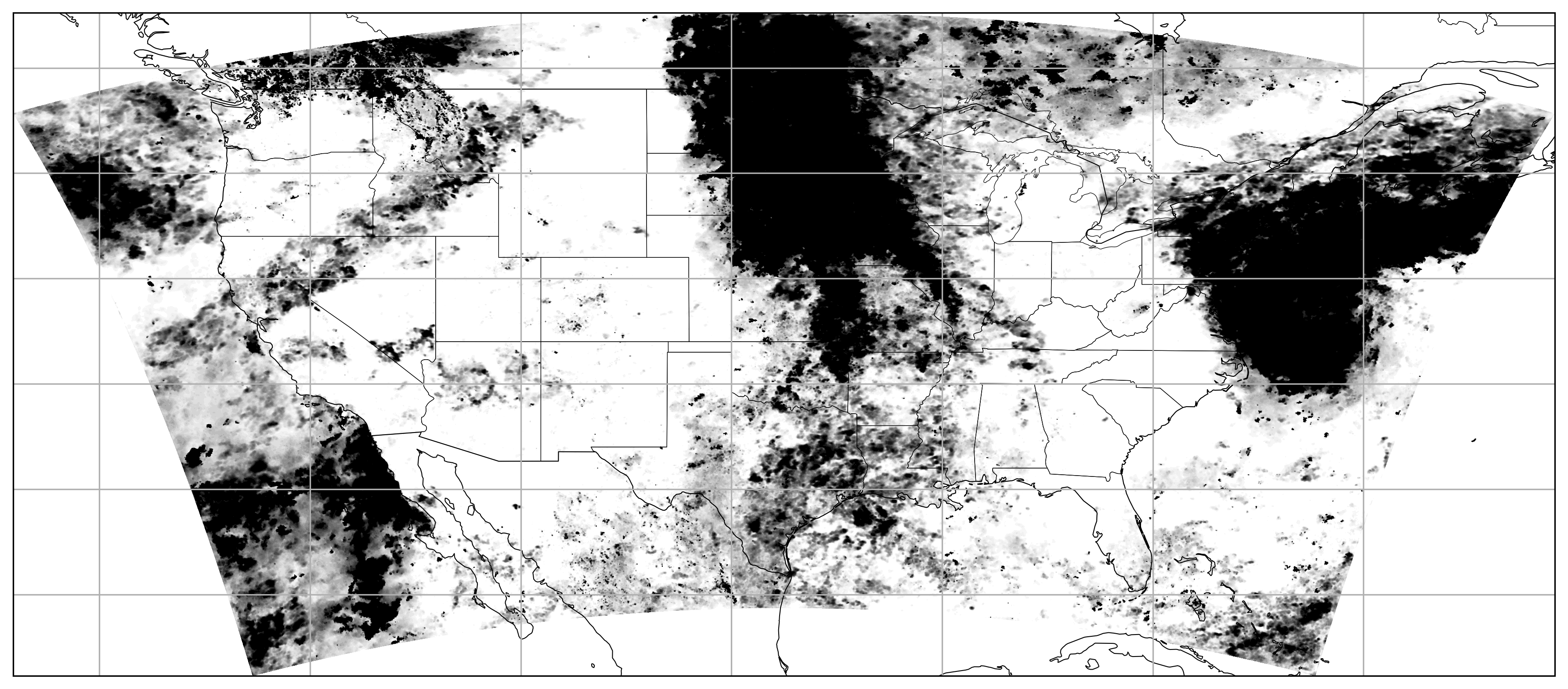} \\
\end{tabular}
\caption{Example HRRRCast forecast evolution for the nor'easter storm on 2026-09-26 at 
1-, 3-, and 6-h forecast leads, showing hourly precipitation, reflectivity, CAPE, 
relative humidity, mean sea-level pressure, 500-hPa geopotential height, and total cloud cover.}
\label{fig:case_study_noreaster}
\end{figure*}

In Appendix~B, we provide a supplemental HRRR-analysis-based RMSE verification for variables not yet supported 
by the observational MATS pipeline (Appendix~\ref{append:analysis_rmse}; Fig.~\ref{fig:rmse_verification}).
We also show day-to-day variability and daily model cycles over a two-week window
(Figs.~\ref{fig:appendix_mats_timeseries_1}--\ref{fig:appendix_mats_timeseries_2}),
along with a representative forecast example illustrating the newly added fields 
(Appendix~\ref{append:case_study}; Fig.~\ref{fig:case_study_noreaster}).

\subsection{Focused ensemble comparison against HREF}
\label{sec:href_comparison}
Here, we examine whether HRRRCast's probabilistic products are competitive with an established 
ensemble forecast system. To address that question, we perform a focused MATS verification of
HRRRCast ensemble performance relative to the operational High-Resolution Ensemble Forecast
system (HREF) over a two-and-a-half-month period (2026-07-27 through 2026-10-09). HREF is a
multi-model, convection-allowing ensemble that combines different operational forecast
systems with varying initial conditions, model physics, and perturbation strategies, and it also includes several
time-lagged members. Because the full HREF is still computationally expensive, it is
produced on a 6-h cycle and provides probabilistic guidance for precipitation and
severe-convection hazards. This comparison is narrower in scope than the
HRRR-based verification and is limited to 3-h accumulated precipitation over a two-and-a-half-month period.

\begin{figure*}[tbp]
\centering
\begin{subfigure}[t]{0.33\textwidth}
\centering
\includegraphics[width=\textwidth]{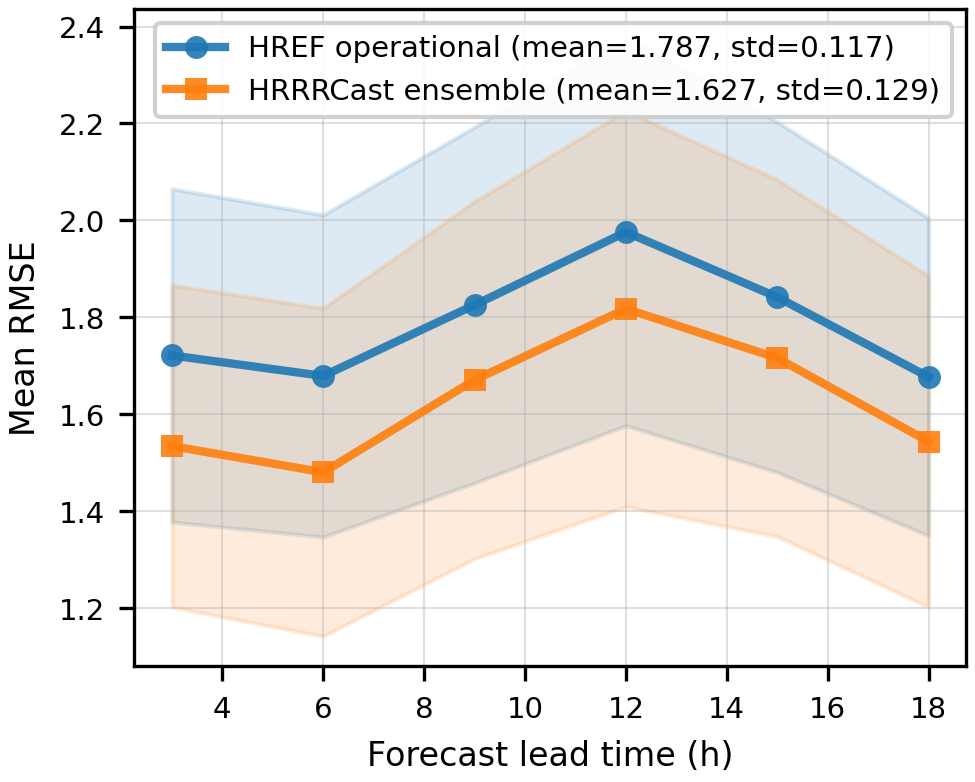}
\caption{RMSE}
\end{subfigure}\hfill
\begin{subfigure}[t]{0.33\textwidth}
\centering
\includegraphics[width=\textwidth]{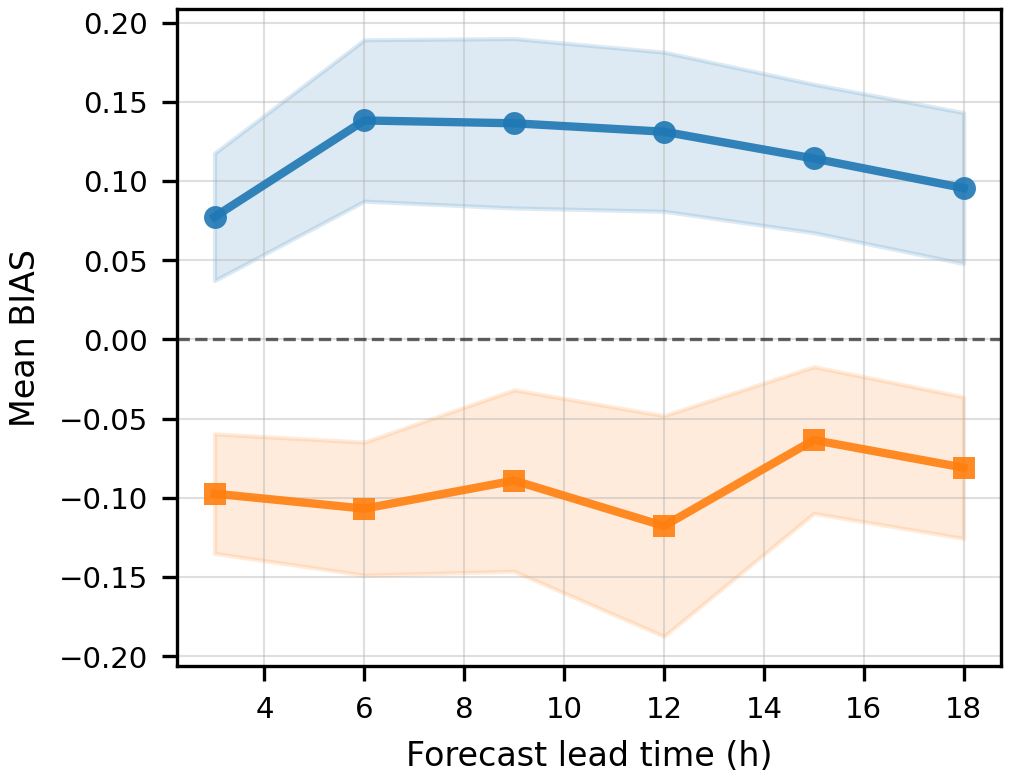}
\caption{Bias}
\end{subfigure}\hfill
\begin{subfigure}[t]{0.33\textwidth}
\centering
\includegraphics[width=\textwidth]{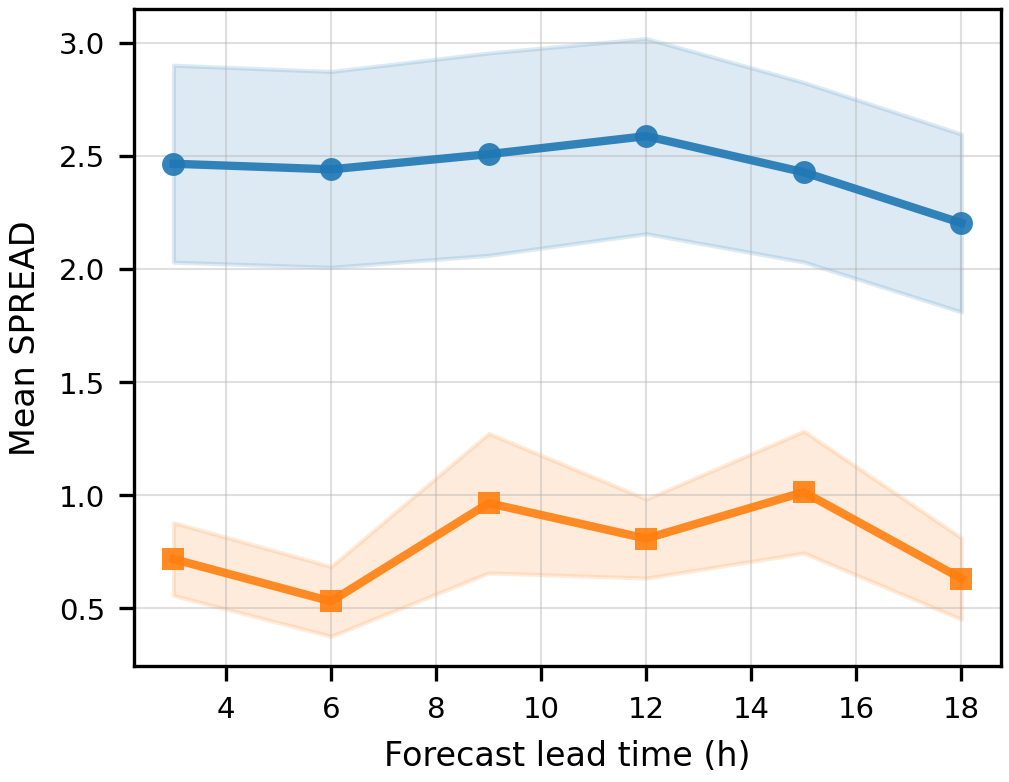}
\caption{Spread}
\end{subfigure}\hfill
\begin{subfigure}[t]{0.33\textwidth}
\centering
\includegraphics[width=\textwidth]{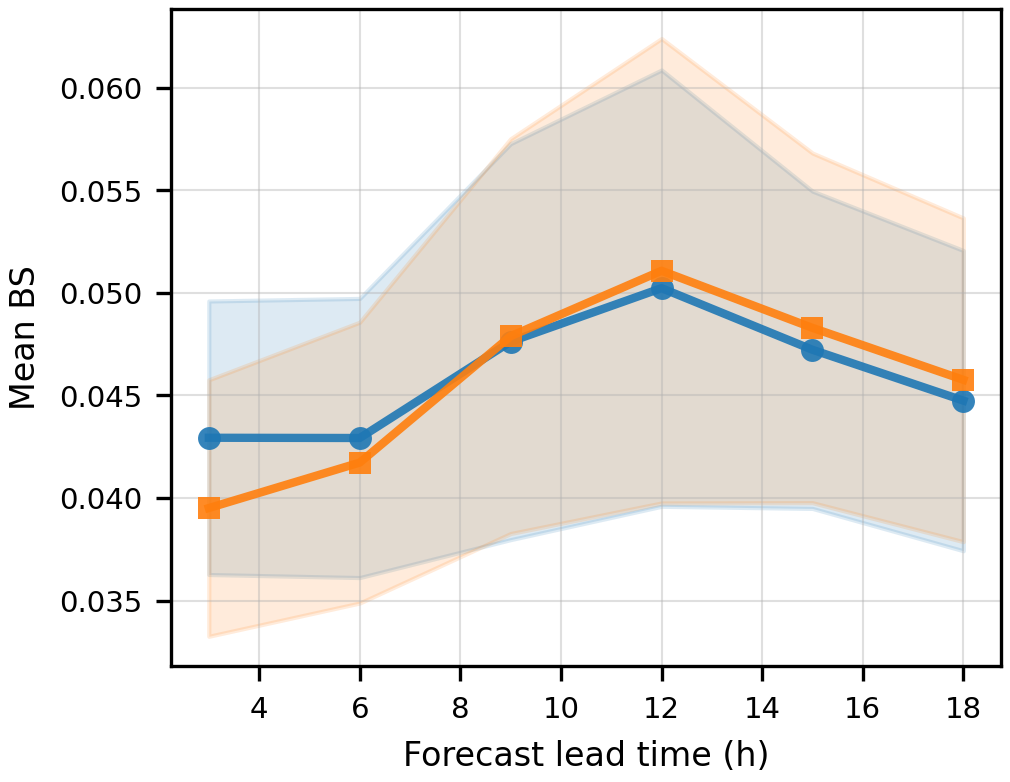}
\caption{Brier Score at 0.635 mm}
\end{subfigure}\hfill
\begin{subfigure}[t]{0.33\textwidth}
\centering
\includegraphics[width=\textwidth]{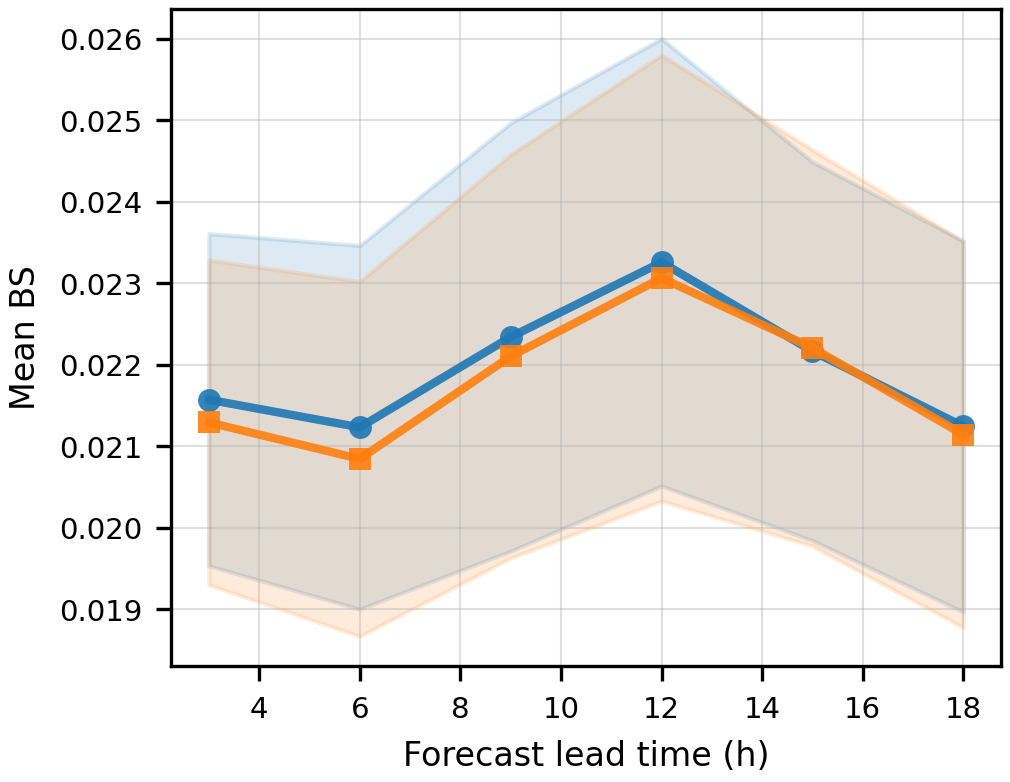}
\caption{Brier Score at 7.62 mm}
\end{subfigure}\hfill
\begin{subfigure}[t]{0.33\textwidth}
\centering
\includegraphics[width=\textwidth]{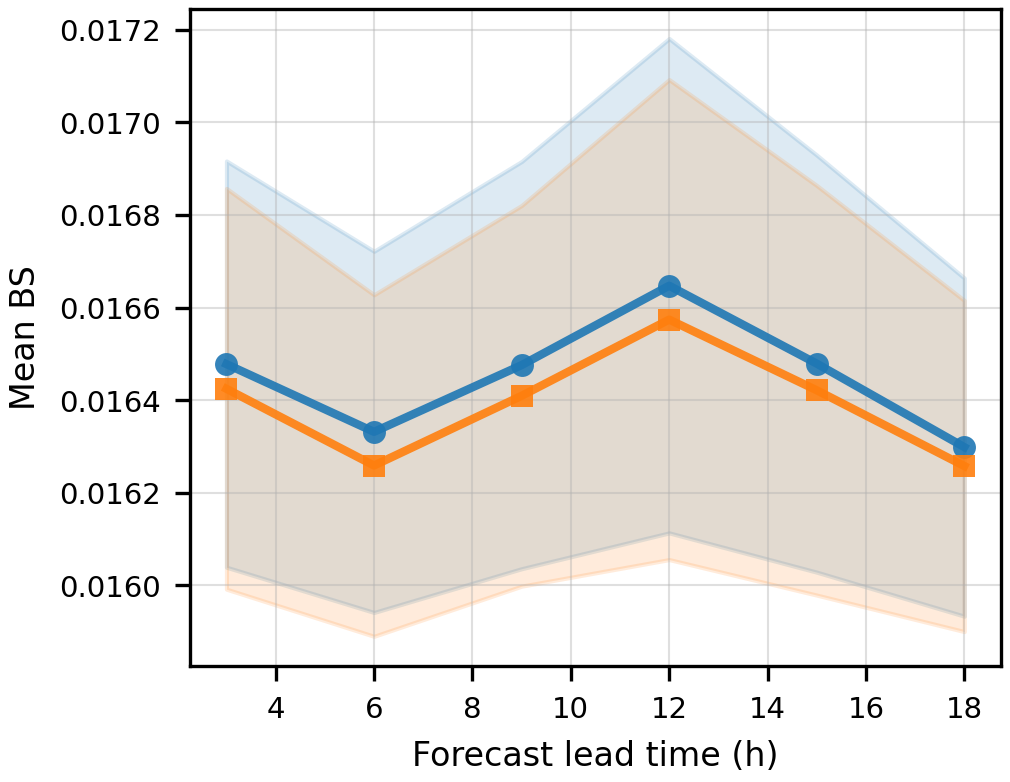}
\caption{Brier Score at 25.4 mm}
\end{subfigure}\hfill
\begin{subfigure}[t]{0.33\textwidth}
\centering
\includegraphics[width=\textwidth]{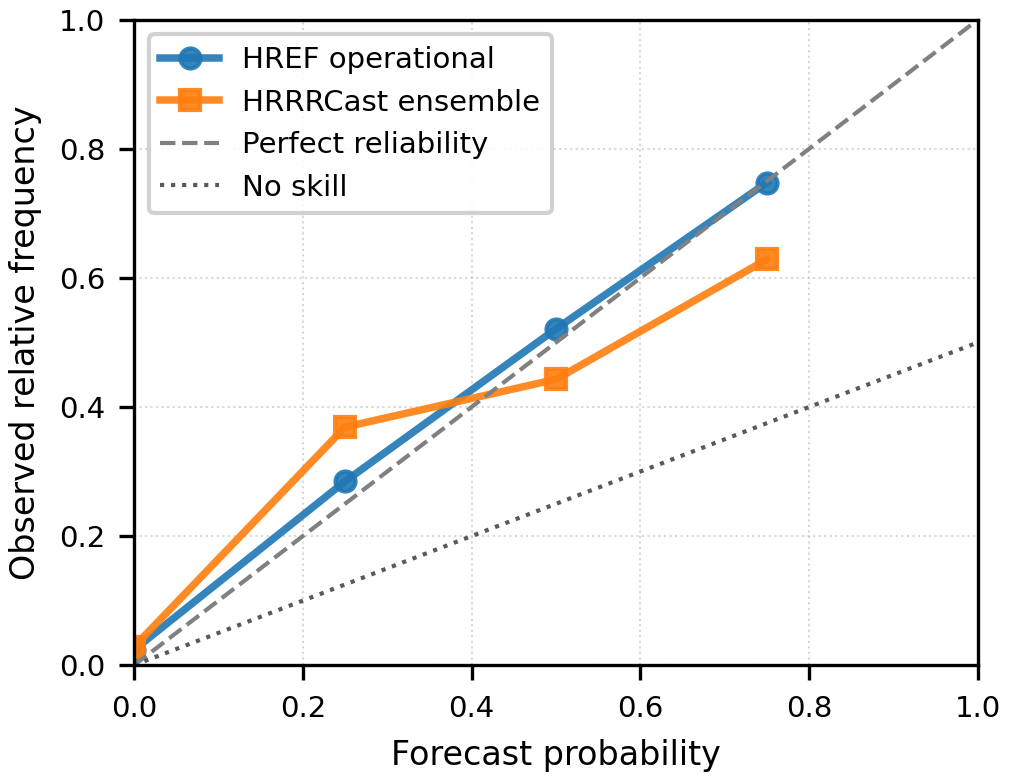}
\caption{Reliability at 0.635 mm}
\end{subfigure}\hfill
\begin{subfigure}[t]{0.33\textwidth}
\centering
\includegraphics[width=\textwidth]{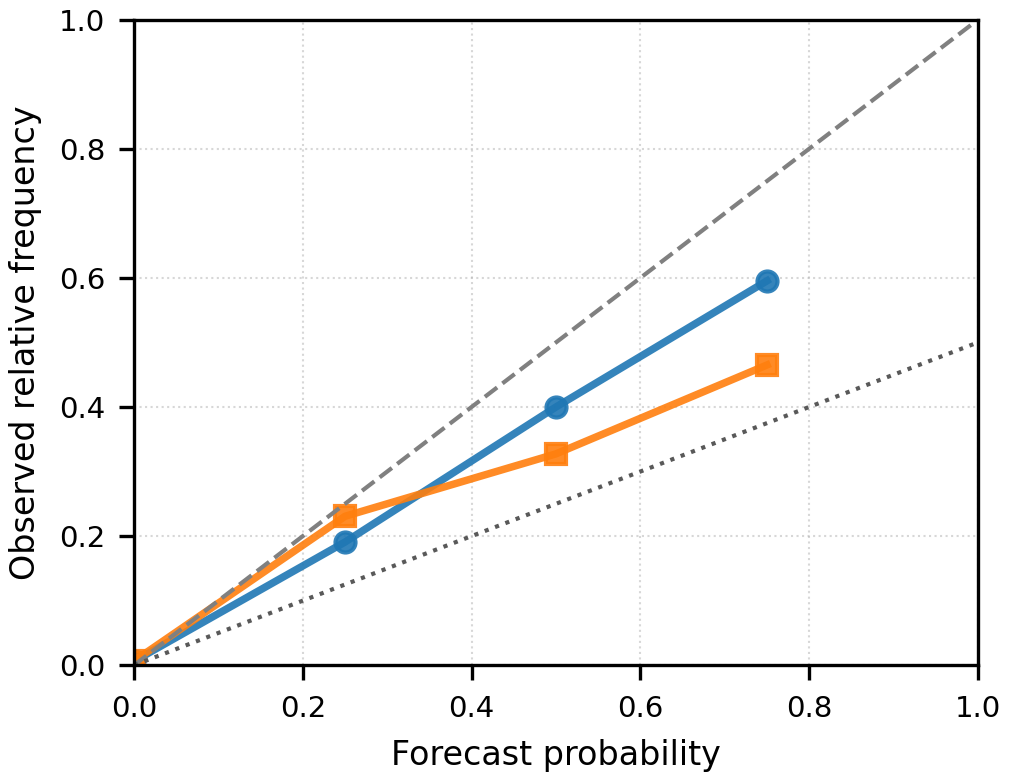}
\caption{Reliability at 7.62 mm}
\end{subfigure}\hfill
\begin{subfigure}[t]{0.33\textwidth}
\centering
\includegraphics[width=\textwidth]{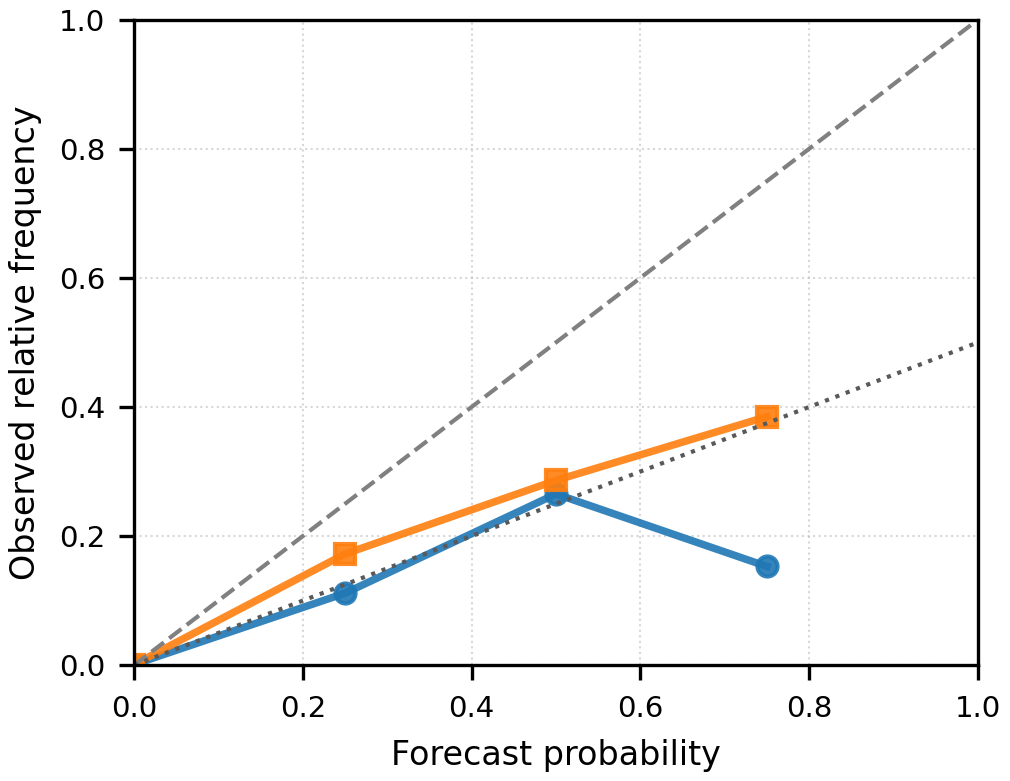}
\caption{Reliability at 25.4 mm}
\end{subfigure}
\vspace{0.5em}
\begin{subfigure}[t]{\textwidth}
\centering
\includegraphics[width=\textwidth]{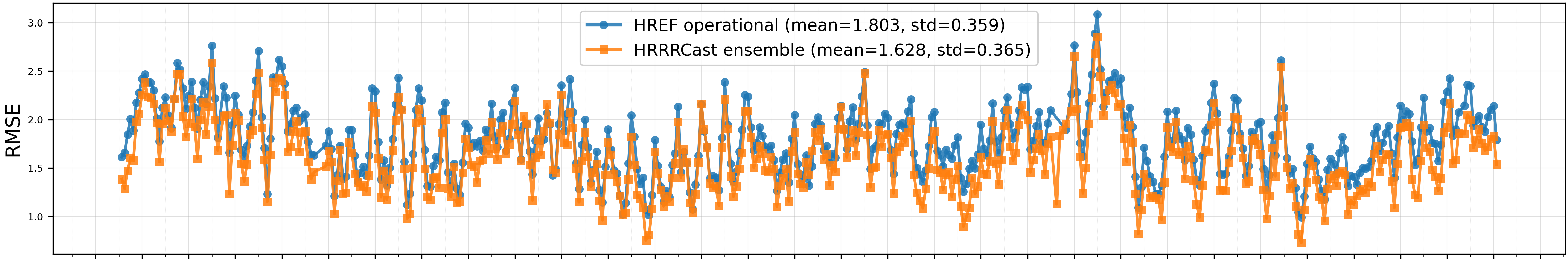}
\includegraphics[width=\textwidth]{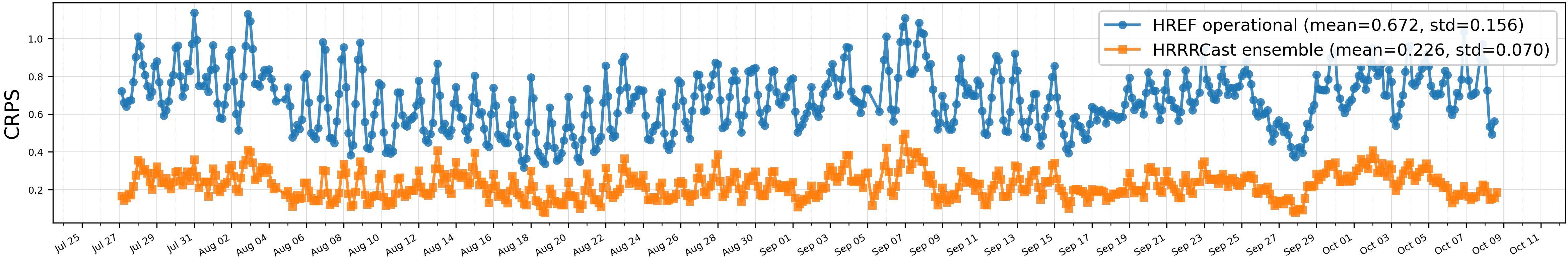}
\caption{Time series of RMSE/CRPS over the full verification period}
\end{subfigure}
\caption{Focused HREF comparison for 3-h accumulated precipitation. 1st row: lead-time dependence of mean RMSE,
mean bias, and ensemble spread for HREF operational and the HRRRCast ensemble. 2nd row: Brier Score lead-time
curves for event thresholds of 0.635, 7.62, and 25.4 mm. 3rd row: corresponding reliability diagrams. Final two panels:
RMSE and CRPS time series over the full verification period. Shaded bands indicate variability. HRRRCast has lower RMSE across
the shared lead times; the spread panel shows HREF is more dispersive, while calibration differences depend strongly
on event threshold.}
\label{fig:mats_href_comparison}
\end{figure*}

Lead-time verification for 3-h accumulated precipitation (top row of Fig.~\ref{fig:mats_href_comparison}) shows that
HRRRCast has lower RMSE than HREF at every shared lead, with mean RMSE values of 1.627 and 1.787, respectively.
The bias comparison is more nuanced: HREF maintains a persistent wet bias, whereas HRRRCast maintains a persistent dry
bias of comparable magnitude. The spread diagnostic reveals a clear calibration contrast. HREF spread exceeds its RMSE
at all leads, indicating an over-dispersive ensemble, whereas HRRRCast spread remains far smaller than its RMSE,
indicating under-dispersion. A complementary CRPS comparison (bottom panel of Fig.~\ref{fig:mats_href_comparison}) also favors HRRRCast, consistent with its
lower mean error and narrower ensemble distribution, although that advantage does not imply uniformly better
threshold-specific calibration.

Threshold-dependent probabilistic skill is more mixed. At 0.635 mm, HRRRCast has slightly lower Brier score at the
shortest leads, but HREF is slightly better from 12 to 18 h. At 7.62 mm, HRRRCast is modestly better at nearly all
shared leads, with only minimal separation at 15 h. At 25.4 mm, HRRRCast has the lower Brier score at every shared
lead.

Reliability behavior (bottom row of Fig.~\ref{fig:mats_href_comparison}) explains this threshold dependence. At the
lightest threshold (0.635 mm), HREF remains close to the one-to-one line, whereas HRRRCast transitions from mild
underconfidence at the 25\% bin to clear overconfidence at the 50\% and 75\% bins. At 7.62 mm, both systems are
overconfident, but HREF stays closer to perfect reliability across the occupied bins. At 25.4 mm, the ranking
reverses: HREF calibration degrades sharply, with its highest-probability bin dropping to an observed relative
frequency of only about 0.16, whereas HRRRCast remains monotonic and reaches about 0.39. Overall, HREF is better
calibrated for light-to-moderate 3-h precipitation, whereas HRRRCast is more reliable for the heaviest events.

We also compare ensemble calibration with a separately trained model optimized using a CRPS
objective in Appendix~\ref{append:crps_comparison} (Fig.~\ref{fig:ser_ablation}). The CRPS model is
better calibrated on most variables than HRRRCast, whose diffusion-based ensemble remains somewhat under-dispersive.
However, calibration of APCP and REFC remains challenging even for the CRPS-based model, highlighting the difficulty
of rare-event probabilistic forecasting.

\section{Ablation Studies}
\label{sec:ablation_studies}

Having established near-real-time deterministic and ensemble performance, we next isolate two
contributors to these gains: native-resolution training and moderate loss reweighting. These
attribution analyses also clarify the scale dependence of HRRRCast skill, including stronger
40-dBZ performance when neighborhood-based FSS metrics are used.

The attribution analyses in the next two subsections use single-member forecasts.
For clarity, we use the following model labels: HRRR 3~km denotes native operational
HRRR output; HRRR 6~km denotes the same HRRR forecast field subsampled to 6~km for
verification; HRRRCast 3~km denotes the native-grid HRRRCast model; and HRRRCast 6~km
denotes the separately trained 6-km HRRRCast model, everything else being the same.

\subsection{Attribution of skill gains: native-resolution training}
\label{sec:native_training_attribution}

Native-grid training provides consistent gains over the 6-km baseline and
competitive reflectivity skill against operational HRRR, with the largest
benefits at finer scales.

\begin{figure*}[bp]
\centering
\begin{subfigure}[t]{0.32\textwidth}
\centering
\includegraphics[width=\textwidth]{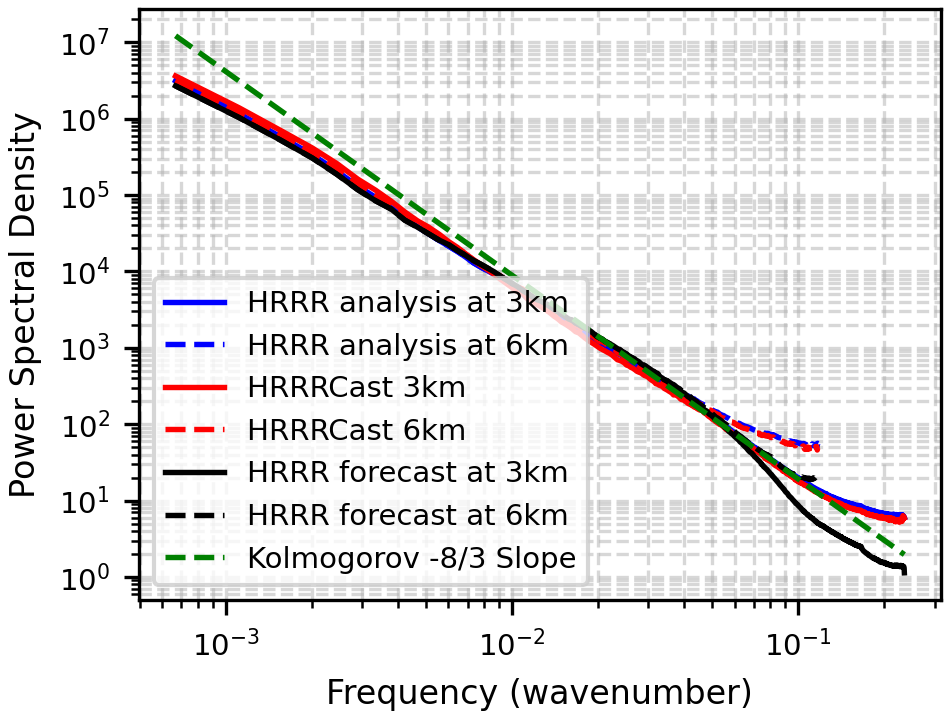}
\caption{Composite reflectivity}
\end{subfigure}\hfill
\begin{subfigure}[t]{0.32\textwidth}
\centering
\includegraphics[width=\textwidth]{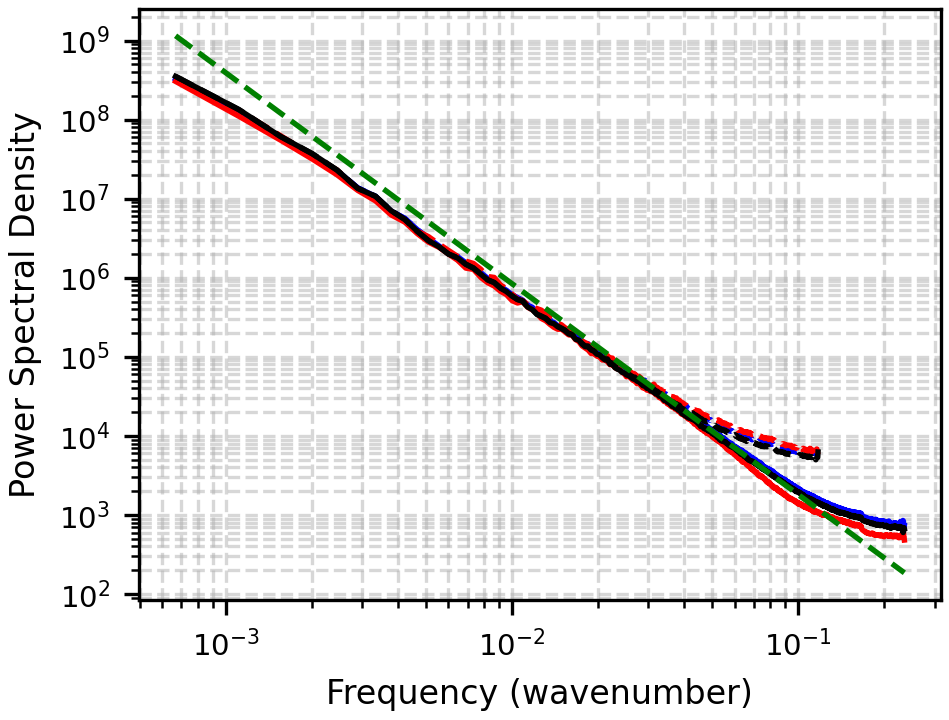}
\caption{Convective inhibition}
\end{subfigure}\hfill
\begin{subfigure}[t]{0.32\textwidth}
\centering
\includegraphics[width=\textwidth]{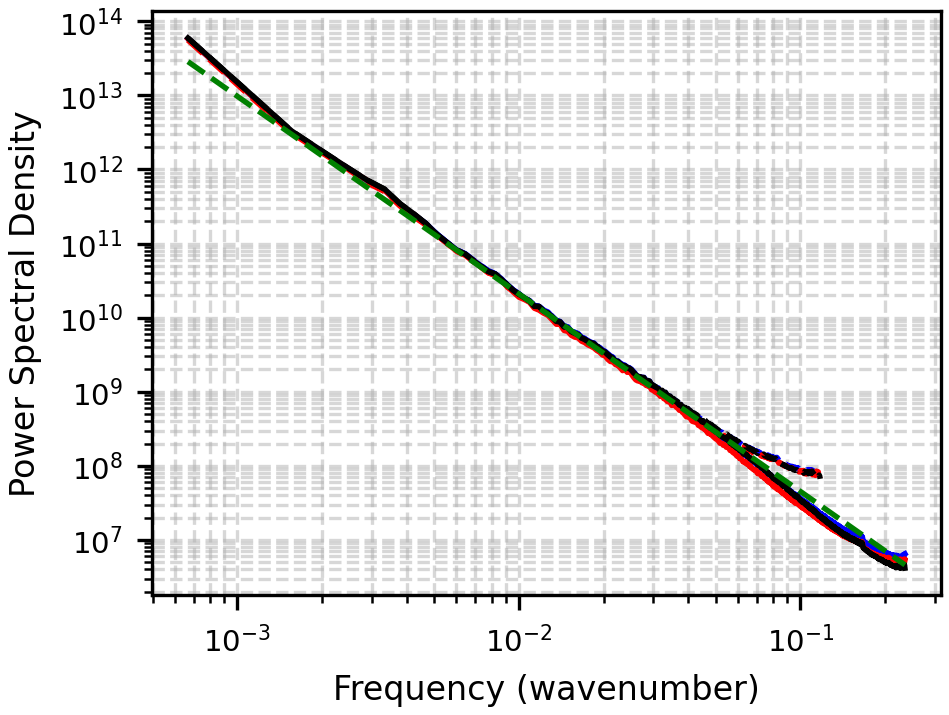}
\caption{Surface visibility}
\end{subfigure}
\caption{Representative PSD comparisons illustrating the native-grid advantage.
At the highest resolved wavenumbers, the 6-km curves commonly turn upward relative to the
native 3-km HRRR analysis, consistent with subsampling-induced aliasing. Native-grid
HRRRCast retains more small-scale variance and remains closer to the 3-km analysis across
these variables.}
\label{fig:psd_native_training}
\end{figure*}

When the HRRR analysis is sub-sampled from its native 3-km grid to a coarser
resolution for training, the absence of a proper anti-aliasing filter folds sub-grid
energy back across the spectrum, corrupting fine-scale spectral structure in the training
targets. A model trained on this data learns these aliasing-contaminated spectral
characteristics rather than the physically consistent fine-scale structure of the native
grid. Training HRRRCast directly on the native 3-km HRRR grid avoids these subsampling-induced aliasing artifacts.
Consistent spectral evidence for this aliasing  behavior and improved small-scale variance 
retention is shown in Fig.~\ref{fig:psd_native_training} and Appendix~B (Fig.~\ref{fig:appendix_psd_all}).

Verification against observations using the fractions skill score (FSS) confirms
substantial improvement at fine spatial scales. At a 6-km pooling window, native-grid
HRRRCast improves markedly over the 6-km baseline across all reflectivity thresholds
examined. For single-member forecasts, FSS is generally better than operational HRRR at
20 and 30 dBZ, while 40-dBZ skill is more nuanced, with near parity at intermediate
leads and slight degradation at longer leads.

\begin{figure*}[tbp]
\centering
\includegraphics[width=0.32\textwidth]{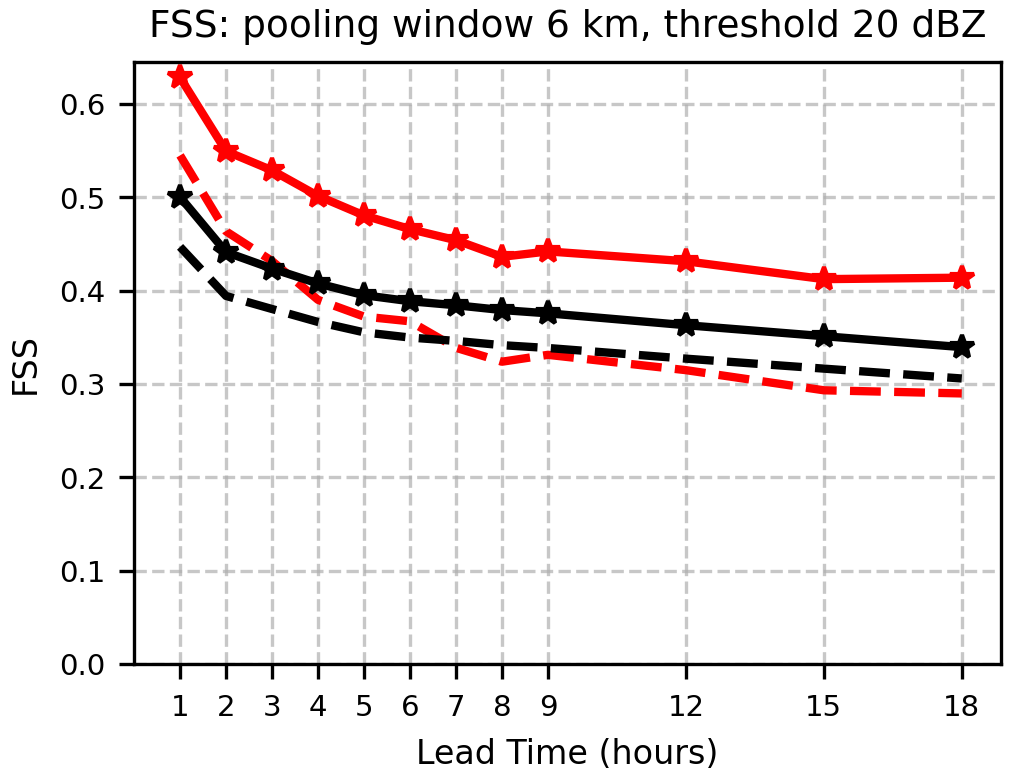}
\includegraphics[width=0.32\textwidth]{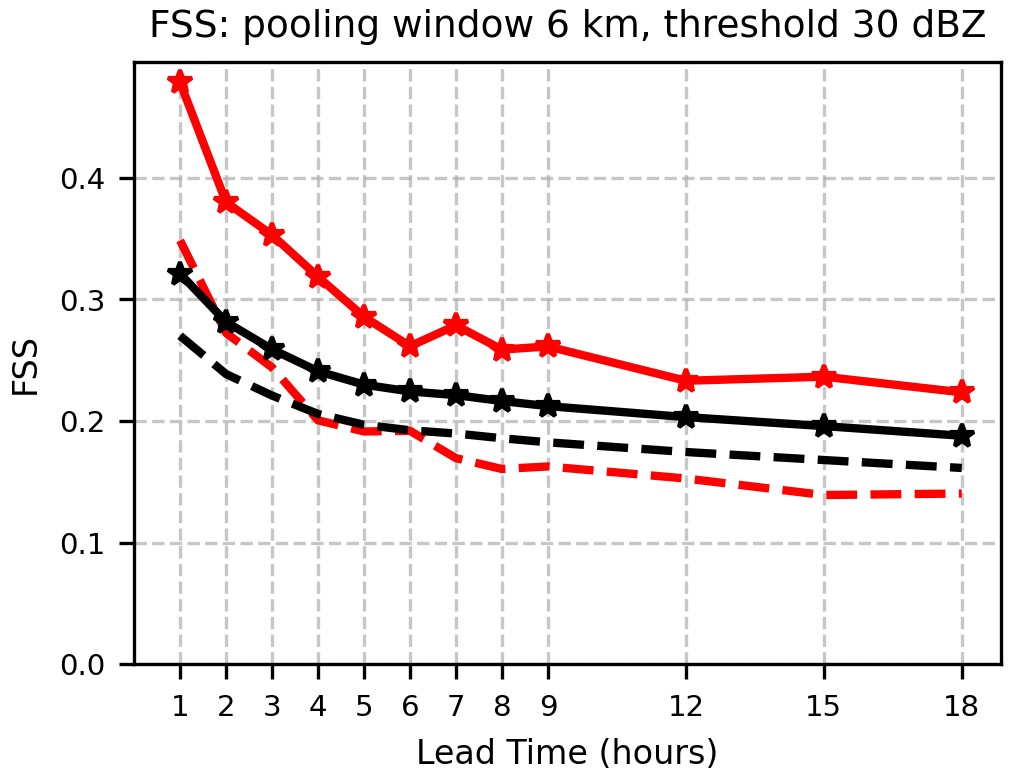}
\includegraphics[width=0.32\textwidth]{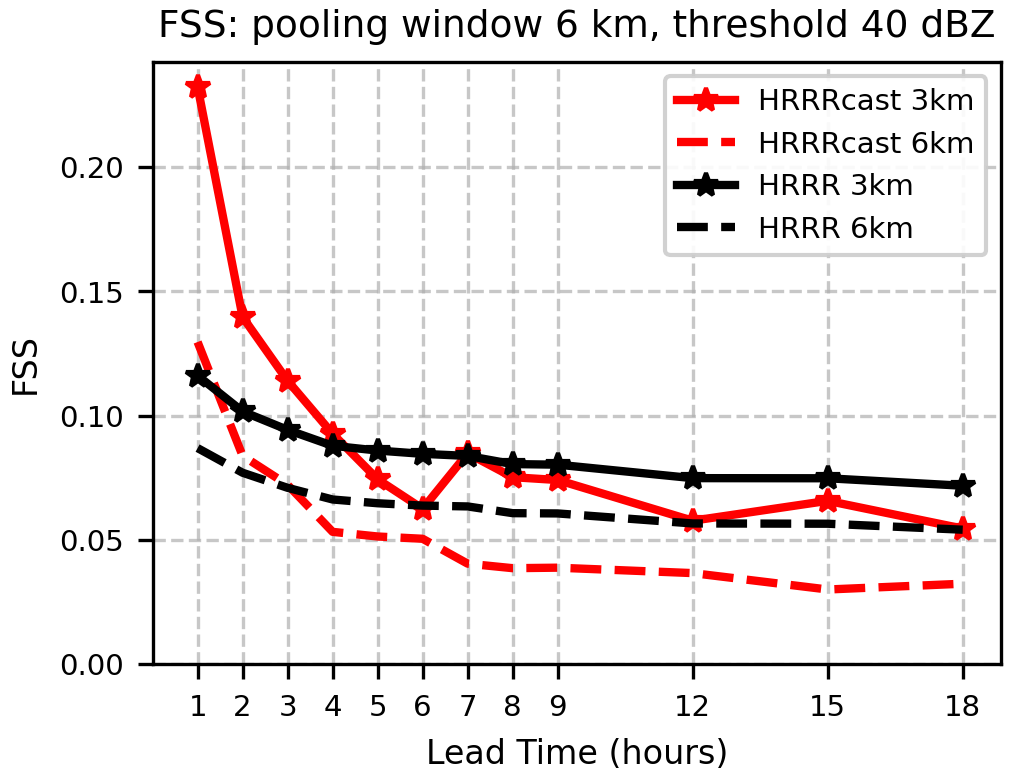}
\\[0.4em]
\includegraphics[width=0.32\textwidth]{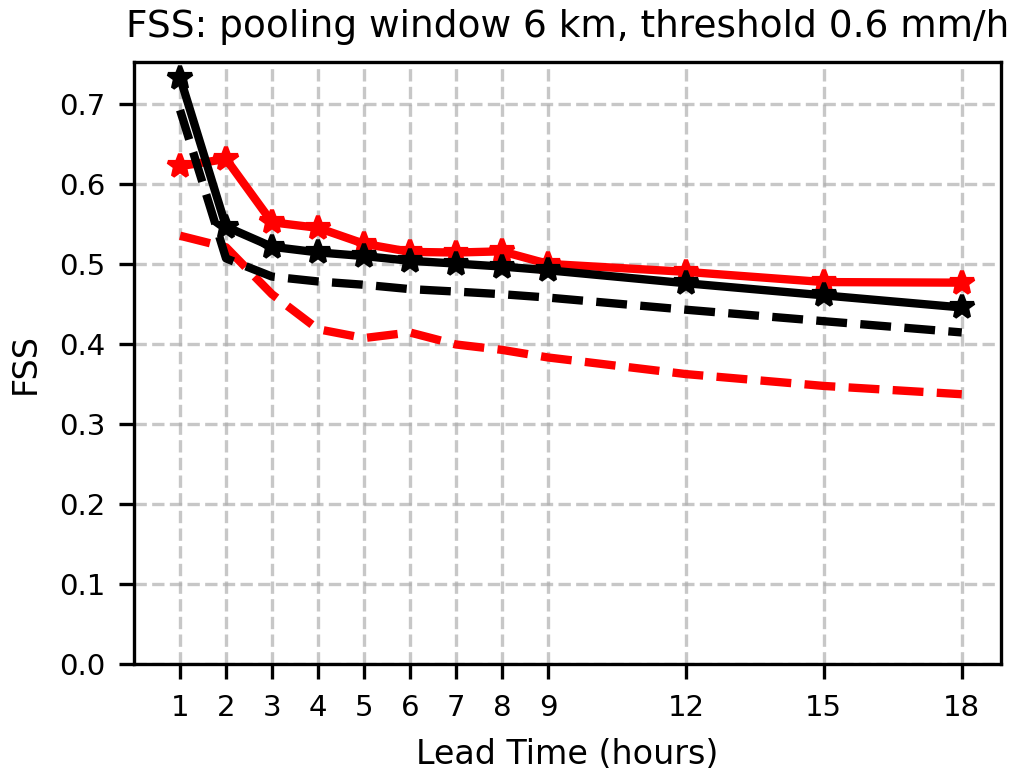}
\includegraphics[width=0.32\textwidth]{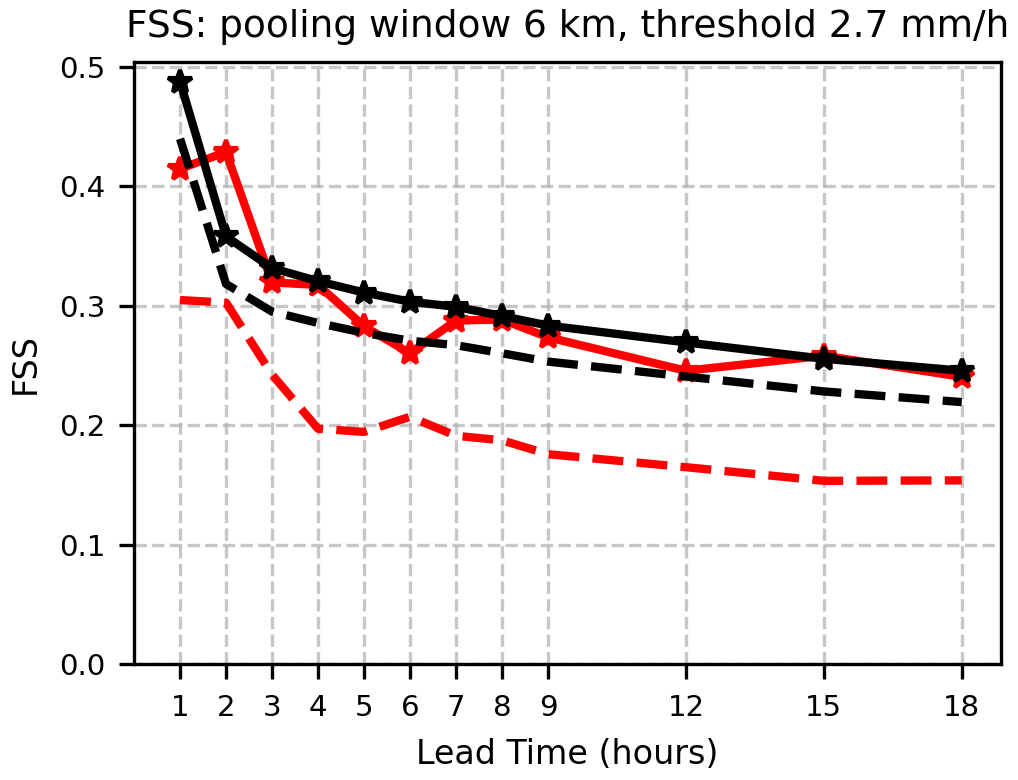}
\includegraphics[width=0.32\textwidth]{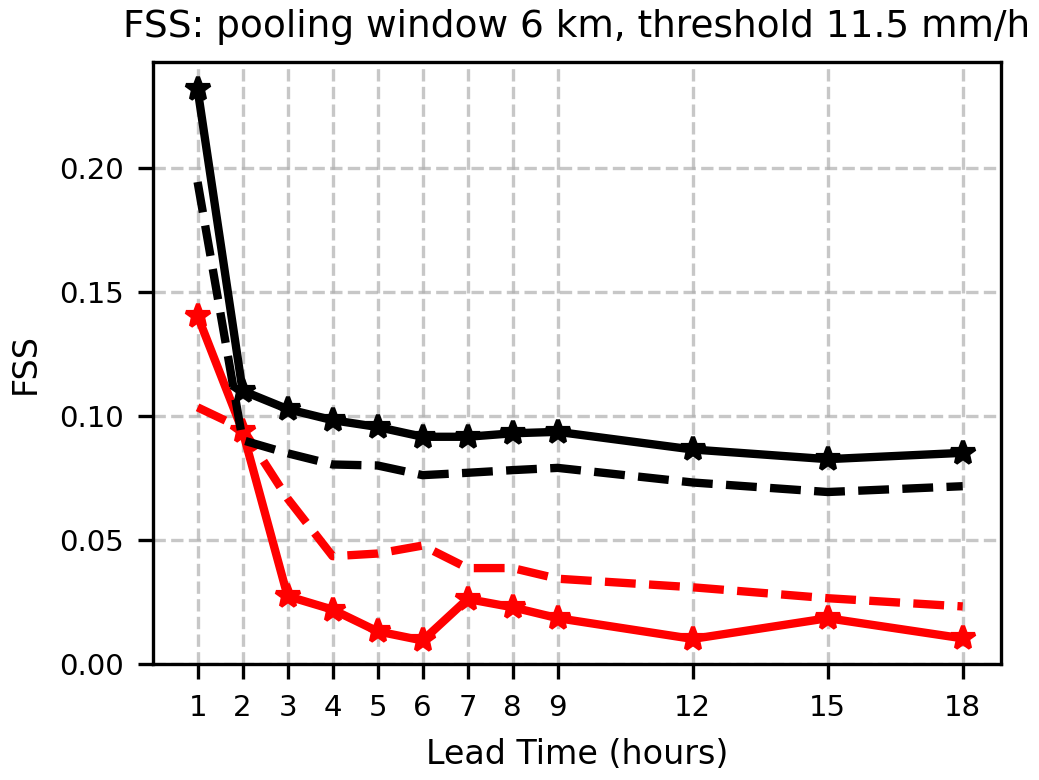}
\caption{Fractions skill score (FSS) lead-time curves comparing native-grid HRRRCast
(3 km), 6-km HRRRCast, and HRRR (each evaluated on both 3-km and 6-km grids). Native
3-km training improves HRRRCast skill at convection-permitting scales and retains
benefits across reflectivity and 1-h precipitation thresholds. Top row: reflectivity
thresholds (20/30/40 dBZ). Bottom row: 1-h precipitation thresholds
(0.6/2.7/11.5 mm h$^{-1}$).}
\label{fig:fss_native3km}
\end{figure*}

Figure~\ref{fig:fss_native3km} illustrates representative lead-time FSS behavior from the
3-km versus 6-km comparison set. In the top row, HRRRCast at native 3 km (solid red)
remains above the 6-km HRRRCast baseline (dashed red) across lead times at both 20 and
30 dBZ, with the clearest separation at 30 dBZ; it also generally outperforms HRRR at
both evaluation resolutions (black curves). At 40 dBZ (top-right panel), native-grid
training still improves on the 6-km baseline, but the comparison against HRRR-3km
narrows to near parity at intermediate leads and slightly favors HRRR at longer leads.

The corresponding 1-h precipitation curves in the bottom row show a similar overall
pattern: native-grid training generally improves HRRRCast relative to the 6-km
baseline. The clearest and most consistent gains appear at light-to-moderate thresholds
(0.6 and 2.7 mm h$^{-1}$), although relative to HRRR-3km these curves are often near
equal to slightly lower at mid-to-late leads. At the heaviest threshold
(11.5 mm h$^{-1}$; bottom-right panel), deterministic HRRRCast-3km degrades
substantially after the first few lead times.

\begin{figure*}[tbp]
\centering
\includegraphics[width=0.32\textwidth]{fss_3km_vs_6km/fss_comp_hi_6km_40dbz_analysis.png}
\includegraphics[width=0.32\textwidth]{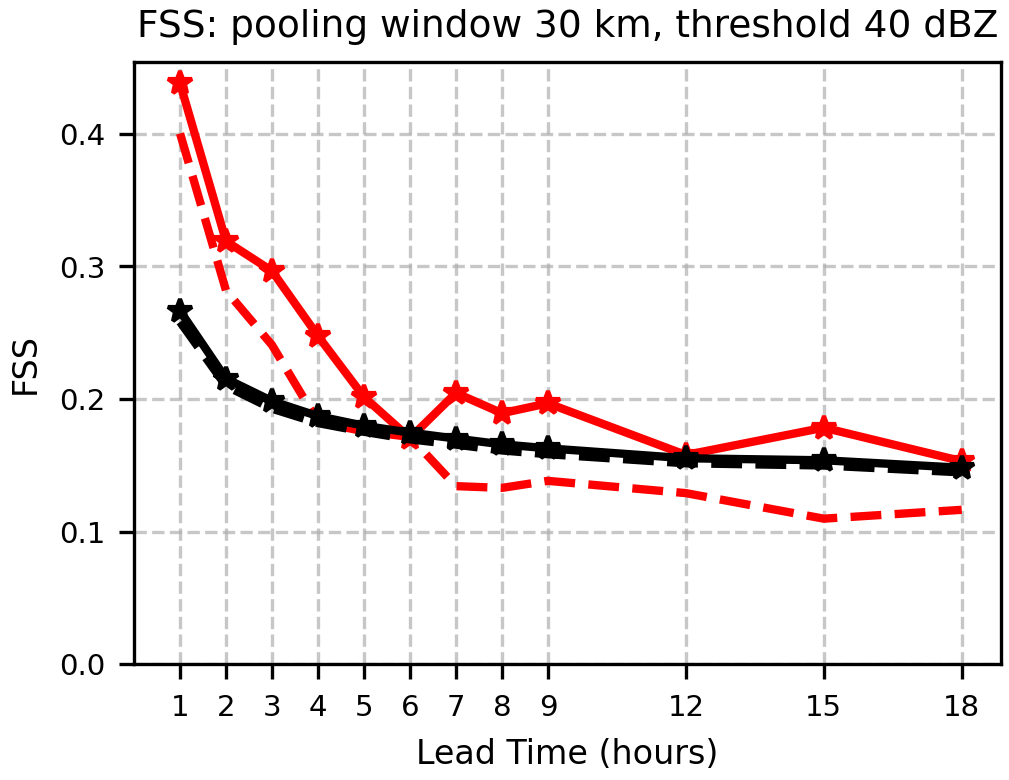}
\includegraphics[width=0.32\textwidth]{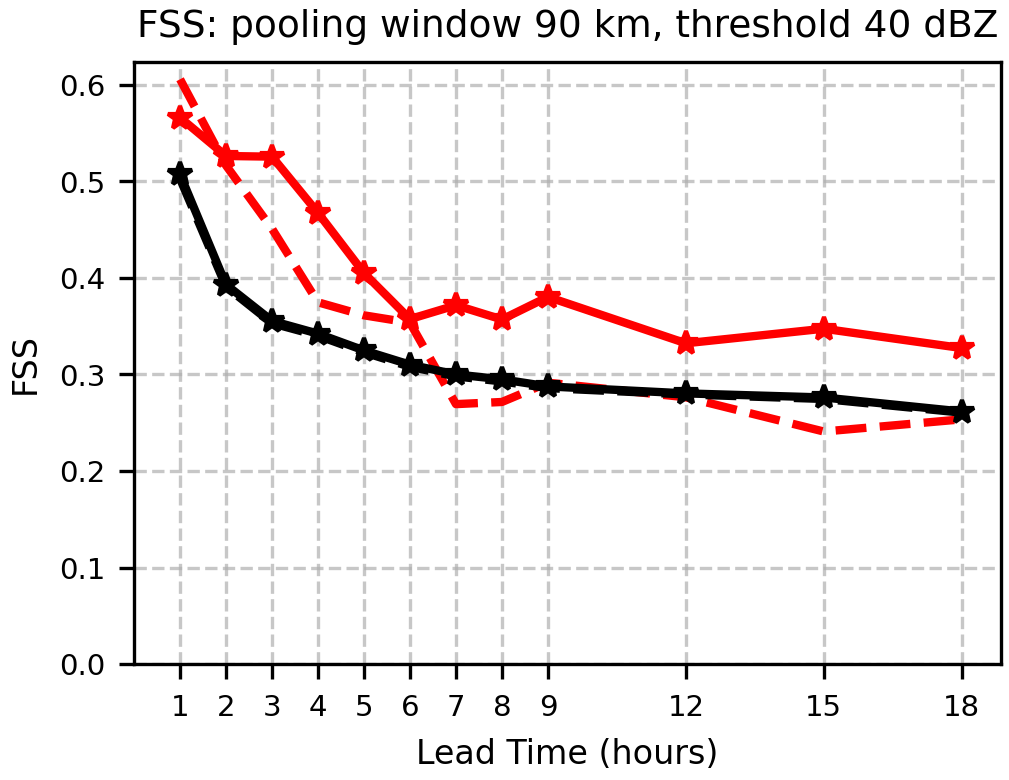}
\\[0.4em]
\includegraphics[width=0.32\textwidth]{fss_3km_vs_6km/fss_comp_hi_6km_2.7mm_per_h_analysis.png}
\includegraphics[width=0.32\textwidth]{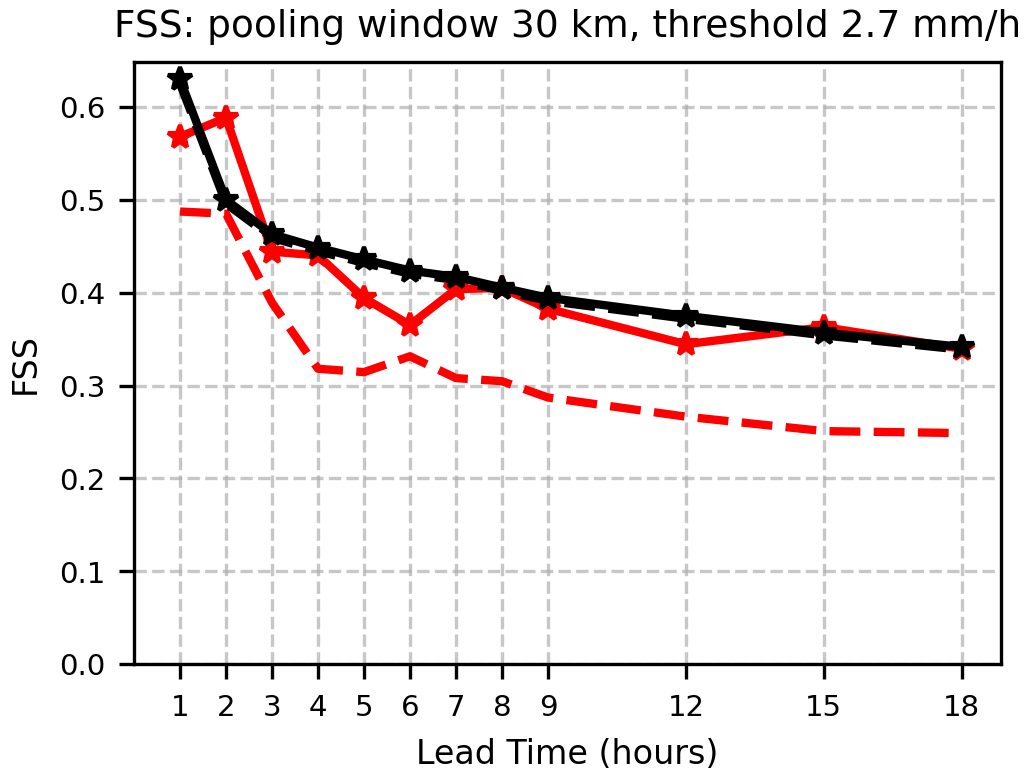}
\includegraphics[width=0.32\textwidth]{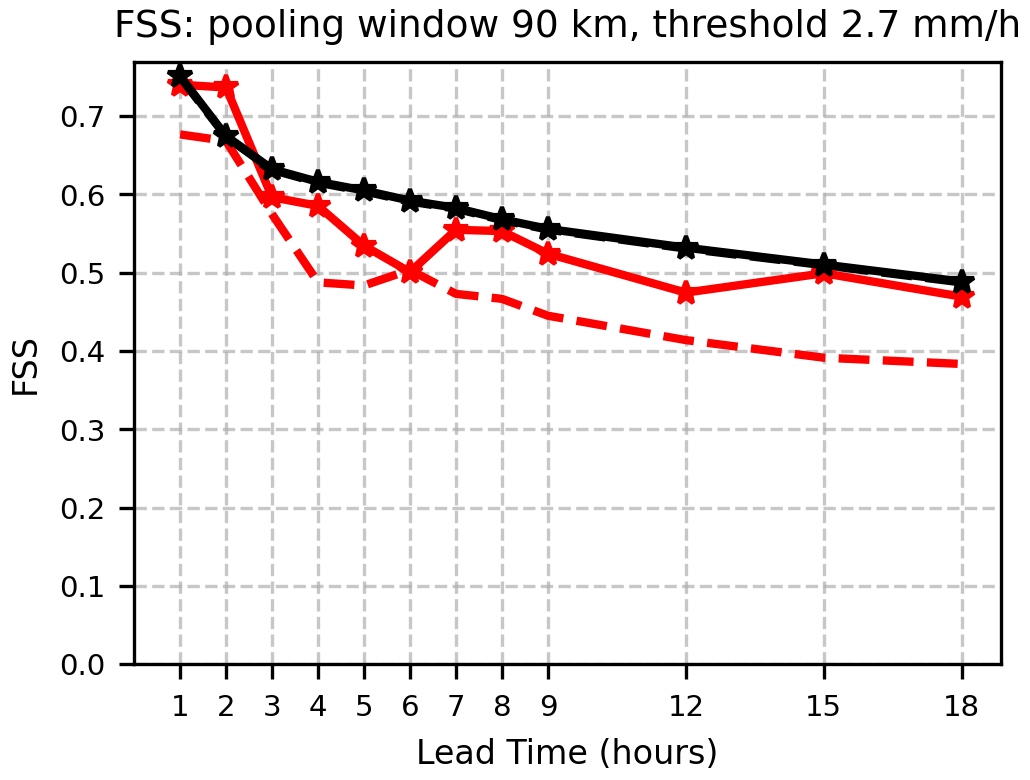}
\caption{Pooling-radius dependence of native-grid benefits. Top row: 40 dBZ reflectivity
for 6-, 30-, and 90-km pooling windows. Bottom row: 1-h precipitation at
2.7 mm h$^{-1}$ for 6-, 30-, and 90-km pooling windows. At larger pooling windows,
HRRR skill is relatively insensitive to evaluation grid spacing, while HRRRCast
continues to show a native-resolution advantage.}
\label{fig:fss_pooling_dependence}
\end{figure*}

To examine scale dependence directly, Fig.~\ref{fig:fss_pooling_dependence} compares
pooling radii from 6 to 90 km at fixed thresholds. For 40-dBZ reflectivity,
single-member HRRRCast is near parity with HRRR at 6-km pooling, but clearly
outperforms HRRR at larger pooling radii (30 and 90 km). A similar, though weaker,
pattern appears for moderate 1-h precipitation (2.7 mm h$^{-1}$).

This behavior is consistent with the verification setup. HRRR's FSS is nearly
insensitive to evaluation grid spacing for pooling radii $\geq 30$ km because the HRRR
6-km curve is simply a subsampled version of the same HRRR 3-km forecast field,
whereas HRRRCast 6 km is a separately trained model. Even so, HRRRCast retains a
persistent native 3-km advantage across pooling scales. This supports a real benefit
from native-resolution training, although the 3-km versus 6-km HRRRCast gap still
reflects both resolution and model-configuration differences.

\subsection{Attribution of extreme-threshold gains: loss reweighting}
\label{sec:extreme_precip}

This subsection shows that moderate, targeted loss reweighting improves
high-threshold skill without destabilizing performance on common events.

As described in Section~\ref{sec:sigmoid_loss}, the sigmoid-weighted balanced MSE loss
applies a capped sigmoid ramp that increases the event weight from approximately $1\times$
to $3\times$ for anomalies in the $4\sigma$--$10\sigma$ range, restricted to REFC and
APCP. The weighting scheme is intentionally moderate---substantially below the weights
of up to $800\times$ used in some nowcasting systems \citep[see e.g.,][]{shi2017deep,jin2024ureslham,cao2025enhancing}.
The candidate formulations explored during loss design are summarized in Fig.~\ref{fig:weighted_mse}.

\begin{figure*}[tbp]
\centering
\includegraphics[width=0.32\textwidth]{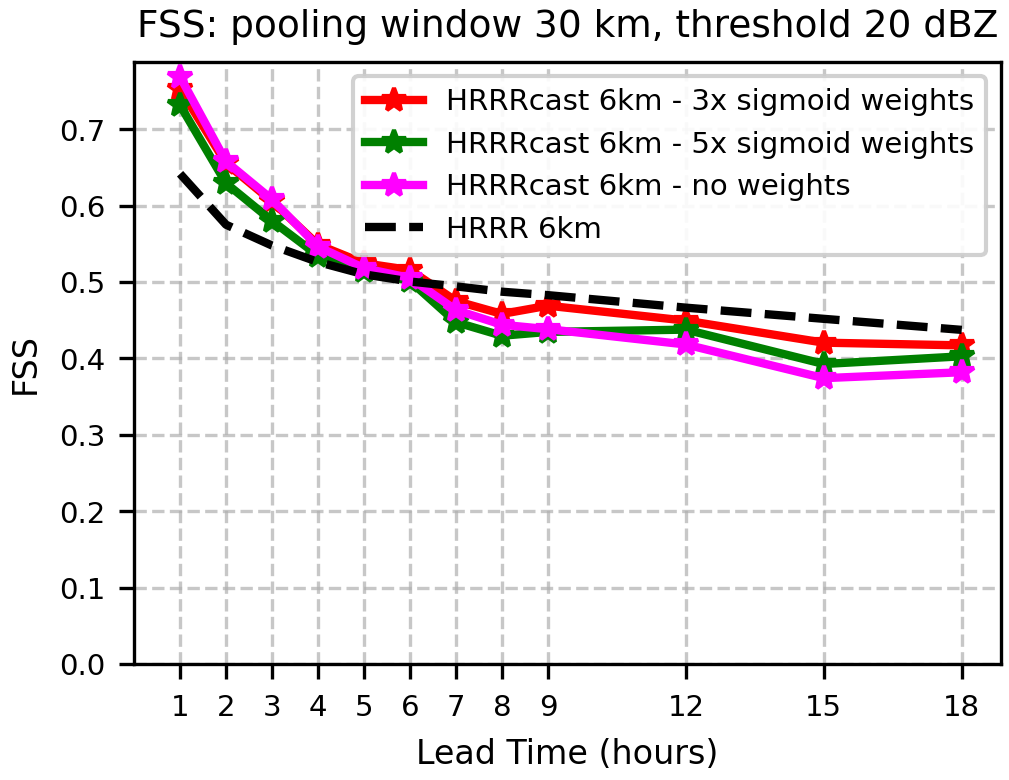}
\includegraphics[width=0.32\textwidth]{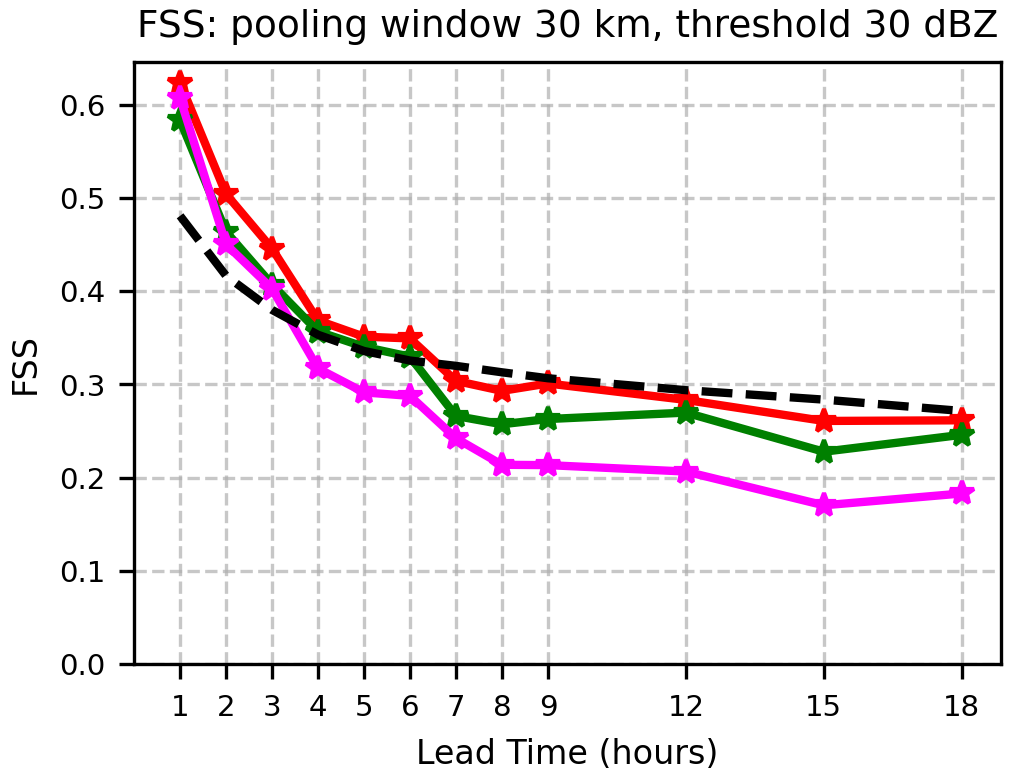}
\includegraphics[width=0.32\textwidth]{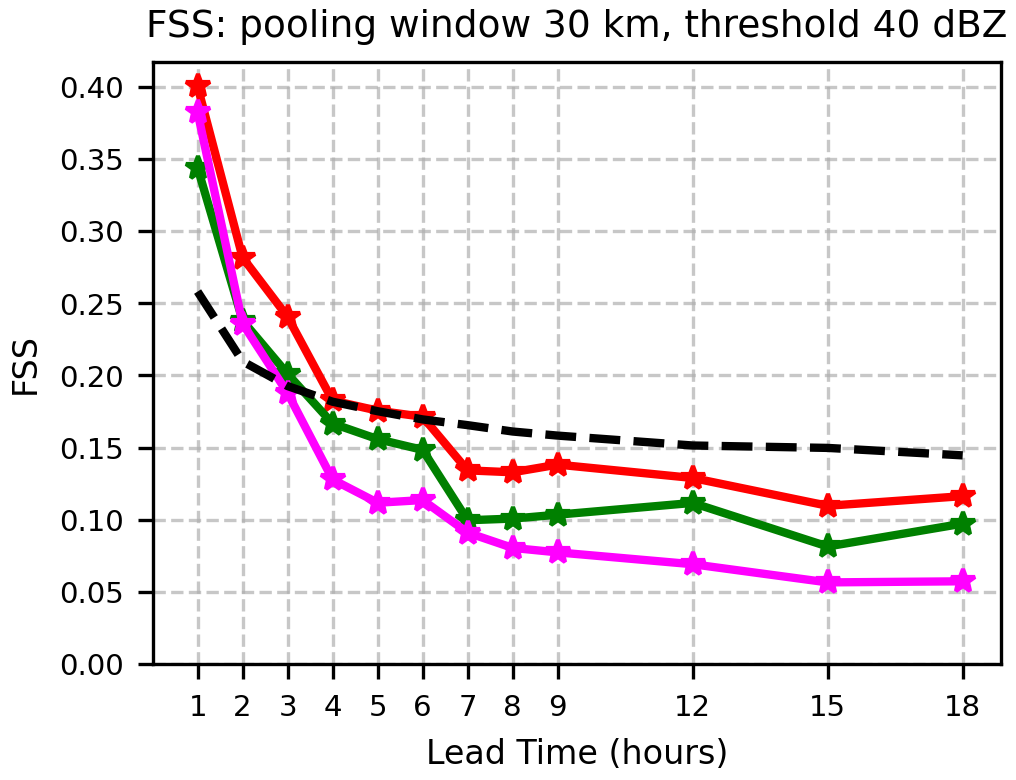}
\\[0.4em]
\includegraphics[width=0.32\textwidth]{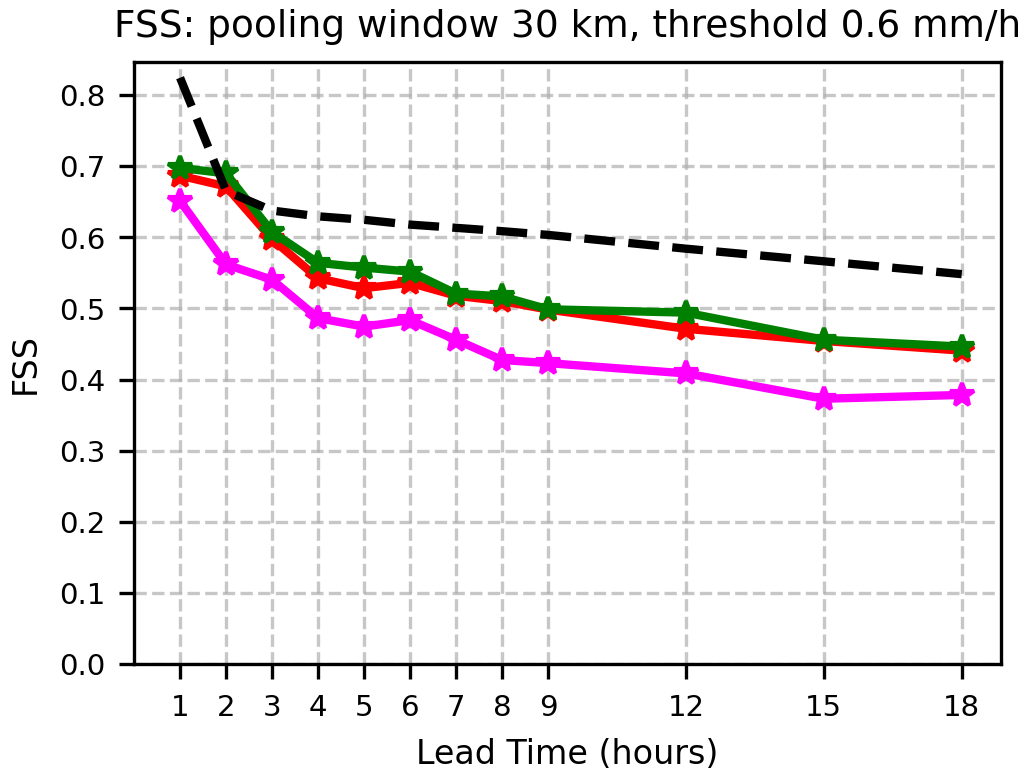}
\includegraphics[width=0.32\textwidth]{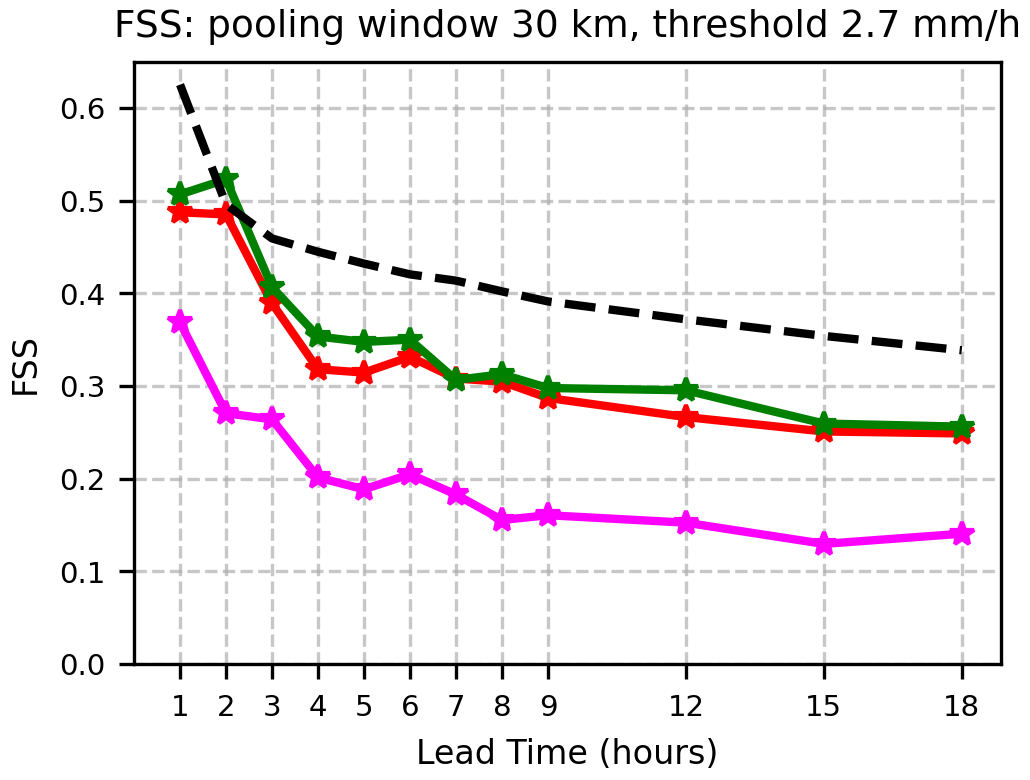}
\includegraphics[width=0.32\textwidth]{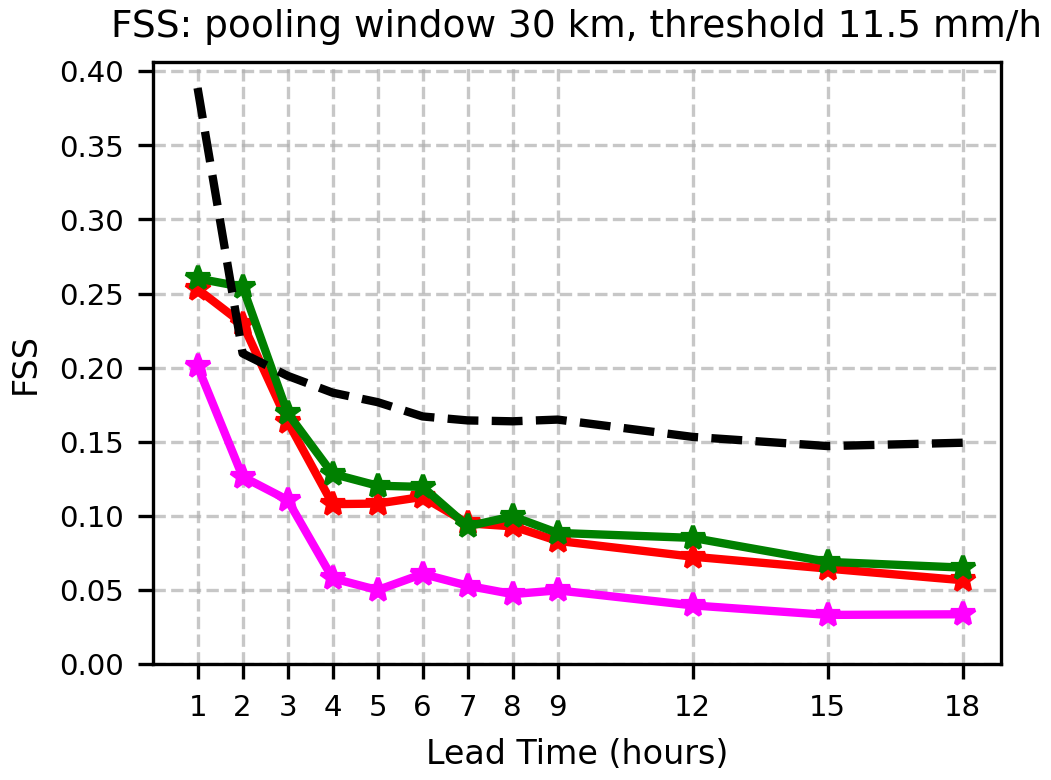}
\caption{Representative 6-km HRRRCast tuning experiment for extreme-event weighting
in the training loss. Top row: reflectivity thresholds (20/30/40 dBZ). Bottom row: 1-h
precipitation thresholds (0.6/2.7/11.5 mm h$^{-1}$). Weighted-loss variants improve FSS
relative to the unweighted model and move performance closer to HRRR; for reflectivity,
the $3\times$ variant is generally best, while for precipitation the $3\times$ and
$5\times$ curves are generally very similar, supporting the use of capped sigmoid
weighting for extreme-event forecasting.}
\label{fig:fss_weighted6km}
\end{figure*}

The 6-km tuning experiments in Fig.~\ref{fig:fss_weighted6km} show that introducing
sigmoid weighting consistently improves FSS over an unweighted loss, especially after
the first few lead times where rare-event penalties become most important. This
behavior appears in both reflectivity and 1-h precipitation verification. For the
reflectivity thresholds, the $3\times$ configuration is generally the best-performing
weighted variant, especially for the higher thresholds.
For the precipitation thresholds, the $3\times$ and $5\times$ variants are very similar,
indicating that increasing beyond moderate weighting provides little additional benefit.
These tests also reinforced the need for moderation: stronger weighting can improve
threshold-based skill while simultaneously making the forecast distribution less
realistic if the loss growth is too aggressive. Based on this behavior, we retained the
capped global sigmoid strategy for the production 3-km training to improve extremes
while limiting degradation on frequent, lower-intensity events. Mechanistically, the
sigmoid weighting improves high-threshold skill by redistributing optimization weight
across the reflectivity spectrum: because values below 20 dBZ dominate the training
distribution, uniform weighting tends to prioritize low-reflectivity skill and
under-emphasize higher-reflectivity events. Increasing the loss weight above 20 dBZ
rebalances this objective, improving higher-reflectivity forecasts while preserving
useful low-reflectivity behavior. The REFC 18-h forecast CDF comparison
(Fig.~\ref{fig:weighted_mse}, right panel) is consistent with this mechanism and
supports selecting the moderate $3\times$ configuration over the more aggressive
$5\times$ option.

\section{Storm-scale behavior}
\label{sec:storm_scale}

\subsection{Storm displacement}
\label{sec:storm_displacement}

Object-based and field-based displacement metrics consistently indicate
improved storm placement by HRRRCast relative to HRRR.

\begin{figure*}[h]
\centering
\includegraphics[width=0.32\textwidth]{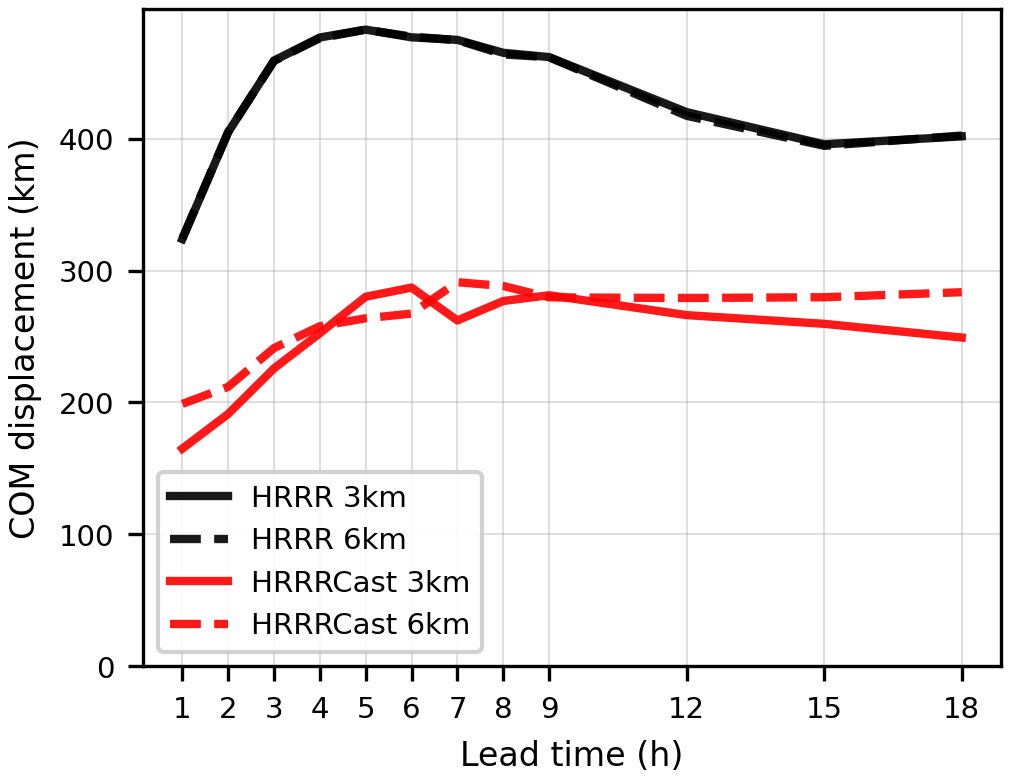}
\includegraphics[width=0.32\textwidth]{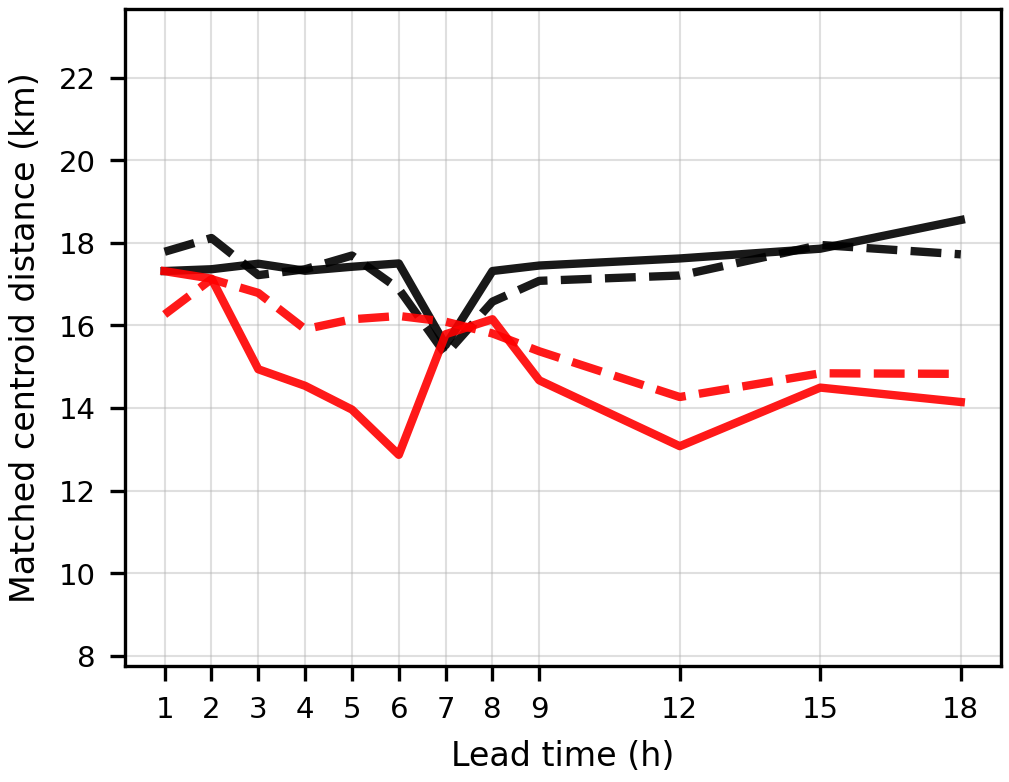}
\includegraphics[width=0.32\textwidth]{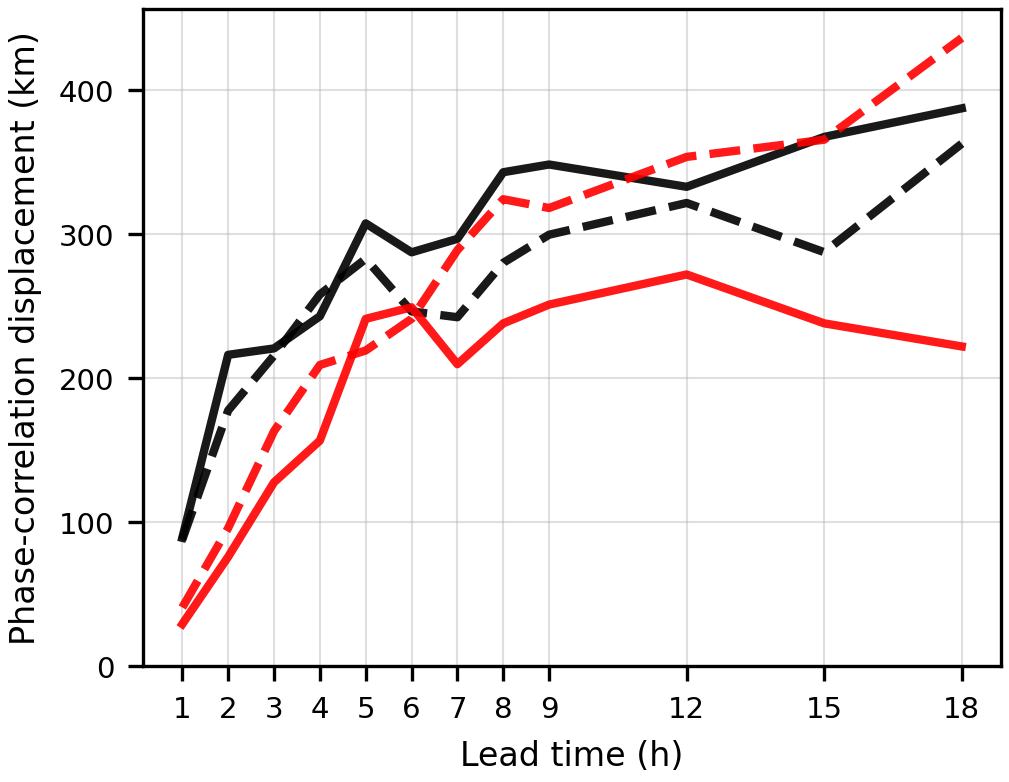}
\caption{Storm-placement diagnostics at 40 dBZ (left to right): center-of-mass (COM)
displacement, matched-pair centroid distance, and phase-correlation displacement.
Lower values indicate better placement fidelity. The first two panels are object-based,
whereas phase-correlation displacement is a field-based translational metric. Across
panels, HRRRCast (red) is generally better than HRRR (black) across lead times.}
\label{fig:structure_com_displacement}
\end{figure*}

We focus here on an object-based diagnostic that is directly interpretable for
storm-placement behavior. Figure~\ref{fig:structure_com_displacement} shows the 40 dBZ
results using two complementary object-based diagnostics: center-of-mass (COM)
displacement and matched-pair centroid distance.
Together, these metrics quantify bulk centroid placement and object-to-object placement
after matching. Across lead times, both metrics are generally lower for HRRRCast
than HRRR, indicating improved storm-location fidelity.
The phase-correlation displacement metric (third panel), which is not object-based and
instead estimates bulk translational offset from the full reflectivity field, shows the
same ranking: HRRRCast generally has smaller displacement than HRRR. Similar
results have also been reported in \citet{smith2026}.

\begin{figure*}[tbp]
\centering
\includegraphics[width=\textwidth]{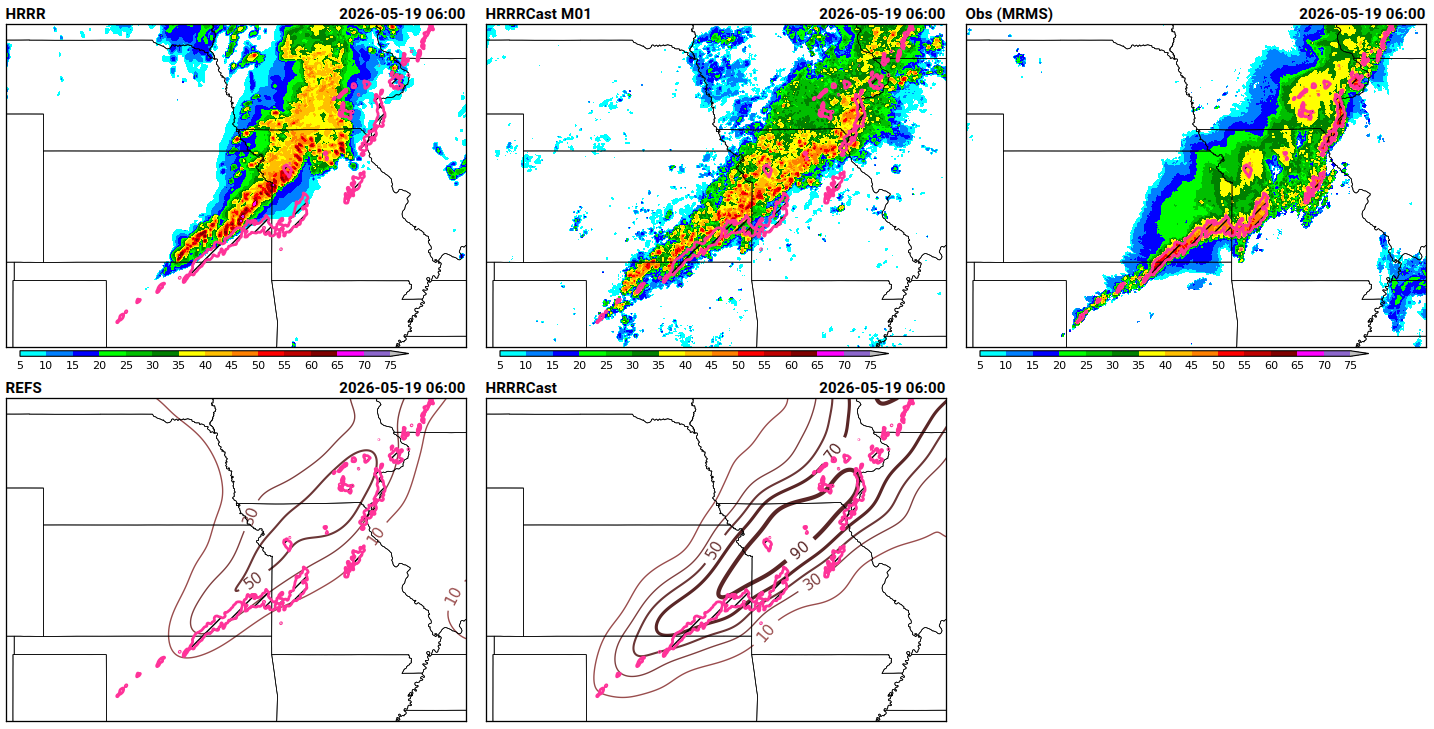}
\caption{Representative HWT real-time example from the 2026 Spring Forecasting Experiment
\citep{hwt2026} valid 2026-05-19 06:00 UTC, corresponding to a 30-h forecast initialized 
from the previous day's 00Z cycle, comparing HRRR, HRRRCast, REFS, and MRMS observations.
Top row shows deterministic composite reflectivity. Bottom row shows probability guidance 
for reflectivity $\geq 40$ dBZ. In this case, HRRRCast places the main convective swath 
closer to the observed MRMS line than REFS, consistent with improved storm-placement behavior.}
\label{fig:hwt_sample_displacement}
\end{figure*}

This ranking is consistent with recent real-time Hazardous Weather Testbed (HWT)
evaluations \citep{hwt2026}. In the representative case shown in Fig.~\ref{fig:hwt_sample_displacement},
HRRRCast aligns the convective line more closely with MRMS than HRRR in deterministic reflectivity,
and the $\geq 40$ dBZ probability swath is likewise better co-located than REFS (the ensemble Rapid Refresh Forecast System).
This qualitative case is consistent with the displacement metrics above and highlights
the operational value of probabilistic severe-convection guidance.

\subsection{Storm structure}
\label{sec:storm_structure}

Despite placement gains, structure diagnostics reveal persistent morphology
deficits, particularly at higher reflectivity thresholds and longer lead times.

\begin{figure*}[tbp]
\centering
\includegraphics[width=0.32\textwidth]{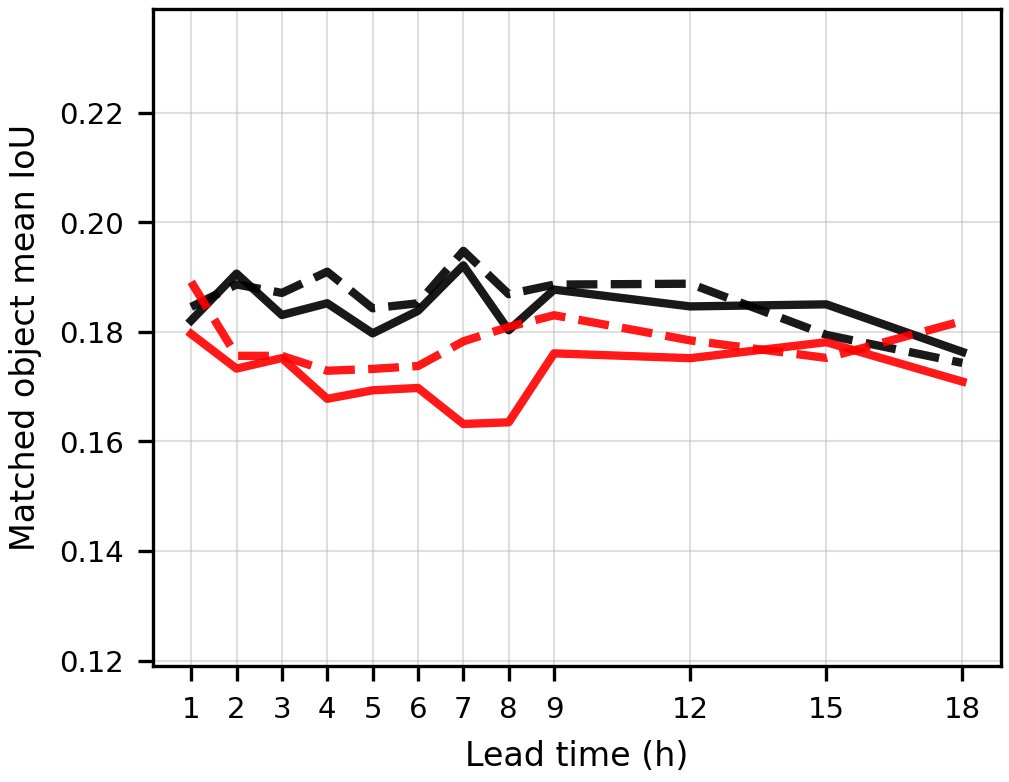}
\includegraphics[width=0.32\textwidth]{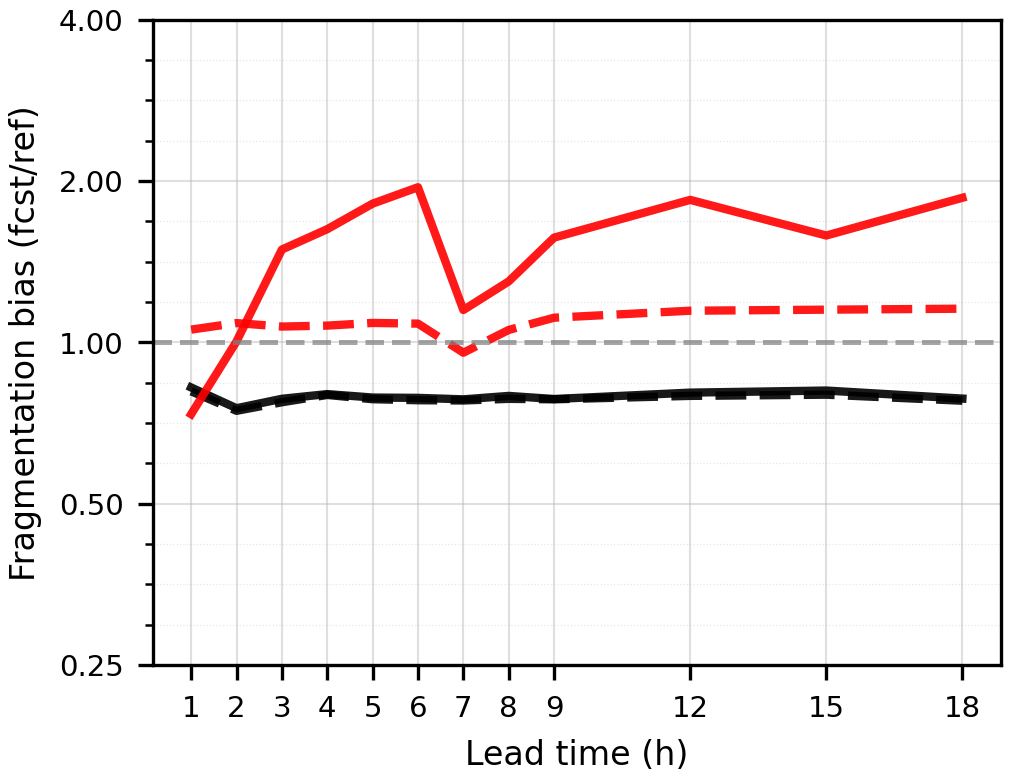}
\includegraphics[width=0.32\textwidth]{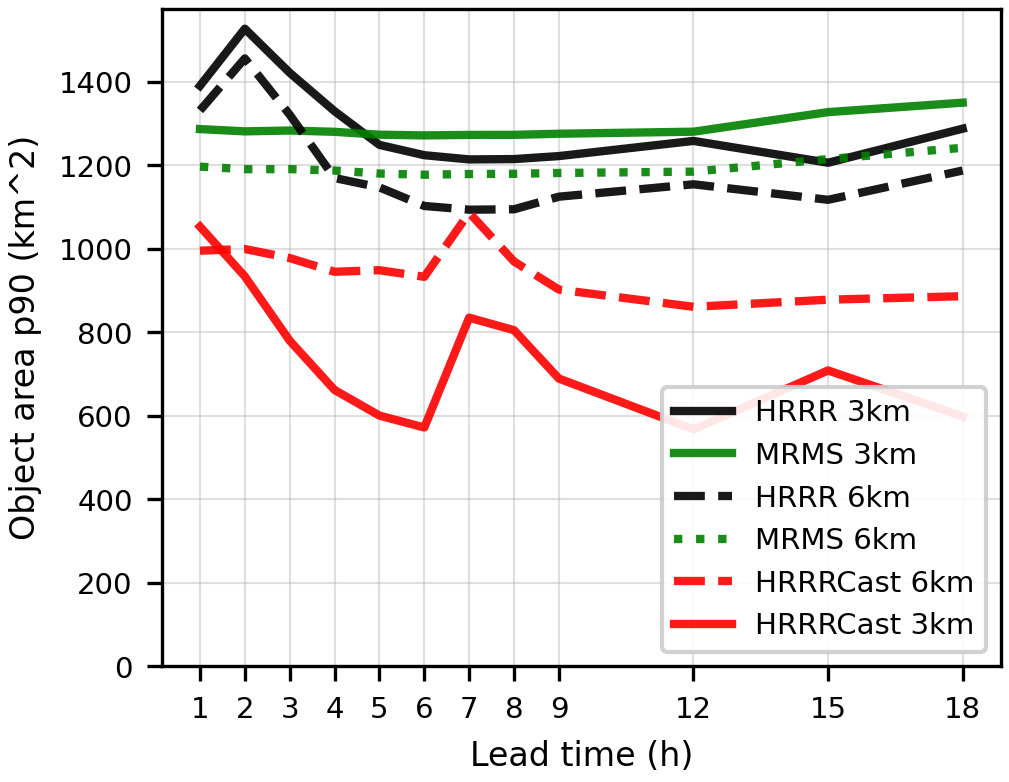}
\\[0.4em]
\includegraphics[width=0.32\textwidth]{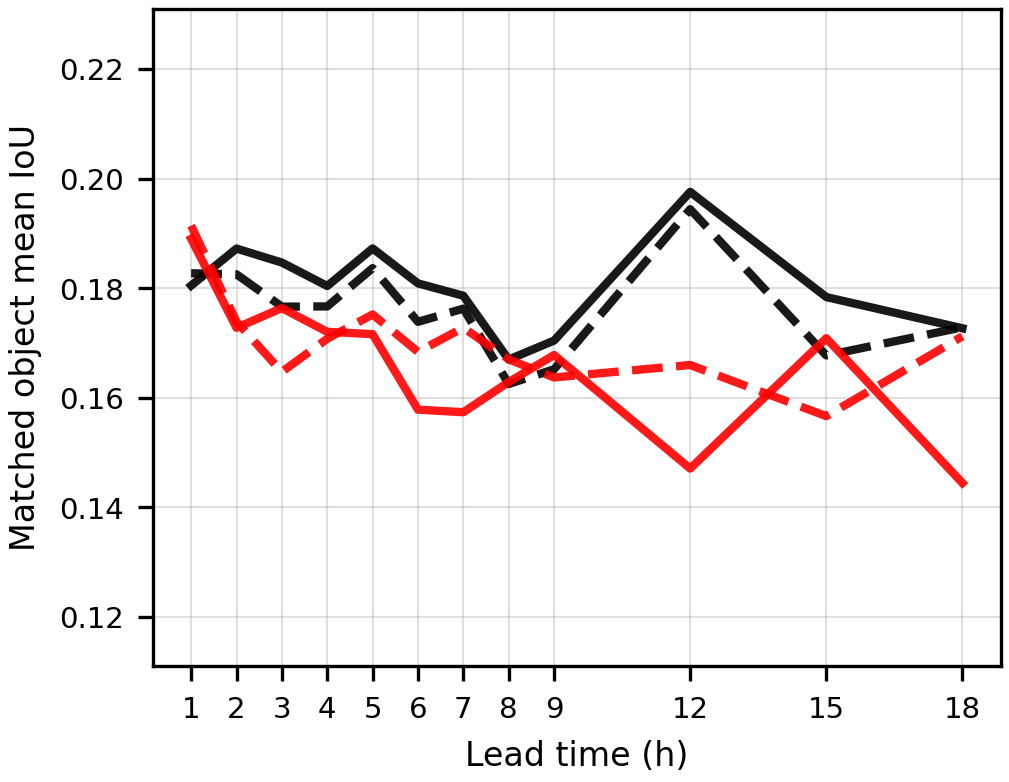}
\includegraphics[width=0.32\textwidth]{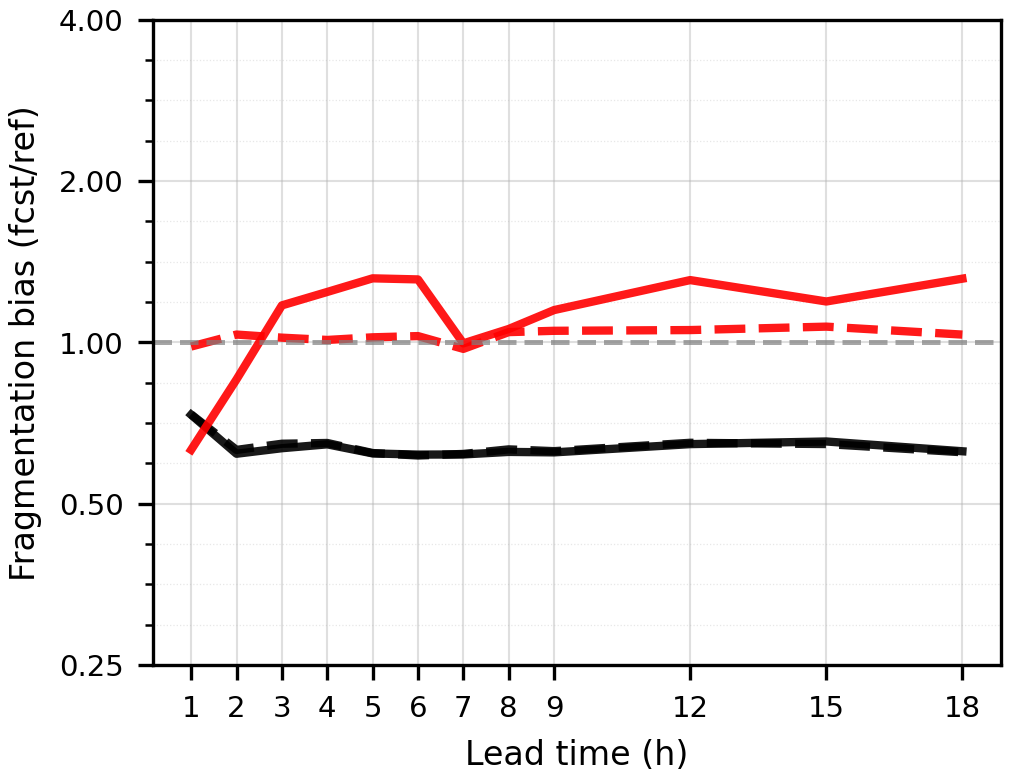}
\includegraphics[width=0.32\textwidth]{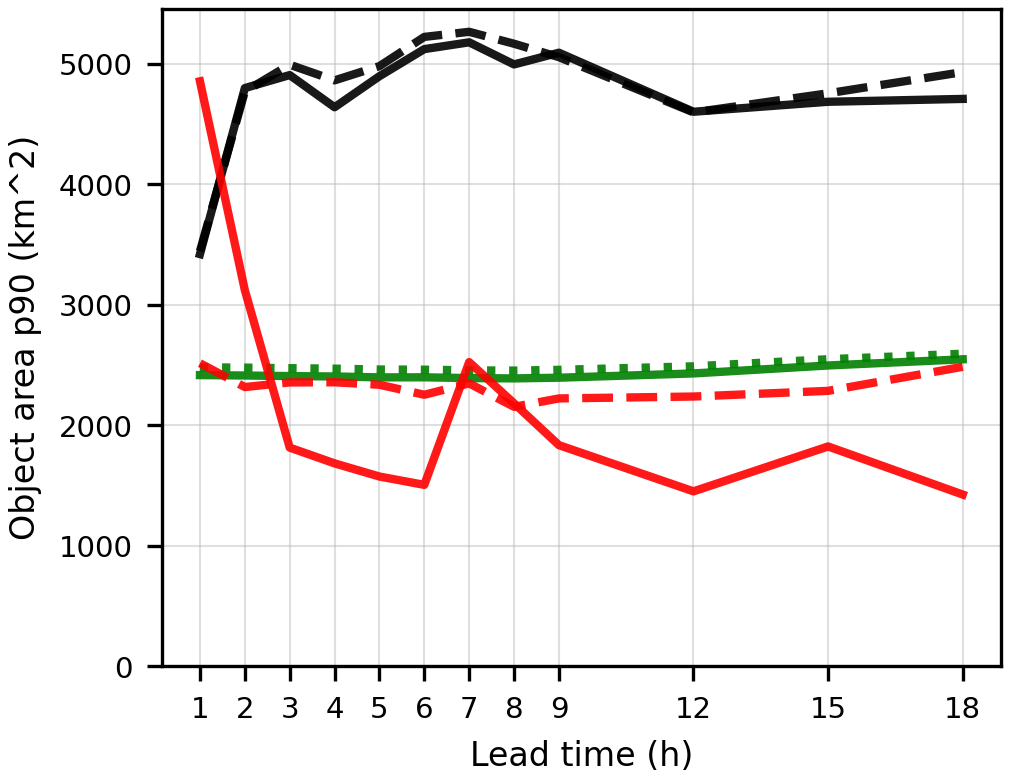}
\caption{Objective storm-structure metrics at 40~dBZ shown for two object-size
filters. Columns are (left to right): matched-object intersection over union
(IoU; higher is better), fragmentation bias (fcst/ref; 1 is ideal), and forecast
90th-percentile object area. The top row uses a 144~km$^2$ minimum-area threshold
(small-object filter; 4$\times$4 grid-cell minimum), and the bottom row uses 576~km$^2$
(MCS-scale filter; 8$\times$8 grid-cell minimum). Across both thresholds,
HRRRCast 3~km remains over-fragmented while HRRR remains under-fragmented. The
object-area contrast is scale dependent: at 144~km$^2$, HRRRCast 3~km tends to
underestimate object size; at 576~km$^2$, HRRR substantially overestimates object
size, while HRRRCast 6~km is closest to MRMS.}
\label{fig:structure_object_metrics}
\end{figure*}

To complement the storm-displacement diagnostics above, we evaluate storm structure
with object-based metrics designed to isolate morphology (overlap, fragmentation,
and object-size realism). Figure~\ref{fig:structure_object_metrics} summarizes
representative thresholds and lead-time behavior using two minimum object area thresholds:
144~km$^2$ ($4\times4$ grid cells) and 576~km$^2$ ($8\times8$ grid cells).
These thresholds bracket the effective resolution of HRRR, which is approximately 
$7\,\Delta x \approx 21$~km \citep{duda2023}, corresponding to a feature area
of roughly 441~km$^2$. Therefore, the 576~km$^2$ threshold is expected to filter the population 
to mesoscale convective system (MCS)-scale systems that HRRR can resolve.
To avoid altering morphology-sensitive metrics, objects are defined directly from
threshold exceedance regions using 8-connectivity, without pre-smoothing or
convolution. \citet{duda2023} likewise minimized smoothing by using
convolution radius of just 1.

Matched-object Intersection over Union (IoU) at 40~dBZ (first column) lies mostly in the
0.15--0.20 range across both rows. HRRRCast 3~km is generally lower than HRRR, while
HRRRCast 6~km is consistently higher than HRRRCast 3~km and is often closer to HRRR. 
The gap is real but modest, indicating that once objects are matched, overlap quality is 
broadly similar.

Fragmentation bias at 40~dBZ (middle column) tells a more definitive story.
We define fragmentation bias as the ratio of forecast to observed object number density,
that is, object count per unit covered area: $FB_{\rm frag}=(N_f/A_f)/(N_o/A_o)$. Equivalently,
since the mean object area is $\bar A=A/N$, this can be written as
$FB_{\rm frag}=\bar A_o/\bar A_f$. Thus, values above 1 indicate that forecast objects
are too small on average (over-fragmented), values below 1 indicate that forecast objects
are too large or too merged (under-fragmented), and a value of 1 indicates the correct mean object size.
HRRR consistently under-fragments relative to MRMS, with bias generally around
0.65--0.75, meaning it produces fewer but larger objects. HRRRCast 3~km, by contrast,
is over-fragmented at both thresholds: the effect is stronger in the 144~km$^2$ row
(often about 1.3--2.0 after the first few leads) and milder in the 576~km$^2$ row
(roughly 1.0--1.4). HRRRCast 6~km is intermediate, remaining near neutral to mildly
over-fragmented (about 1.0--1.2).

Comparing rows in Fig.~\ref{fig:structure_object_metrics} clarifies how these structure
biases depend on object scale. The largest row-dependent change appears in the object-area 
column: at 144~km$^2$, HRRRCast 3~km yields smaller  90th-percentile (P90) object areas 
(consistent with many small fragments), whereas at 576~km$^2$ HRRR yields substantially 
larger P90 areas than MRMS ($\sim$4500~km$^2$ vs.~$\sim$2500~km$^2$), consistent with 
merged, oversized MCS-scale objects. HRRRCast 6~km is closest to MRMS in the 576~km$^2$ panel. 
Together, the row-wise comparison reinforces a complementary error structure: HRRRCast 3~km is too
fragmented for smaller objects, while HRRR is too merged at MCS scales.

The representative HWT case in Fig.~\ref{fig:hwt_sample_displacement} shows the same
mixed behavior: good line placement but more fragmented high-intensity reflectivity
structure than MRMS. This pattern is consistent with an important limitation of the current
setup---predominantly 1-h supervision, which leaves the model to infer substantial
sub-hour storm evolution and may degrade fine-scale convective organization.

\subsection{Discussion: Storm structure in AI weather models}
\label{sec:storm_structure_discussion}

In this section, we discuss three candidate mechanisms behind storm-structure
limitations in current AI regional weather models: the coarse autoregressive time step,
the choice of training target, and the use of deterministic versus generative
training objectives.

\paragraph{Base autoregressive cadence.}
Operational convection-allowing NWP models (e.g., HRRR) integrate prognostic equations
with internal time steps on the order of seconds, enabling frequent dynamical adjustment
of storm organization.  Most current AI regional models, by contrast, evolve the atmosphere 
at $\geq$1-h intervals and must therefore infer the full sub-hour storm evolution from a single 
autoregressive step rather than resolving it. Within this fixed cadence, subhourly states can be 
generated either by direct lead-time-conditioned queries or by short-step autoregressive rollout: the latter can
improve short-time continuity but may accumulate stepwise error with depth.
High-frequency AI stepping is possible, but stable long-range short-step composition
remains technically difficult. StormScope \citep{pathak2026} uses 10-min stepping for
nowcasting but reverts to 1-h stepping for longer leads, whereas STRATA \citep{hu2026} demonstrated
stable 10-min rollouts up to 24 h only with large model capacity ($\sim$330M parameters),
autoregressive fine-tuning, and multiple stabilization measures, including local spatial
processing, de-aliasing, overlapping-tile blending, and physical/stability constraints.
On the other hand, physics-based CAMs can maintain much  shorter internal time steps over longer 
integrations, which likely contributes to stronger long-lead storm morphology, though they 
also accumulate displacement error (Section~\ref{sec:storm_displacement}).

\begin{figure*}[tbp]
\centering
\setlength{\tabcolsep}{0pt}
\renewcommand{\arraystretch}{0.98}
\begin{tabular}{@{}r@{\hspace{0pt}}c@{}}
 & \makebox[0.86\textwidth][c]{%
\makebox[0.286\textwidth][c]{\textbf{HRRRCast}}%
\makebox[0.286\textwidth][c]{\textbf{MRMS}}%
\makebox[0.286\textwidth][c]{\textbf{HRRR}}} \\
\makebox[0pt][r]{\textbf{15 min}}  & \includegraphics[width=\textwidth]{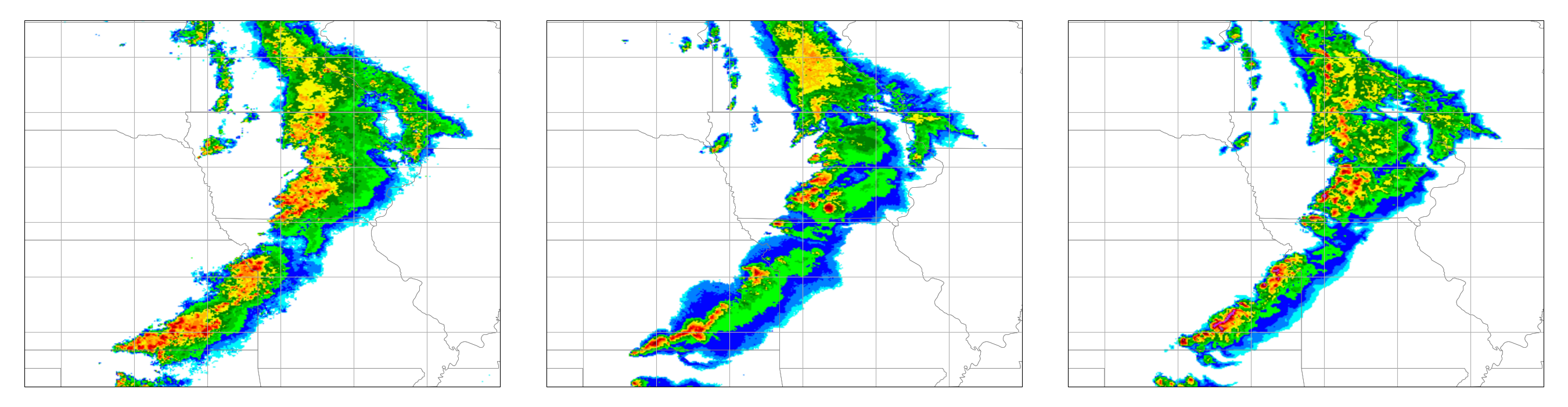} \\[-0.15em]
\makebox[0pt][r]{\textbf{45 min}}  & \includegraphics[width=\textwidth]{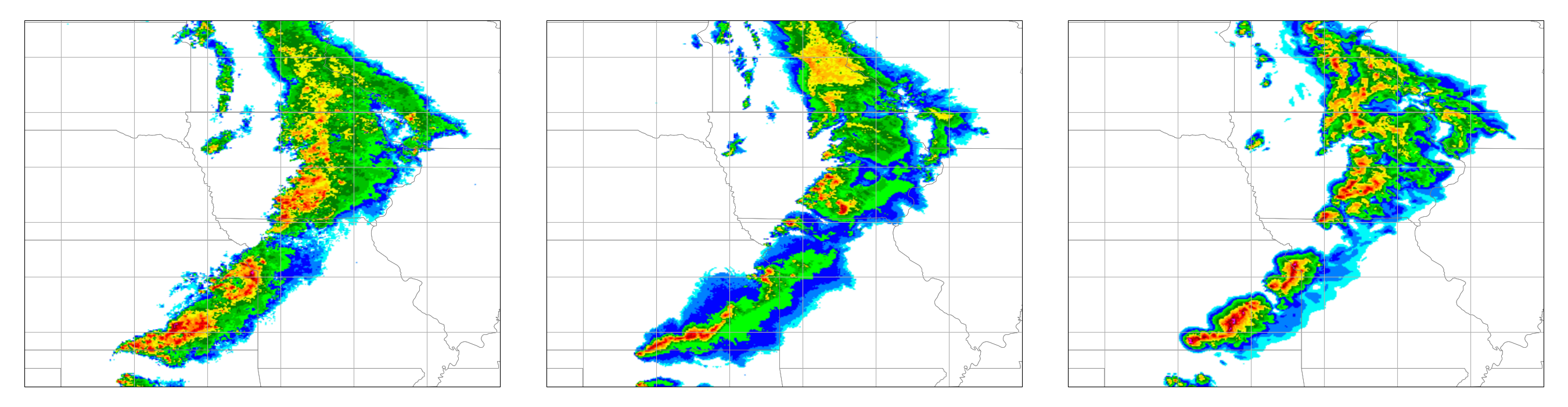} \\[-0.15em]
\makebox[0pt][r]{\textbf{75 min}}  & \includegraphics[width=\textwidth]{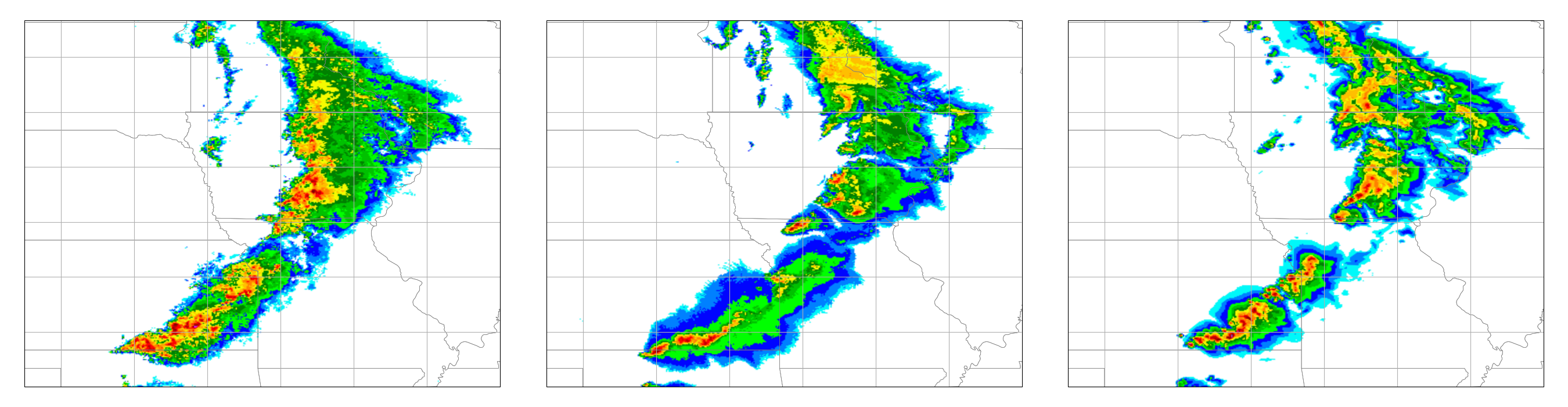} \\[-0.15em]
\makebox[0pt][r]{\textbf{135 min}} & \includegraphics[width=\textwidth]{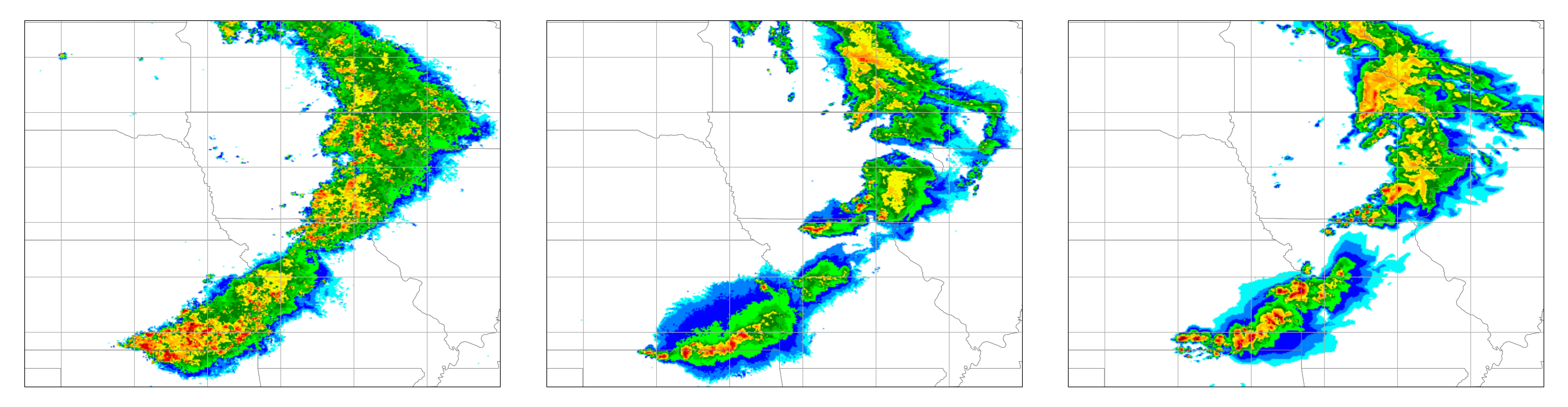} \\[-0.15em]
\makebox[0pt][r]{\textbf{165 min}} & \includegraphics[width=\textwidth]{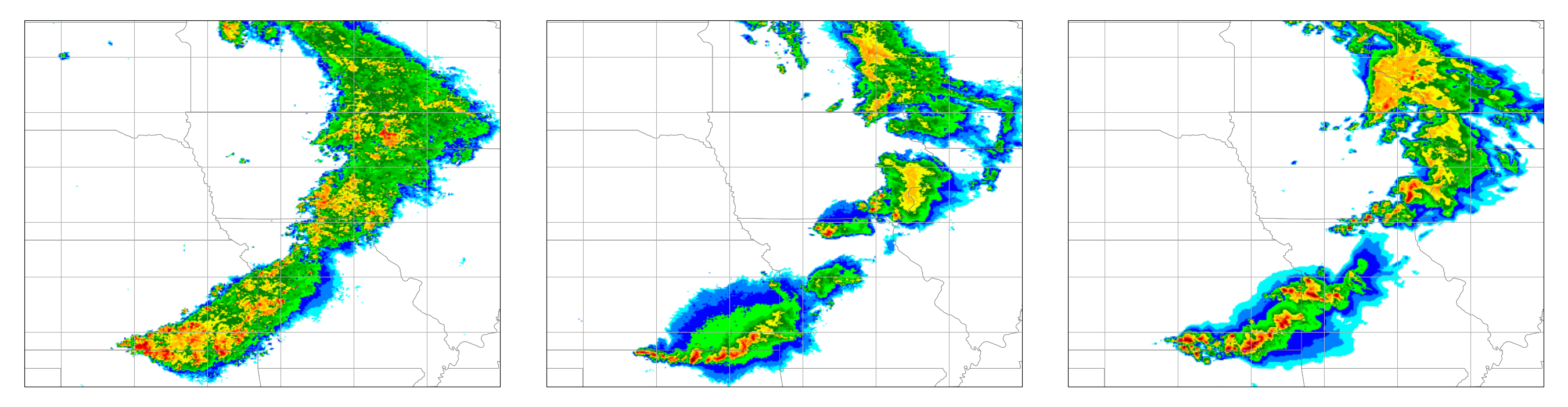} \\
\end{tabular}
\caption{Representative subhourly composite-reflectivity evolution for a 3-h forecast at 2024-05-01T00
comparing diagnostically queried HRRRCast fields with MRMS and HRRR at 15-min intervals.
HRRRCast is temporally smooth but shows broader, more connected reflectivity than MRMS,
whereas HRRR appears sharper and more segmented.}
\label{fig:subhourly_sequence}
\end{figure*}

We run an experiment with an experimental HRRRCast subhourly-inference configuration to
show that subhourly inference is not equivalent to subhourly supervision.
Figure~\ref{fig:subhourly_sequence} shows a representative 3-h case where HRRRCast is
queried diagnostically at fractional lead times rather than run as a separately trained
subhourly forecast system; these queries yield smooth frame-to-frame evolution, yet still
exhibit storm over-connectivity and reduced fine-scale convective realism.
In this example, HRRRCast shows stronger band continuity than both MRMS and HRRR, but
with broader, more diffuse areal coverage and a missed southern separation that
contributes to local over-connection. MRMS retains stronger cell-scale intermittency,
whereas HRRR appears more segmented with sharper compact cores.

More broadly, there are at least three routes to subhourly AI forecasts: explicit
subhourly state evolution, fractional lead queries from a lead-time-conditioned model
as in HRRRCast, and temporal downscaling of a coarser forecast (e.g.,
\citet{ingstad2026}). These approaches differ mainly in how fine-time structure is
generated and constrained. The first and third rely on explicit subhourly
supervision, whereas HRRRCast can generate temporally smooth intermediate states and
some fine-scale variability without directly constraining those states with
subhourly observations or analyses. For the supervised approaches, forecast or
observational training data (e.g., \citealt{ingstad2026,hu2026,pathak2026}) are
likely needed, rather than analyses or reanalyses alone, to construct temporally
consistent fine-time forecast sequences. A fourth, fundamentally different route is
physics-AI hybrid evolution, as in WeatherGFT \citep{xu2024weathergft}, in which a fixed numerical PDE operator
advances the atmospheric state at fine internal time steps and a parallel neural branch
corrects unresolved processes and accumulated error, enabling subhourly states without
direct subhourly supervision .

\paragraph{Training target.}
When an AI model is trained against NWP model output---whether analyses or short-range
forecasts---it may learn to reproduce spatial smoothing, effective-resolution
limits, and structural artifacts the NWP system itself introduces.  Training directly
against radar observations removes this layer of NWP-induced distortion.
Consistent with this mechanism, StormScope results show improved morphology under
MRMS-supervised training relative to NWP-supervised targets \citep{pathak2026}.
HRRRCast trains on HRRR analyses, which are internally consistent but still inherit
NWP effective-resolution and spectral characteristics; radar-supervised or blended
observation/analysis targets are therefore a promising path for improving fine-scale
storm realism.

\paragraph{Deterministic versus generative objective.}
Even with fine spatiotemporal resolution, deterministic models trained under pointwise
losses tend toward conditional means, producing spatially blurred output.
This smoothing occurs because pointwise losses reward averages over plausible
realizations rather than any single sharp realization. Generative training objectives
can mitigate this and recover sharper textures (e.g., diffusion or token-based
generative approaches), but objective choice alone is not sufficient
\citep{andrychowicz2023,franch2025,agrawal2025}. In practice, any benefit from a
generative objective remains coupled to cadence and supervision target: 1-h stepping
limits sub-hour evolution fidelity, and HRRR-analysis supervision constrains the
distribution the model learns.

These three factors expose a fundamental tension in current AI weather modeling:
models operating at coarse autoregressive cadences ($\geq$1~h), such as HRRRCast,
can achieve competitive storm placement out to 48~h but structurally under-resolve
fine-scale convective organization, while short-step AI models that better resolve
morphology are confined to the nowcasting regime by rapid error accumulation
($\lesssim$6~h; e.g., \citealt{pathak2026}).
No present AI strategy simultaneously achieves both goals at extended convective leads.
Absent a fundamental advance that simultaneously enables stable long-range autoregression 
with small timesteps and faithful representation of fine-scale convective dynamics, degradation of 
storm structure at extended leads ($\geq$12~h) remains a fundamental limitation of current 
AI weather models rather than merely an architecture-specific deficiency.

Additionally, given the observed displacement--structure contrast between AIWP and NWP models, 
it is important to  motivate a balanced evaluation framework in which storm-structure 
inspection is paired with storm-displacement metrics and other object-based and scale-aware 
diagnostics (e.g., FSS across thresholds/scales, neighborhood methods, and probabilistic reliability), 
so that model differences are treated fairly and operationally relevant skill is not underestimated.

\section{Conclusions}
\label{sec:conclusions}

This study shows that regional AI weather forecasting can become more operationally useful 
through targeted treatment of extreme events, native-resolution training, expanded output 
coverage, and sustained near-real-time verification. The principal findings are as follows.

\textbf{Native 3-km training substantially improves convective-scale skill.} Training
directly on the native 3-km HRRR grid, rather than sub-sampled 6-km data, avoids
the aliasing effects incurred in the earlier 6-km configuration. Fractions skill score verification shows
consistent gains over the 6-km baseline across reflectivity and precipitation thresholds,
with 20- and 30-dBZ reflectivity generally better than HRRR and 40-dBZ behavior that is
near parity at 6-km pooling but better at larger pooling windows.

\textbf{Sigmoid-weighted loss improves extreme-event forecasting.} Applying a capped
sigmoid weighting schedule selectively to composite reflectivity and hourly precipitation
improves high-threshold FSS relative to an unweighted baseline. Across tuning experiments,
the $3\times$ cap provides the best trade-off between rare-event gains and preserving
common-event behavior.

\textbf{Expanded observational verification confirms broad skill improvements.} Seven
months of near-real-time observational verification using MATS demonstrate that
HRRRCast ensemble means outperform operational HRRR in RMSE for relative humidity,
2-m dewpoint, 10-m wind speed. HRRRCast also outperforms HRRR in CSI at multiple 
reflectivity thresholds. For hourly precipitation CSI, HRRRCast is competitive through
1.00 in and shows especially strong improvement in the final two-month subset of the
near-real-time archive.

\textbf{Storm-scale evaluation reveals a displacement--structure trade-off.}
Across Section~\ref{sec:storm_scale}, displacement diagnostics generally favor HRRRCast, while structure
diagnostics show persistent fragmentation and over-connectivity biases, consistent with limits 
imposed by hourly supervision cadence and training-target characteristics; this distinction is
illustrated in Fig.~\ref{fig:subhourly_sequence}, where temporally smooth fractional
lead-time inference can still coexist with storm-structure limitations.

Overall, HRRRCast shows clear skill gains from native-resolution training and targeted
loss design, while fine-scale storm morphology remains the main outstanding limitation.

\section{Future Directions}
\label{sec:future_directions}

Several avenues for further improvement are identified by the current verification results
and methodological choices.

\textbf{Sub-hourly training cadence.} An important limitation in this context is training
supervision rather than model inference capability. HRRRCast can generate forecasts at
sub-hourly intervals if needed because lead time is provided explicitly as a model
input. However, training is currently dominated by hourly targets, which likely imposes
a ceiling on storm-scale morphology fidelity relative to physics-based models that
integrate at sub-minute time steps. As discussed in the storm-structure section,
training with 10-minute or finer targets could provide a direct test of whether finer 
temporal supervision improves storm-structure realism. 

\textbf{Broader variable coverage in observational verification.} MATS-based
observational verification is currently limited to a subset of HRRRCast outputs.
Extending MATS verification to cloud cover, visibility, CAPE, CIN, and 80-m wind
variables would enable independent benchmarking of the full expanded output set
beyond the HRRR-analysis-relative RMSE diagnostics presented here.

\textbf{Ensemble calibration.} Future work will focus on improving ensemble spread and
threshold-dependent probabilistic calibration, especially for precipitation and other 
rare-event variables.

\section*{Acknowledgment}
 The authors gratefully acknowledge support from the Software Environment for Novel Architectures
(SENA) effort at GSL under grant NA24OARX405C0022. Additional support for the first, third
through sixth, and ninth authors was provided by NOAA cooperative agreement NA22OAR4320151
through the Cooperative Institute for Earth System Research and Data Science (CIESRDS).
Additional support for the third and tenth authors was provided by NOAA cooperative agreement
NA24OARX432C0007-T1-01. This research used resources of the NOAA Research and Development High
Performance Computing System (RDHPCS; http://rdhpcs.noaa.gov). We would like to thank Liaofan Lin
for reviewing the manuscript, David Dowell and Jeffrey Duda for helpful discussions.

\section*{Disclaimer}
The statements, findings, conclusions, and recommendations are those of the author(s) and do not
necessarily reflect the views of NOAA or the U.S. Department of Commerce.

\section*{Availability Statement}
HRRRCast has been running in near-real-time for nearly one year. Forecasts are initialized
with HRRR analyses, disseminated publicly via the NOAA-GSL AWS bucket, archived on NOAA
HPSS, and visualized through DESI. Code, data access, visualization, and verification
resources are available at:
\begin{itemize}
\item \url{https://github.com/NOAA-GSL/HRRRCast-live/tree/master}
\item \url{https://noaa-gsl-experimental-pds.s3.amazonaws.com/index.html#HRRRCast}
\item \url{https://sites.gsl.noaa.gov/desi}
\item \url{https://gsl.noaa.gov/mats}
\end{itemize}

\clearpage

\amsappendixA
\appendixtitle{Methods and Data Details}

\section{Variable Set}
\label{append:variable_set}
The variable selection follows a utility-first expansion strategy similar to recent
operational AI-forecast model updates \citep{moldovan2025}: we aim to cover as much of
the operational forecast space as possible while preserving training consistency.
Practically, this means retaining a physically complete atmospheric core, then adding
surface and integrated fields that are valuable for hazardous-weather interpretation and 
downstream applications, but only when they are consistently available and quality-controlled 
across the multi-year archive.

Table~\ref{inputs} summarizes the full HRRRCast variable configuration used for
training. The dynamic meteorological fields are treated as prognostic variables
(ingested and predicted autoregressively), while static fields (terrain height and land
mask) are included as fixed conditioning inputs.

The atmospheric core consists of six 3D variables (TMP, HGT, UGRD, VGRD, VVEL, SPFH)
on 20 HRRR pressure levels from 1000 to 200 hPa, with denser sampling in the lower
troposphere (1000--800 hPa) to better resolve boundary-layer and convective structure.
GFS conditioning for these 3D variables is provided on four standard levels
(1000, 850, 500, and 250 hPa), supplying large-scale synoptic context without requiring
full vertical matching often used in nested AI models such as \citep{smith2026}.

The surface/integrated block includes near-surface thermodynamics and winds, cloud-cover
fractions, visibility, precipitation, reflectivity, and instability diagnostics. The
loss-weight column reflects practical forecast priorities: highest weights are assigned to
core-impact variables such as composite reflectivity, hourly precipitation, and 2-m
temperature; intermediate weights are used for dewpoint, cloud/visibility and CAPE-like
fields; and lower weights are assigned to noisier or lower-priority channels (e.g.,
80-m winds and pressure fields) to stabilize multi-target training. For the 3D atmospheric
variables, the listed 1.0 weight is distributed across levels with larger weight near the surface \citep{abdi39}.

Unlike recent work that applies variable-specific bounding layers (e.g., ReLU/HardTanh)
to enforce output ranges \citep{moldovan2025}, HRRRCast does not hard-constrain the
network head. Instead, we clip in normalized space using variable-wise min/max limits
estimated from the training dataset. This design choice is tied to the model
formulation: the network predicts residual tendencies that are added to the current
state $x_t$ through an output skip connection, and several targets are represented in
transformed space (log-based transforms and z-normalization). In this setting, a simple
bounded activation at the network output constrains only the residual increment, while
the final prediction after the additive update ($x_t + \Delta x$) can still move outside
the intended range. That makes robust, physically consistent range enforcement difficult
to impose uniformly across variables and can suppress needed large corrections during
autoregressive rollout.

\begin{table*}
\centering
\caption{HRRRCast input variables and pressure levels.}
\begin{tabular}{p{6.4cm}p{3.45cm}p{2.75cm}p{2.1cm}}
\toprule
\textbf{Variables} & \textbf{HRRR levels (hPa)} & \textbf{GFS levels (hPa)} & \textbf{Loss weight} \\
\midrule
\textit{Atmospheric variables:} \newline
Temperature (\textbf{TMP}) \newline
Geopotential height (\textbf{HGT}) \newline
U component of wind (\textbf{UGRD}) \newline
V component of wind (\textbf{VGRD}) \newline
Vertical velocity (\textbf{VVEL}) \newline
Specific humidity (\textbf{SPFH}) &
200, 300, 350, 400, 450, \newline
500, 550, 600, 650, 700, \newline
750, 800, 825, 850, 875, \newline
900, 925, 950, 975, 1000 & 250, 500, 850, 1000 & 1.0 per variable \newline distributed among levels \\
\midrule
\multicolumn{4}{l}{\textit{Surface/Integrated variables}} \\
Composite reflectivity (\textbf{REFC}) & Integrated & Integrated & 1.0 \\
2-meter temperature (\textbf{T2M}) & 2m AGL & 2m AGL & 1.0 \\
2-meter dewpoint temperature (\textbf{D2M}) & 2m AGL & 2m AGL & 0.7 \\
10-meter U wind component (\textbf{UGRD10M}) & 10m AGL & 10m AGL & 0.3 \\
10-meter V wind component (\textbf{VGRD10M}) & 10m AGL & 10m AGL & 0.3 \\
80-meter U wind component (\textbf{UGRD80M}) & 80m AGL & 80m AGL & 0.1 \\
80-meter V wind component (\textbf{VGRD80M}) & 80m AGL & 80m AGL & 0.1 \\
Total cloud cover (\textbf{TCDC}) & Entire atmo. & Entire atmo. & 0.3 \\
High cloud cover (\textbf{HCDC}) & High cloud  & High cloud  & 0.1 \\
Mid cloud cover (\textbf{MCDC}) & Mid cloud & Mid cloud & 0.1 \\
Low cloud cover (\textbf{LCDC}) & Low cloud & Low cloud & 0.1 \\
Visibility (\textbf{VIS}) & Surface & Surface & 0.3\\
Accumulated precipitation (\textbf{APCP}) & Surface (1-h acc.) & Surface (6-h acc.) & 1.0 \\
Cloud ceiling height (\textbf{HGTCC}) & N/A & N/A & 0.3 \\
Convective available potential energy (\textbf{CAPE}) & Surface & Surface & 0.3 \\
Convective inhibition (\textbf{CIN}) & Surface & Surface & 0.2 \\
Mean sea level pressure (\textbf{MSLET} or \textbf{PRMSL}) &  Sea level & Sea level & 0.1\\
Surface pressure (\textbf{PRES}) & Surface & Surface & 0.02 \\
\midrule
\multicolumn{4}{l}{\textit{Static variables}} \\
Orography height (\textbf{OROG}) & Surface & N/A & N/A \\
Land mask (\textbf{LAND}) & Surface & N/A & N/A \\
\bottomrule
\end{tabular}
\label{inputs}
\end{table*}

Beyond the prognostic fields, HRRRCast computes several classes of post-processed diagnostic
variables from the native model output. These include near-surface thermodynamic diagnostics such
as 2-m relative humidity, 2-m potential temperature; column and precipitation diagnostics such as precipitable water, conditional rain or freezing-rain masks, and
warm- or cold-layer depth; wind diagnostics such as gust estimates and column-maximum wind;
severe-convection diagnostics such as low-level shear, storm-relative helicity; and freezing-level diagnostics. This post-processing step broadens the usable
forecast variable set without requiring the network to predict every downstream hazard-oriented
field directly.

\section{Dataset Curation and Harmonization}
\label{append:dataset_curation}
The dataset is expanded to nearly a decade of HRRR analyses and GFS forecasts, spanning
October 2016 to October 2025. This period covers three HRRR versions (v2, v3, and v4) and three GFS
configurations (pre-FV3 spectral-element GFS, v15, and v16). Incorporating multiple model
generations introduces nontrivial distributional shifts arising from changes in physical
parameterizations, data assimilation systems, and -- most notably for GFS -- the transition to a
fundamentally different dynamical core.

A separate challenge arises from temporal inconsistency in GFS outputs. Versions prior to v16
provide forecasts at 3-hour intervals in the first 120 hours. This reflects a design choice tied to
downstream applications (e.g., boundary condition updates for limited-area models such as RAP),
rather than a deficiency in model fidelity. However, our training setup requires temporally aligned
HRRR-GFS pairs at hourly resolution to construct a consistent learning target for the HRRR
emulator. To achieve this temporal alignment, we apply interpolation to the 3-hourly GFS data.

Missing variables present an additional complication. HRRR v1 is excluded due to the absence of
several key predictors required by the model. While older GFS versions also lack certain variables,
this is less critical because HRRR serves as the primary source of high-resolution target fields;
missing predictors in GFS can be partially mitigated through learned mappings, whereas missing
targets in HRRR would directly degrade supervision. Specifically, pre-v16 GFS data lacks the
following variables: REFC, TCDC, LCDC, MCDC, HCDC, VIS, and HGTCC. 

We address these three issues—distributional shifts, missing variables, and coarse temporal
resolution—using a unified strategy: explicit conditioning on model version via temporal metadata.
Specifically, the network is provided with two binary indicator masks denoting whether a sample
originates from HRRR v3 or v4, allowing it to implicitly infer v2. Another mask is used to indicate
older GFS versions that use a spectral dycore instead of FV3. This approach enables the model to
learn version-dependent corrections without requiring separate training pipelines. These auxiliary
conditioning channels along with others for encoding temporal phase, and rollout-specific context 
are shown in Table~\ref{tab:conditioning_channels}.

\begin{table*}[t]
\centering
\caption{Auxiliary conditioning channels.}
\begin{tabular}{p{3.3cm}p{6.2cm}p{5.6cm}}
\toprule
\textbf{Category} & \textbf{Channels} & \textbf{Purpose} \\
\midrule
Seasonal/diurnal phase & $\sin(\mathrm{doy}),\ \cos(\mathrm{doy}),\ \sin(\mathrm{hour}),\ \cos(\mathrm{hour})$ & Encodes the annual and daily cycle. \\
HRRR version masks & v3 mask, v4 mask & Identifies the HRRR version. \\
GFS version mask & v15 mask & Distinguishes older GFS configurations. \\
Forecast controls & lead time, diffusion step & Forecast conditioning. \\
\bottomrule
\end{tabular}
\label{tab:conditioning_channels}
\end{table*}

While these are the primary issues encountered in constructing the expanded dataset, additional
challenges are expected if native model-level (rather than pressure-level) data are used, because
of inconsistencies in vertical coordinates and evolving model configurations.

\section{Accumulated Precipitation and Transforms}
\label{append:apcp_transforms}
HRRRCast predicts only the last-hour accumulated precipitation at each lead time. For example, for
an 18-h forecast, the model predicts the 17--18 h accumulation at lead 18 h. Longer accumulations
are reconstructed by summing hourly forecasts.

In the near-real-time semi-operational setup, APCP is initialized by the previous cycle's 1-h forecast accumulation (F01) since it is
trained as a prognostic variable; however, it may be more convenient to design the network to treat 
accumulated variables diagnostically. Additionally, we apply logarithmic transforms to precipitation 
and other highly skewed variables including visibility, CAPE, CIN, and cloud-ceiling height, 
followed by z-normalization. The implemented transforms are
\begin{equation*}
\log(1+\max(x,0)) \quad \text{for nonnegative variables,}
\end{equation*}
and
\begin{equation*}
\mathrm{sign}(x)\log(1+|x|) \quad \text{for CIN.}
\end{equation*}
Hard clipping based on min/max limits computed from the training dataset is applied during rollout
to reduce unphysical drift.

\section{Training at Native 3-km Resolution}
\label{append:native_3km_training}
Training at the native HRRR resolution (1059$\times$1799) increases data volume by roughly
4$\times$ compared to the prior 6-km setup. To manage this, normalized data are stored as
\texttt{float16} for I/O efficiency, while mixed-precision training is used with selective
\texttt{float32} compute in sensitive layers.

\begin{figure*}[t]
\centering
\includegraphics[width=0.49\textwidth]{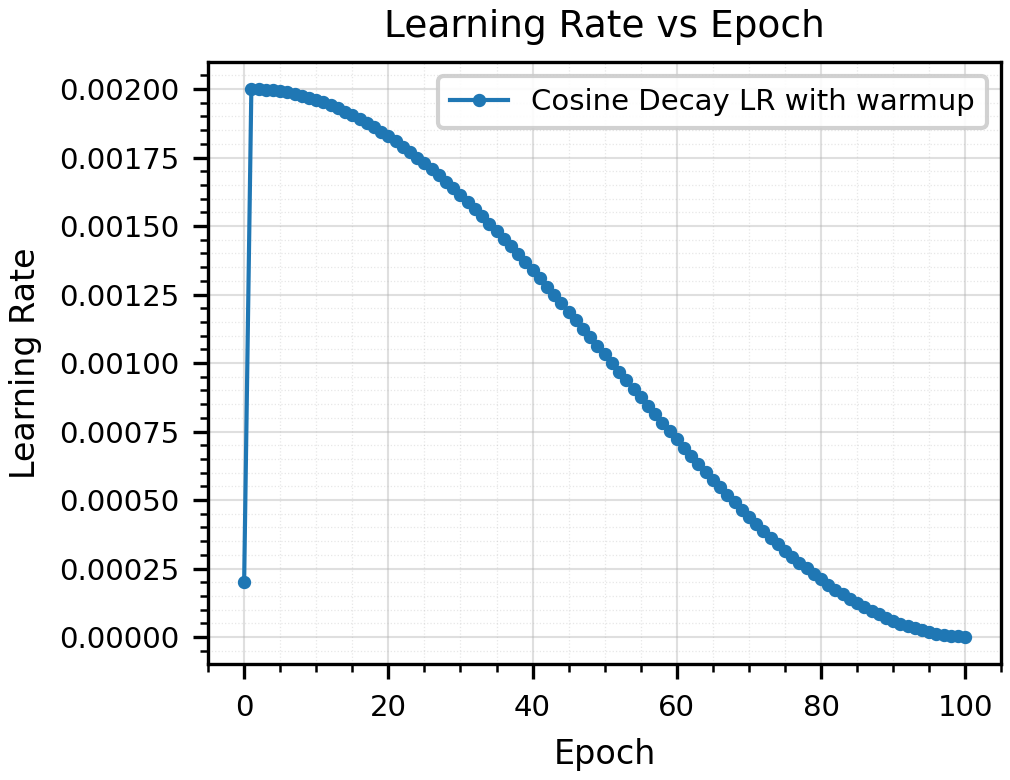}
\hfill
\includegraphics[width=0.49\textwidth]{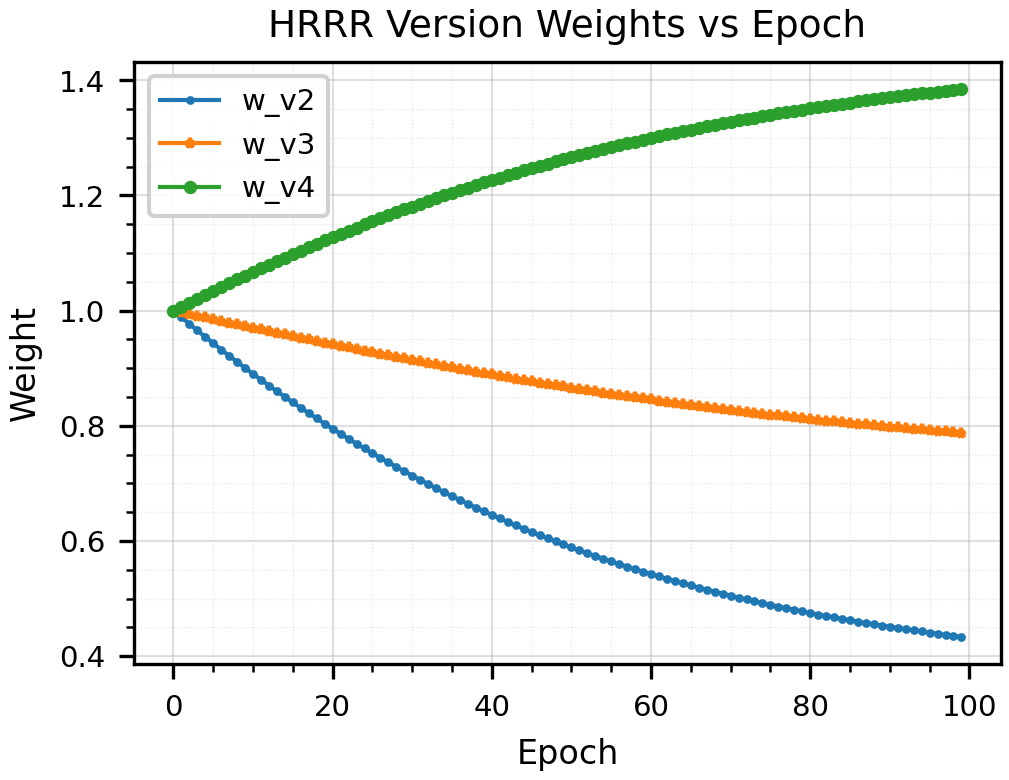}
\caption{Left: learning-rate schedule used for native 3-km training (1-epoch warmup
to a peak of $2\times 10^{-3}$ followed by cosine annealing over 100 epochs). Right:
epoch-dependent HRRR version weights used during training. Samples from the newest
HRRR version (v4) retain the largest weight, while older versions (v3 and especially
v2) are progressively down-weighted. The weights are normalized so that all HRRR
versions continue to contribute while emphasizing the most recent model climatology;
the schedule keeps the total version weight fixed each epoch (v2+v3+v4 = 3) so the
loss scale remains comparable over training, while using smooth decay to avoid
over-dominating updates with v4 alone.}
\label{fig:version_weights}
\end{figure*}

Training uses TensorFlow distributed data parallelism, with the effective global batch
size scaling as the per-node batch size times the number of MPI workers. For the
native-grid production run, we trained on four nodes with two NVIDIA H100 GPUs per
node (eight H100 GPUs total). Completing training over the full 9-year dataset took
approximately 1.5 months of wall-clock time.

For the native 3-km training, we use the AdamW optimizer with the learning-rate schedule 
that uses a 1-epoch warmup followed by cosine annealing over a 100-epoch run 
(see left panel of Figure~\ref{fig:version_weights}).
The learning rate ramps to a peak value of $2\times 10^{-3}$ at the end
of warmup and then decays smoothly to nearly zero, reaching approximately
$10^{-6}$ by the end of training. In addition, version-dependent sample weights are
applied during training to gradually de-emphasize older HRRR versions while preserving
their contribution to the expanded historical dataset. Figure~\ref{fig:version_weights}
summarizes both the learning-rate schedule and the epoch-dependent HRRR version-weight
schedule used in this study.

\section{Architecture Details}
\label{append:architecture_details}
Relative to the original HRRRCast model \citep{abdi39}, the core network architecture was updated
in several ways to improve skill at native 3-km resolution while preserving numerical stability
during training. 

\subsection{Model configuration and memory optimization}
\label{append:model_config_memory}

As in the original HRRRCast model \citet{abdi39}, the model follows a U-Net-style encoder--processor--decoder 
architecture with skip connections, built entirely from 3$\times$3 residual blocks.

\noindent The full stage configuration is
\begin{align*}
\texttt{resolution} &= [\text{3 km},\ \text{6 km},\ \text{12 km},\ \text{24 km},\ \text{12 km},\ \text{6 km},\ \text{3 km}], \\
\texttt{blocks}     &= [2,\ 3,\ 4,\ 14,\ 4,\ 3,\ 2], \\
\texttt{filters}    &= [288,\ 256,\ 256,\ 256,\ 256,\ 256,\ 288].
\end{align*}
The resulting model has approximately 50 million trainable parameters.

To make native-grid training tractable, we use gradient checkpointing (recomputation) in selected
preprocessing, encoder, decoder, and output-head stages. We also use spatially grouped convolutions
(2$\times$2 tiling with overlap) in highest-resolution stages. Overlap-and-crop stitching preserves
equivalence to full-domain \texttt{same}-padding convolutions while reducing memory pressure.

Because training uses normalized \texttt{float16} data and mixed precision, numerically sensitive
operations are explicitly computed in \texttt{float32} (e.g., normalization, key dense/attention
operations, and selected residual/output operations) to avoid instability.

\subsection{Input preprocessing branches}
\label{append:input_preprocessing_branches}
We introduced dedicated input preprocessing branches before the main
encoder--decoder path so that HRRR, GFS, and (for diffusion training) noised-HRRR channel groups
are processed by separate residual stems and then fused. This design allows each input type to
undergo specialized feature extraction before entering the shared encoder, improving the model's
ability to integrate multi-source information.

\subsection{Attention mechanism upgrade: CBAM}
\label{append:cbam_upgrade}
We replaced the squeeze-and-excitation (SE) blocks used in the original model with Convolutional Block Attention Modules 
(CBAM) \citep{woo2018cbam}. While SE blocks provide channel-wise attention that adaptively recalibrates feature 
responses by modeling inter-channel relationships, they do not explicitly capture spatial attention 
patterns. CBAM extends this capability by sequentially applying both channel and spatial attention mechanisms,
enabling the network to focus on both \emph{what} features are important (channel attention) and
\emph{where} they are located (spatial attention).

The CBAM module consists of two sequential sub-modules: a channel attention module that uses both
average-pooled and max-pooled features to generate channel-wise weights, followed by a spatial
attention module that uses channel-pooled features to produce a 2D spatial attention map. This
sequential refinement is applied at multiple scales throughout the encoder-decoder architecture,
allowing the network to adaptively modulate both feature channels and spatial locations at each
resolution level.

\section{Sigmoid-Weighted Loss: Design Details}
\label{append:sigmoid_loss}

Prior work has explored similar strategies in the nowcasting context, with some published schedules 
applying weights as large as $800\times$ for the most extreme events
\citep{shi2017deep,jin2024ureslham,ravuri2021skillful,cao2025enhancing}. Our schedule is intentionally
moderate to preserve skill on common events while improving performance at high thresholds.

The same weighting approach can in principle be applied to other forecast variables with
skewed or heavy-tailed behavior, but weighting is restricted here to REFC and APCP because
these variables drive the largest practical deficits in operational utility. To validate
training stability and tune the weighting schedule, experiments were first conducted on the
6-km model before their adoption in the 3-km training.

\begin{figure*}[tbp]
\centering
\includegraphics[width=0.49\textwidth]{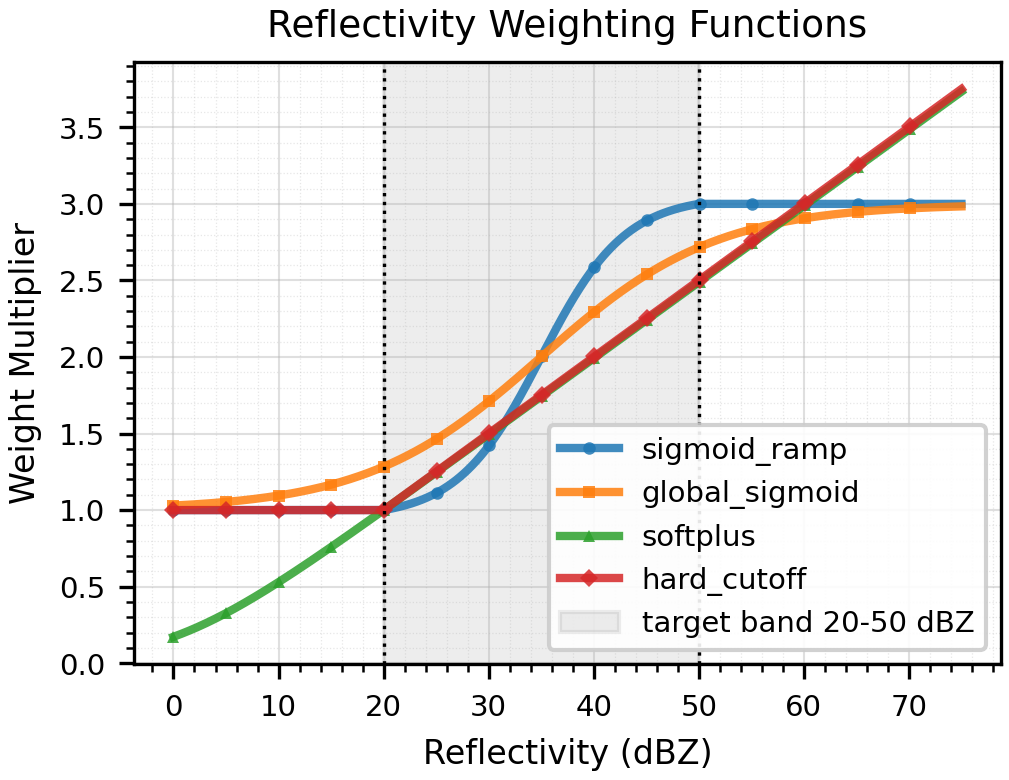}
\hfill
\includegraphics[width=0.49\textwidth]{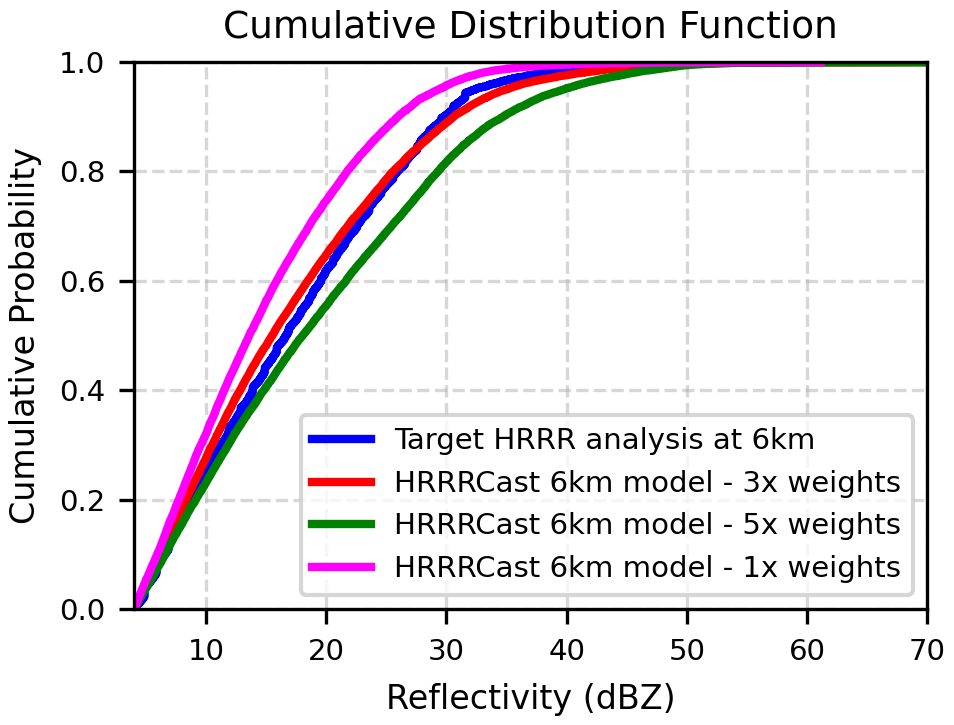}
\caption{Left: candidate weighting functions explored for extreme-event training.
A softplus-based formulation increased weights too aggressively and led to
unrealistically large reflectivity and precipitation amplitudes. Right:
REFC cumulative-distribution comparison at 18-h forecast lead for different
weight caps. The $3\times$ weighting produces a distribution that remains much
closer to the reference behavior than the more aggressive $5\times$ option,
while still providing stronger extreme-event emphasis than $1\times$. This
supports the final choice of a global sigmoid weighting function capped at
$3\times$. For REFC, the ramp onset near 20 dBZ corresponds approximately to
the $4\sigma$ threshold used in Section~\ref{sec:sigmoid_loss}
($\sigma \approx 5$ dBZ).}
\label{fig:weighted_mse}
\end{figure*}

This design choice should be interpreted in the context of domain size and sampling.
Over a CONUS-scale training domain, most grid points and times correspond to weak or
non-convective conditions, so extreme reflectivity and heavy precipitation events form
a small fraction of the optimization signal. In that regime, explicit reweighting helps
counteract the dominance of non-storm examples and preserves sensitivity to rare,
operationally critical extremes.

By contrast, models trained on smaller, storm-focused domains (for example, WoFS-like
windows centered on active convection) can have a much higher effective event density.
When the training distribution is already concentrated on convective states, specialized
extreme-event weighting may be less necessary to recover strong high-reflectivity skill.
This interpretation is consistent with WoFSCast results \citep{Flora2025}, which
are trained on storm-focused windows rather than broad CONUS domains and therefore
do not face the same class-imbalance regime (a large background ``sea of zeros'')
that motivates stronger imbalance correction in our setup. Our weighting
strategy can be viewed as a partial solution to extreme-weather prediction
difficulties, and newer benchmarks such as Extreme Weather Bench
\citep{mandelbaum2026ewb} help the community measure and improve progress
on this challenge.

To implement this loss design, we evaluated several candidate weighting functions in early experiments.
A softplus-based weighting function was initially preferred, but its rapidly
increasing tail encouraged unrealistically large reflectivity and precipitation values
in the generated forecasts. We therefore replaced it with a capped sigmoid family that
still emphasizes the targeted extreme-event range while preventing uncontrolled growth
in the loss assigned to the most intense samples. Figure~\ref{fig:weighted_mse}
summarizes the candidate weighting functions explored during this tuning process; we
selected a global sigmoid formulation with a maximum weight of $3\times$, which provided
the best balance between stronger extreme-event emphasis and stable forecast amplitudes.
The REFC CDF comparison at 18 h (right panel of Fig.~\ref{fig:weighted_mse}) further
reinforced this decision: the $3\times$ configuration stayed substantially closer to
the baseline distribution than $5\times$, while still shifting the tail more than
$1\times$.

\section{Near-Real-Time Configuration}
\label{sec:near_real_time_config}
The near-real-time HRRRCast system runs hourly on the NOAA Research and Development High
Performance Computing System (RDHPCS). It produced a 9-member ensemble forecast for most of
the near-real-time archive, but since July 1, 2026, it has produced an 18-member ensemble forecast.
Extended 48-h forecasts are initialized at 00, 06, 12, and 18 UTC, while 18-h forecasts
are generated at all other initialization times. Model inference is distributed across
three 94-GB NVIDIA H100 GPUs. Total wall-clock compute time is approximately 48 min for
the 18-h cycles and 1 h 56 min for the extended 48-h cycles. Final model output is
delivered in standard GRIB2 format, and the full list of core output variables is
provided in Table~\ref{inputs}.

Ensemble members start from the same initial state but differ through the diffusion noise 
used to initialize the forecast \citep{abdi39}. To add diversity and avoid ensemble collapse, 
the GFS boundary-condition inputs are phase-shifted for each member; this shift is zero for 
the first 12 forecast hours and then increases by 1h in 12-hour blocks. The diffusion noises are 
also correlated across forecast hours, following common video diffusion practice, to avoid 
jumpiness during rollout.

\amsappendixB
\appendixtitle{Additional Verification and Sensitivity Results}

\section{Power spectral density comparisons}
\label{append:psd_comparisons}

\begin{figure*}[tbp]
\centering

\begin{subfigure}[t]{0.32\textwidth}
\centering
\includegraphics[width=\textwidth]{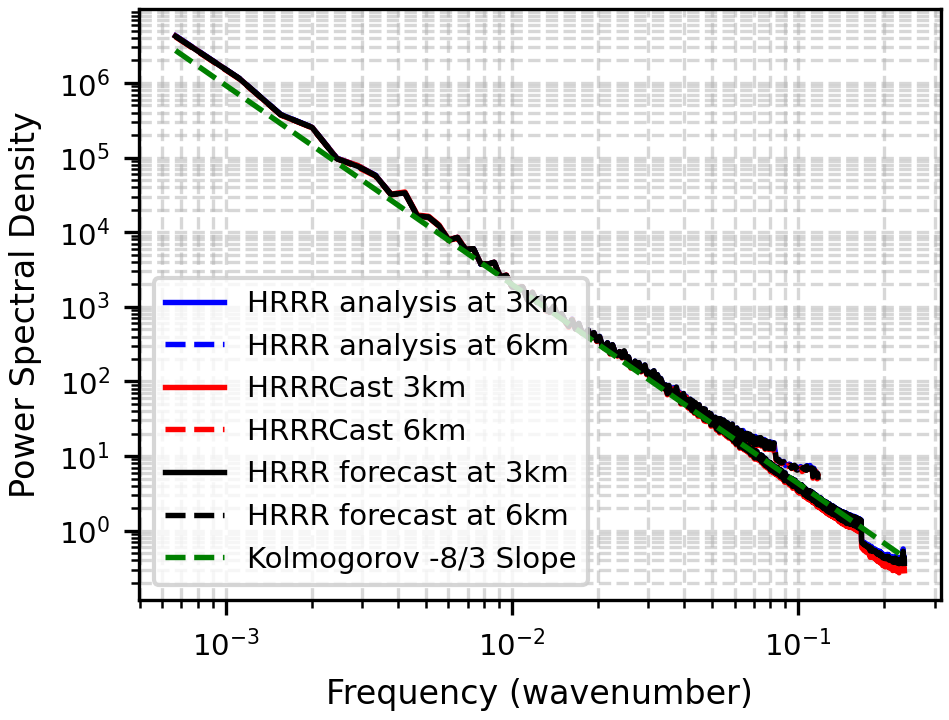}
\caption{2-m temperature}
\end{subfigure}\hfill
\begin{subfigure}[t]{0.32\textwidth}
\centering
\includegraphics[width=\textwidth]{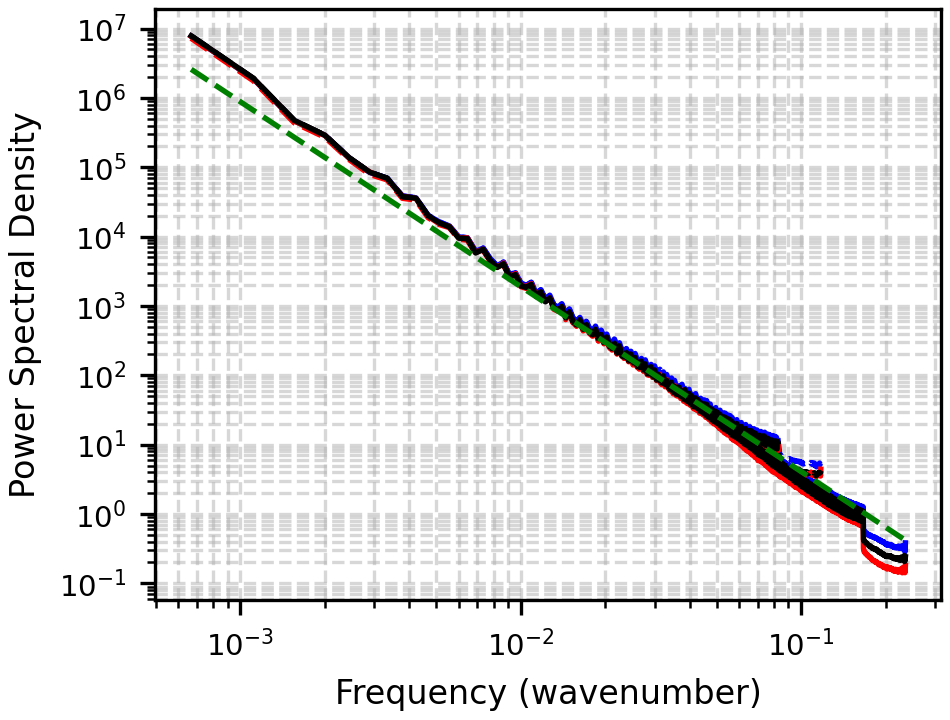}
\caption{2-m dewpoint}
\end{subfigure}\hfill
\begin{subfigure}[t]{0.32\textwidth}
\centering
\includegraphics[width=\textwidth]{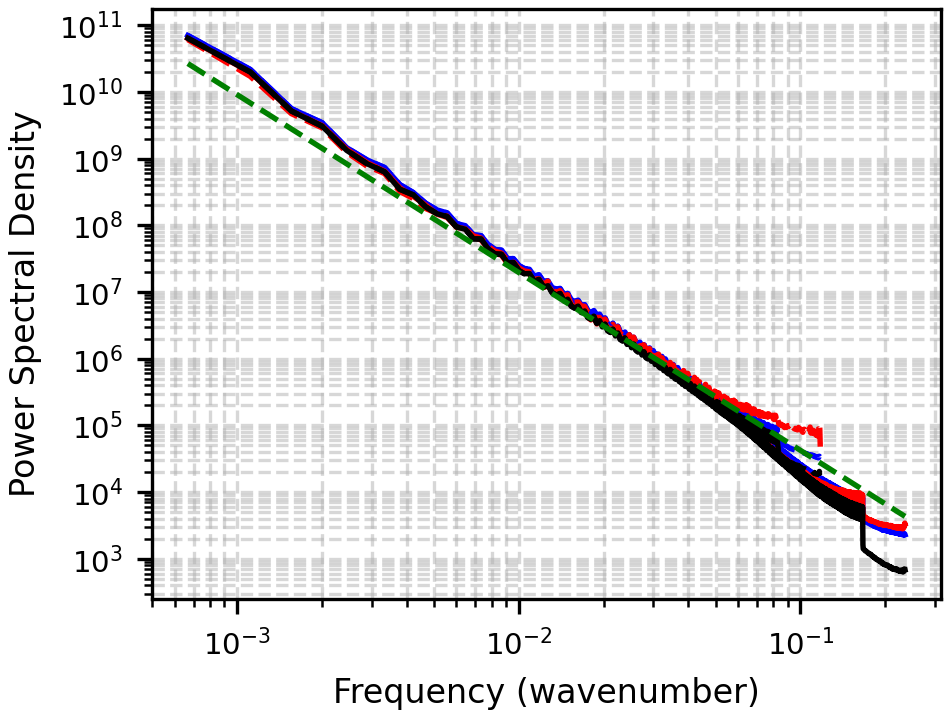}
\caption{Surface CAPE}
\end{subfigure}

\vspace{0.35em}

\begin{subfigure}[t]{0.32\textwidth}
\centering
\includegraphics[width=\textwidth]{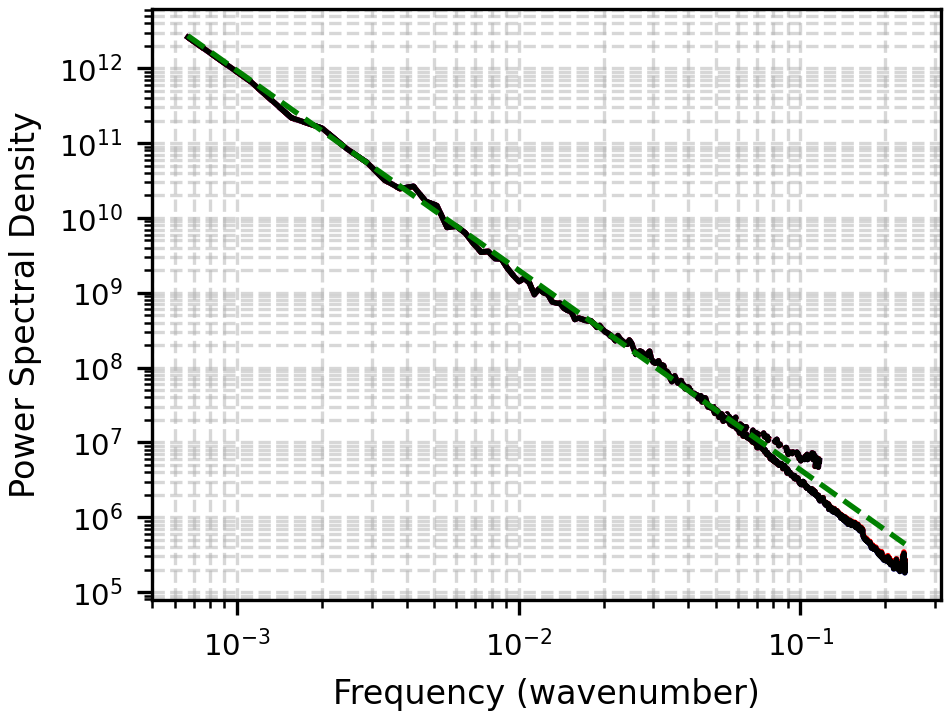}
\caption{Surface pressure}
\end{subfigure}\hfill
\begin{subfigure}[t]{0.32\textwidth}
\centering
\includegraphics[width=\textwidth]{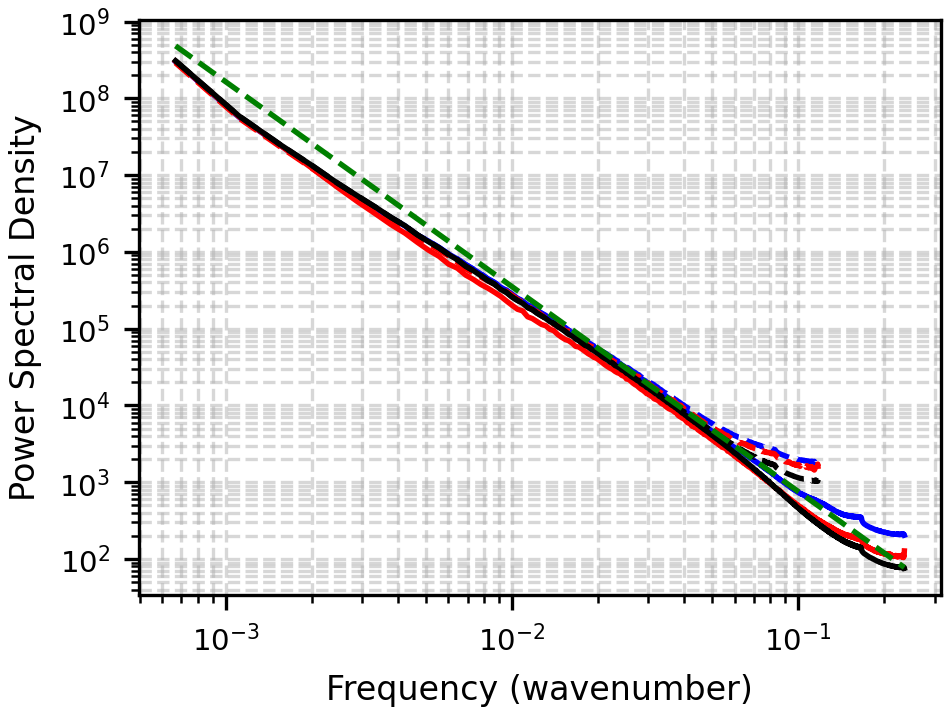}
\caption{Total cloud cover}
\end{subfigure}\hfill
\begin{subfigure}[t]{0.32\textwidth}
\centering
\includegraphics[width=\textwidth]{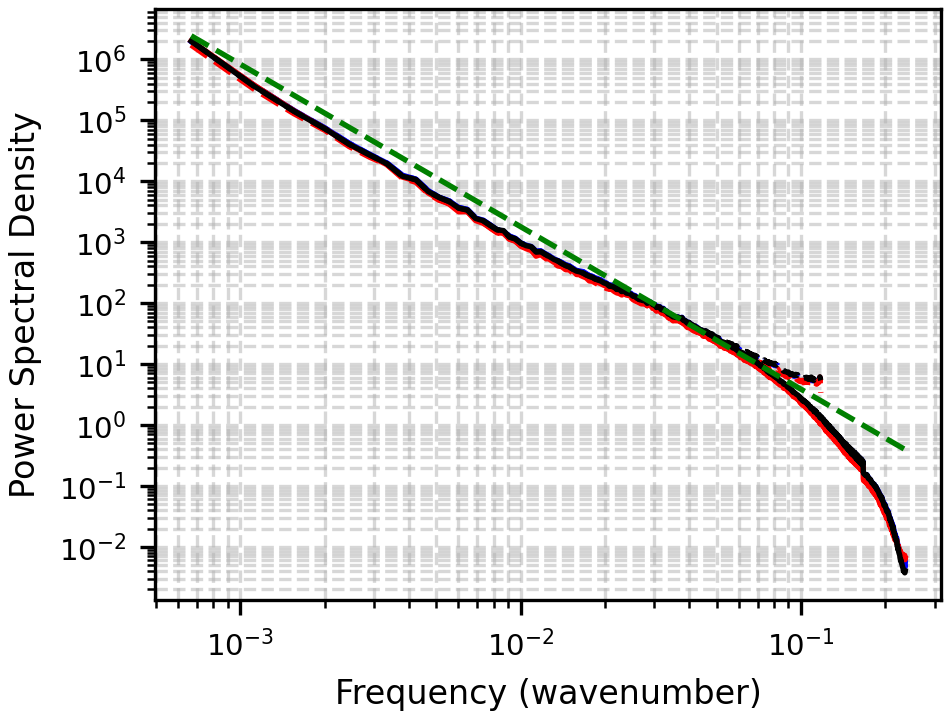}
\caption{10-m u wind}
\end{subfigure}

\vspace{0.35em}

\begin{subfigure}[t]{0.32\textwidth}
\centering
\includegraphics[width=\textwidth]{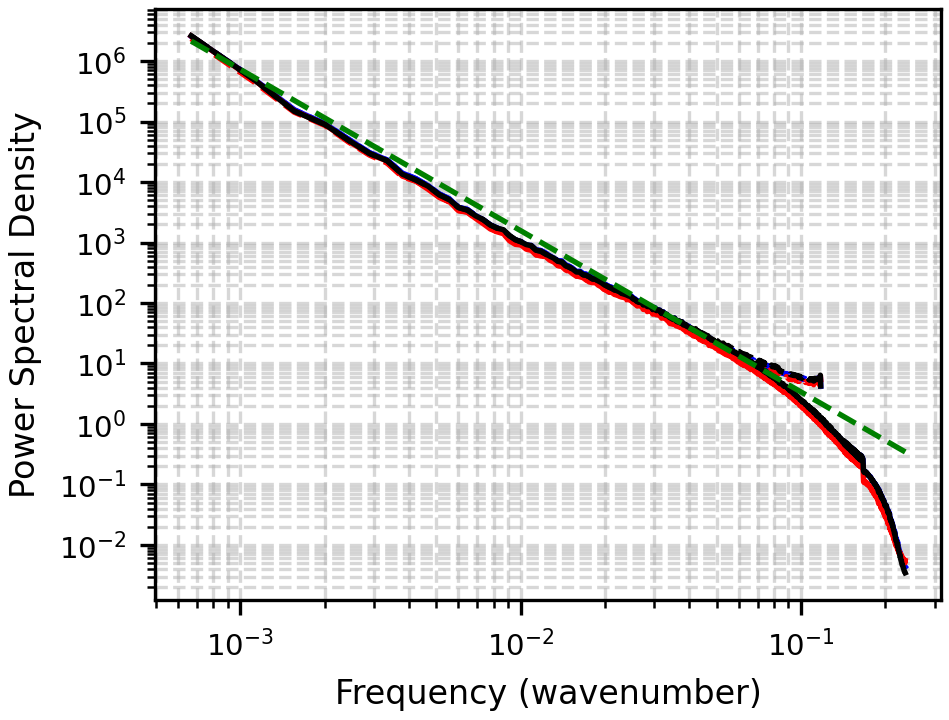}
\caption{10-m v wind}
\end{subfigure}\hfill
\begin{subfigure}[t]{0.32\textwidth}
\centering
\includegraphics[width=\textwidth]{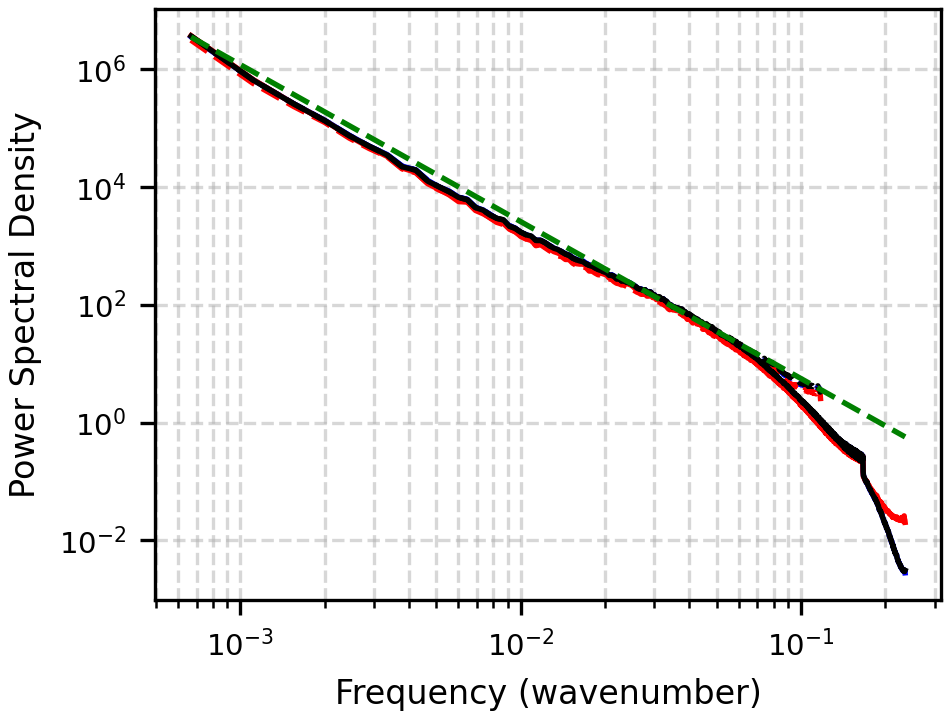}
\caption{80-m u wind}
\end{subfigure}\hfill
\begin{subfigure}[t]{0.32\textwidth}
\centering
\includegraphics[width=\textwidth]{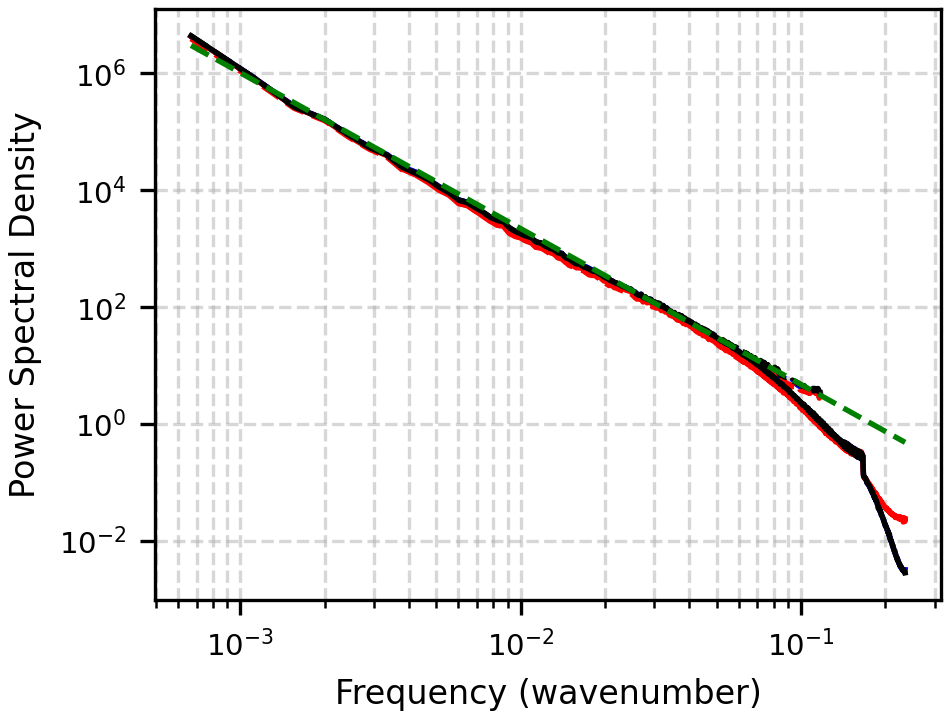}
\caption{80-m v wind}
\end{subfigure}
\caption{Power spectral density comparisons for selected variables for 1h lead time}
\label{fig:appendix_psd_all}
\end{figure*}

Figure~\ref{fig:appendix_psd_all} compares isotropic power spectral density (PSD)
across selected near-surface and storm-related variables. Across variables, all
forecast systems show the expected reduction in variance toward smaller scales
relative to the analyses, with consistently shorter small-scale tails in the 6-km
curves than in their 3-km counterparts. At the highest resolved wavenumbers, the
6-km spectra also commonly turn upward relative to the 3-km HRRR-analysis
reference, a pattern consistent with aliasing introduced by subsampling. This signal
appears not only in 6-km HRRRCast and HRRR forecasts, but also in the 6-km
HRRR-analysis curves, indicating that subsampling alone can imprint high-wavenumber distortion.

These PSD diagnostics reinforce the rationale for native 3-km training: avoiding
6-km subsampling better preserves physically meaningful small-scale variance and
limits aliasing artifacts before they are learned by the model. Consistent with
this, HRRRCast generally stays closer than HRRR to the 3-km HRRR-analysis spectra
at intermediate wavelengths, especially for reflectivity and near-surface winds,
indicating improved mesoscale structure representation rather than only large-scale
amplitude correction.

\section{Supplemental verification against HRRR analyses}
\label{append:analysis_rmse}

The near-real-time verification in Section~\ref{sec:mats_nrt_verification} provides 
robust observational benchmarking for variables with established MATS infrastructure 
(surface thermodynamic and wind fields, reflectivity, and precipitation). However, MATS
verification has not yet been implemented for all variables included in
HRRRCast, particularly the expanded set of diagnostic fields and
cloud variables introduced to support hazardous-weather applications. To assess
these newly added variables, we provide supplemental verification against HRRR 
analyses using RMSE as the primary metric.

\begin{figure*}[tbp]
\centering

\begin{subfigure}[t]{0.23\textwidth}
\centering
\includegraphics[width=\textwidth]{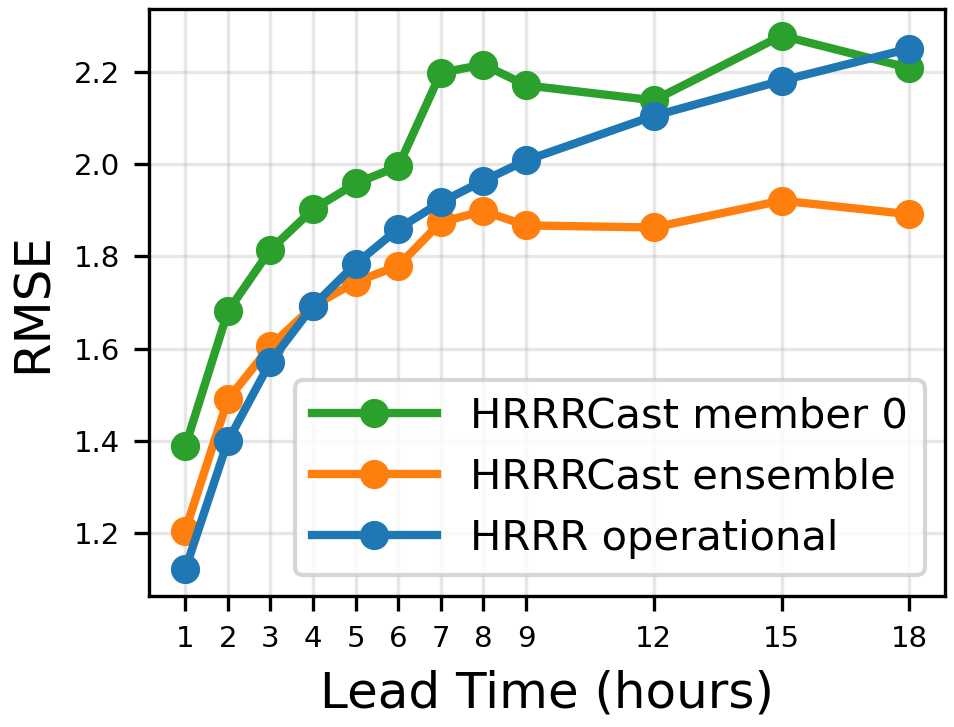}
\caption{80-m zonal wind (m s$^{-1}$)}
\end{subfigure}\hfill
\begin{subfigure}[t]{0.23\textwidth}
\centering
\includegraphics[width=\textwidth]{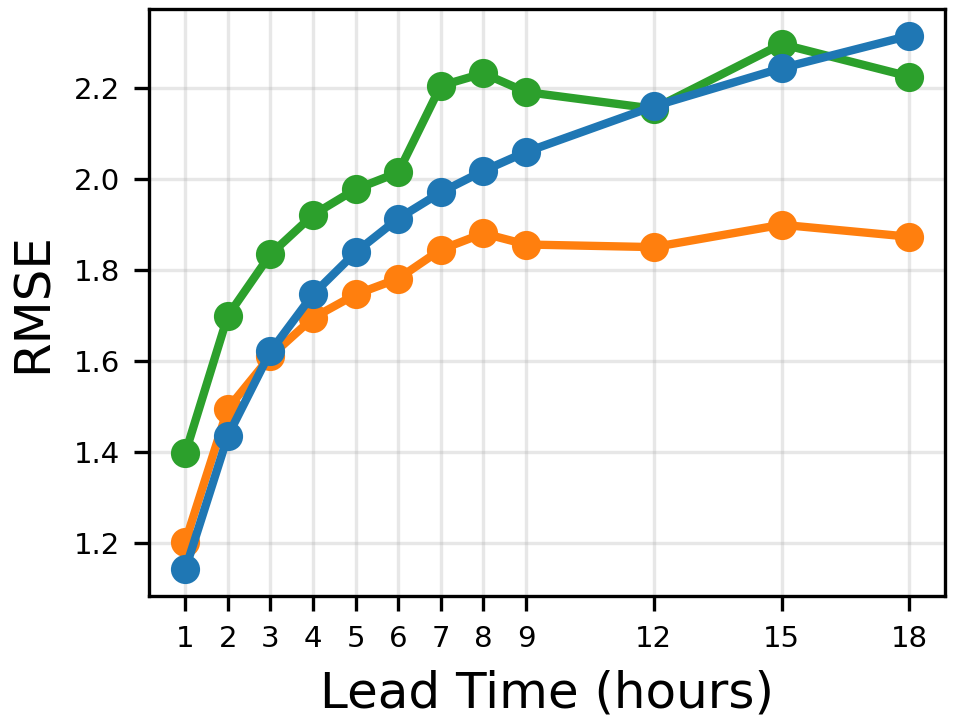}
\caption{80-m merid. wind (m s$^{-1}$)}
\end{subfigure}\hfill
\begin{subfigure}[t]{0.23\textwidth}
\centering
\includegraphics[width=\textwidth]{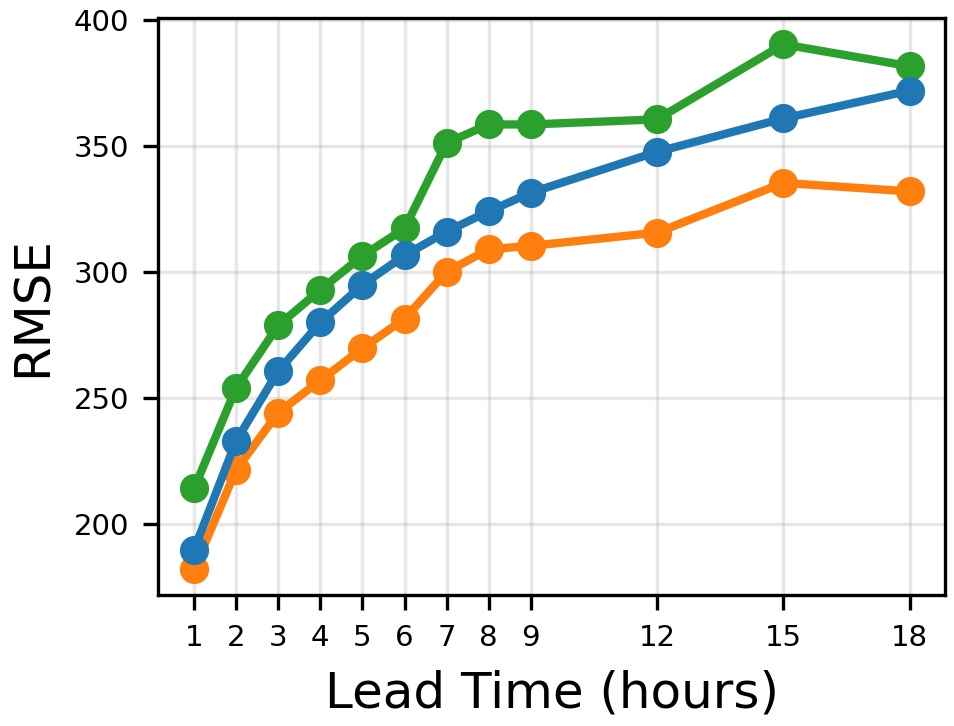}
\caption{CAPE (J kg$^{-1}$)}
\end{subfigure}
\begin{subfigure}[t]{0.23\textwidth}
\centering
\includegraphics[width=\textwidth]{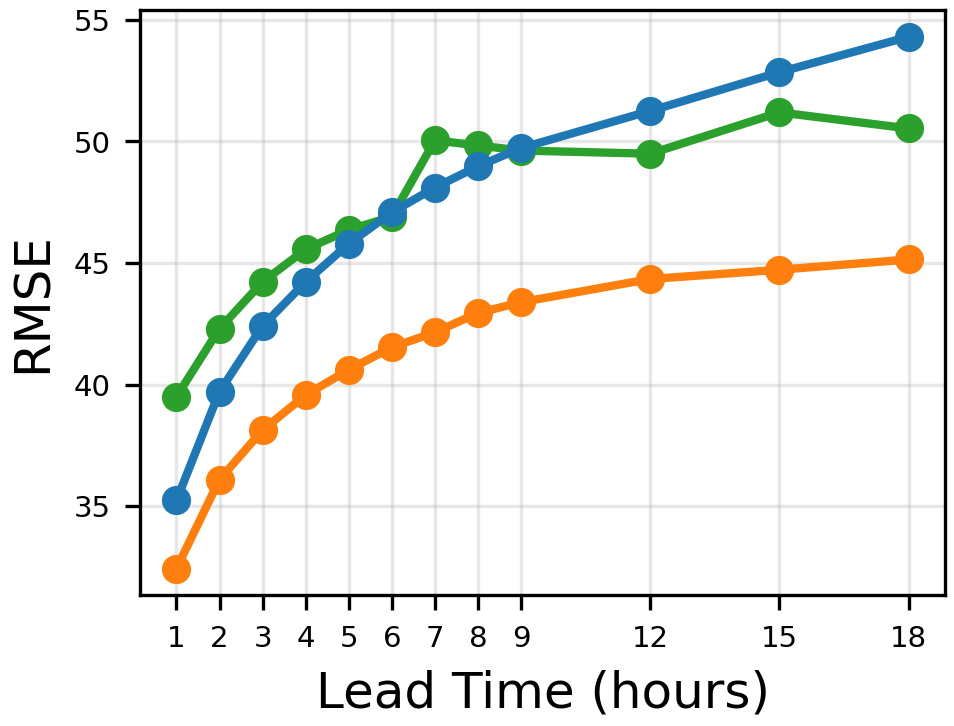}
\caption{CIN (J kg$^{-1}$)}
\end{subfigure}\hfill

\vspace{0.6em}

\begin{subfigure}[t]{0.23\textwidth}
\centering
\includegraphics[width=\textwidth]{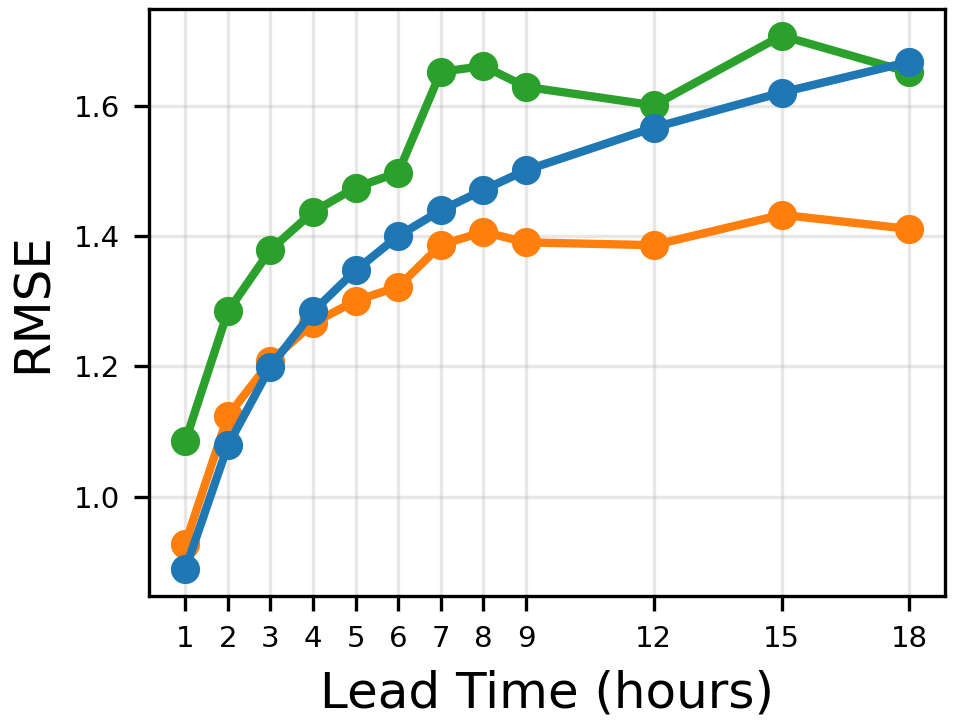}
\caption{10-m zonal wind (m s$^{-1}$)}
\end{subfigure}\hfill
\begin{subfigure}[t]{0.23\textwidth}
\centering
\includegraphics[width=\textwidth]{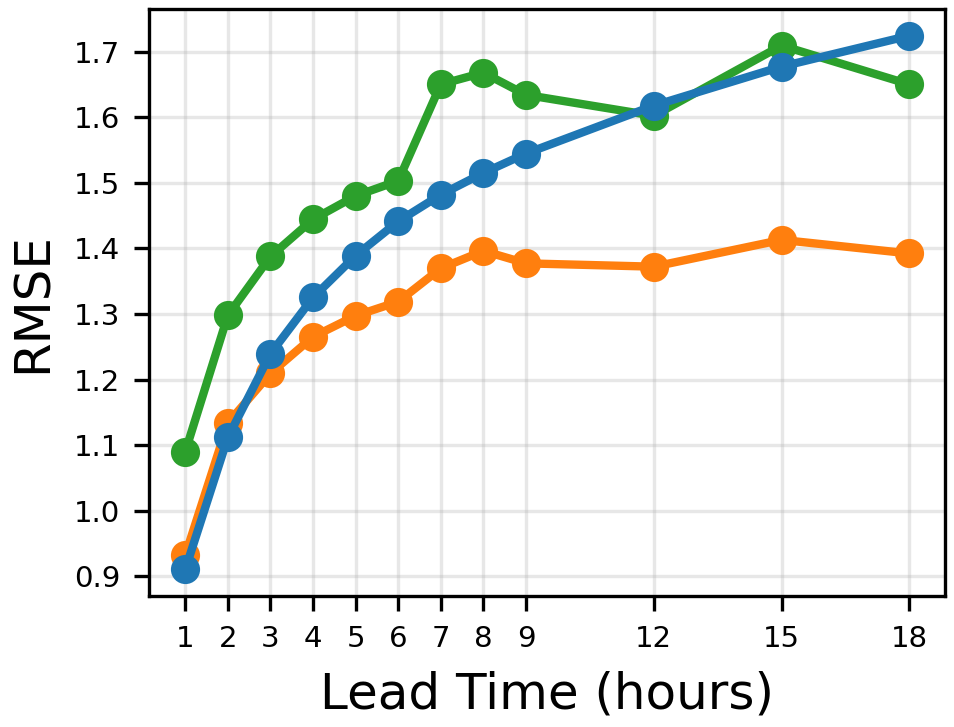}
\caption{10-m merid. wind (m s$^{-1}$)}
\end{subfigure}
\begin{subfigure}[t]{0.23\textwidth}
\centering
\includegraphics[width=\textwidth]{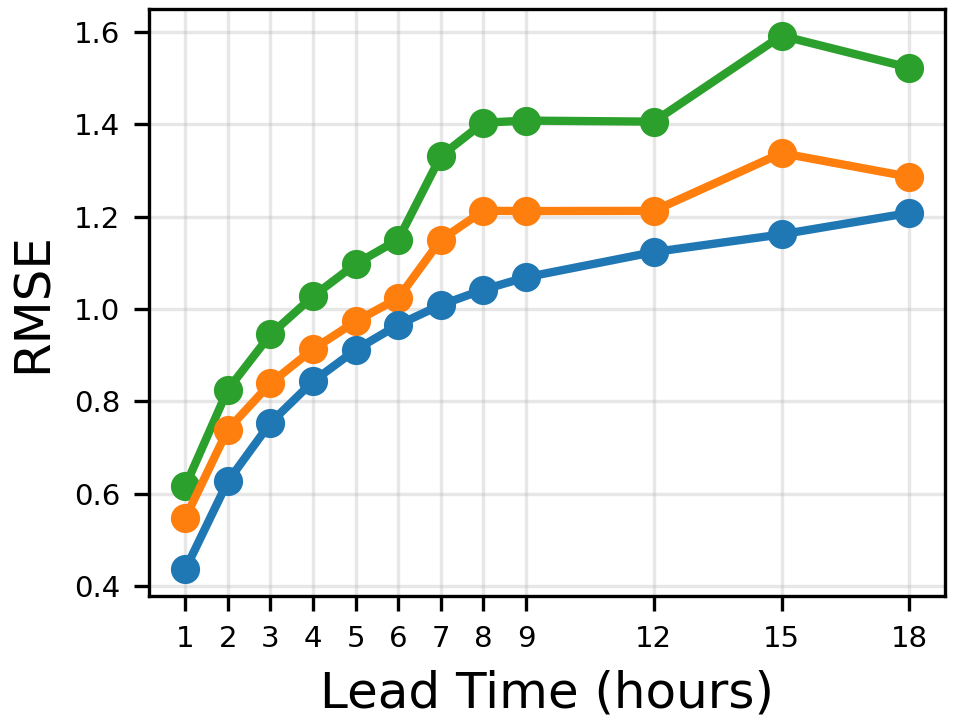}
\caption{2-m temperature (K)}
\end{subfigure}\hfill
\begin{subfigure}[t]{0.23\textwidth}
\centering
\includegraphics[width=\textwidth]{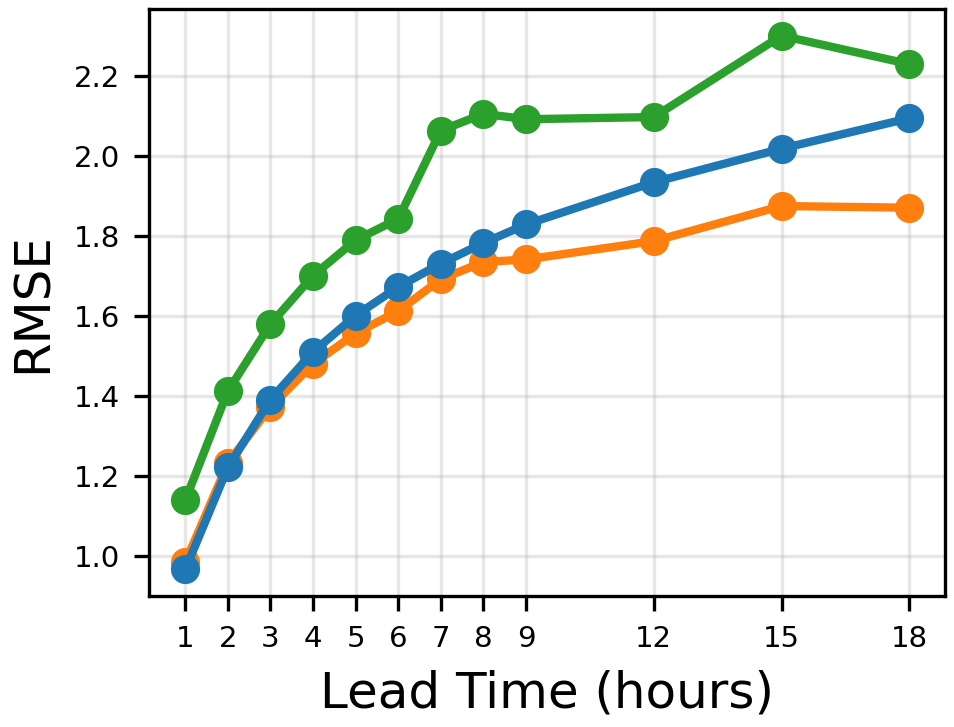}
\caption{2-m dewpoint (K)}
\end{subfigure}\hfill

\vspace{0.6em}

\begin{subfigure}[t]{0.23\textwidth}
\centering
\includegraphics[width=\textwidth]{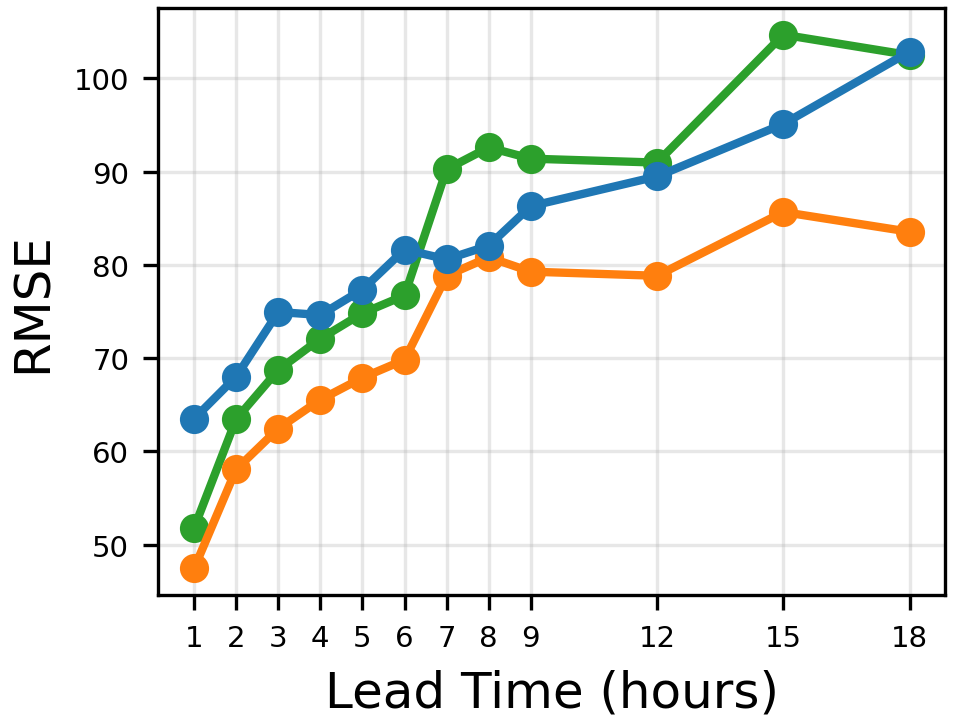}
\caption{Mean sea-level pressure (Pa)}
\end{subfigure}\hfill
\begin{subfigure}[t]{0.23\textwidth}
\centering
\includegraphics[width=\textwidth]{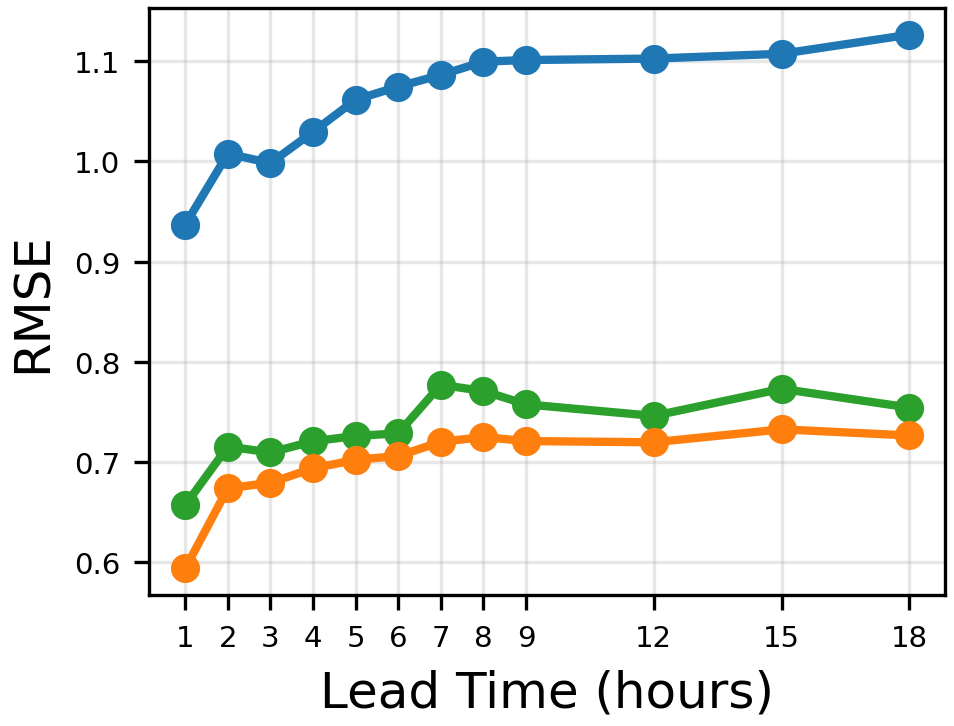}
\caption{Hourly precip. (mm)}
\end{subfigure}\hfill
\begin{subfigure}[t]{0.23\textwidth}
\centering
\includegraphics[width=\textwidth]{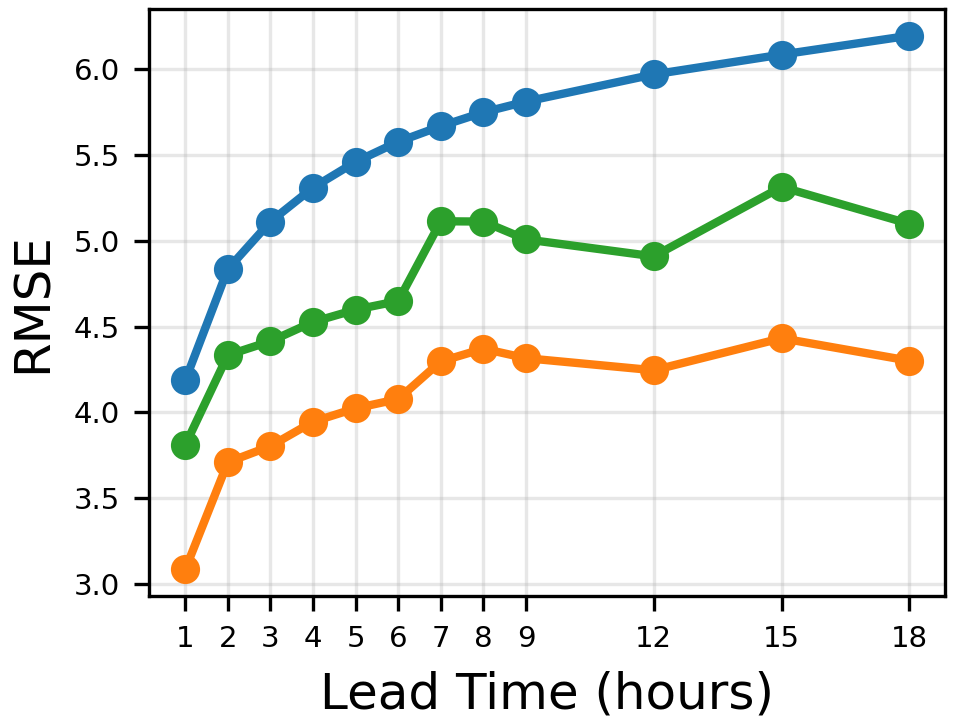}
\caption{Composite refl. (dBZ)}
\end{subfigure}
\begin{subfigure}[t]{0.23\textwidth}
\centering
\includegraphics[width=\textwidth]{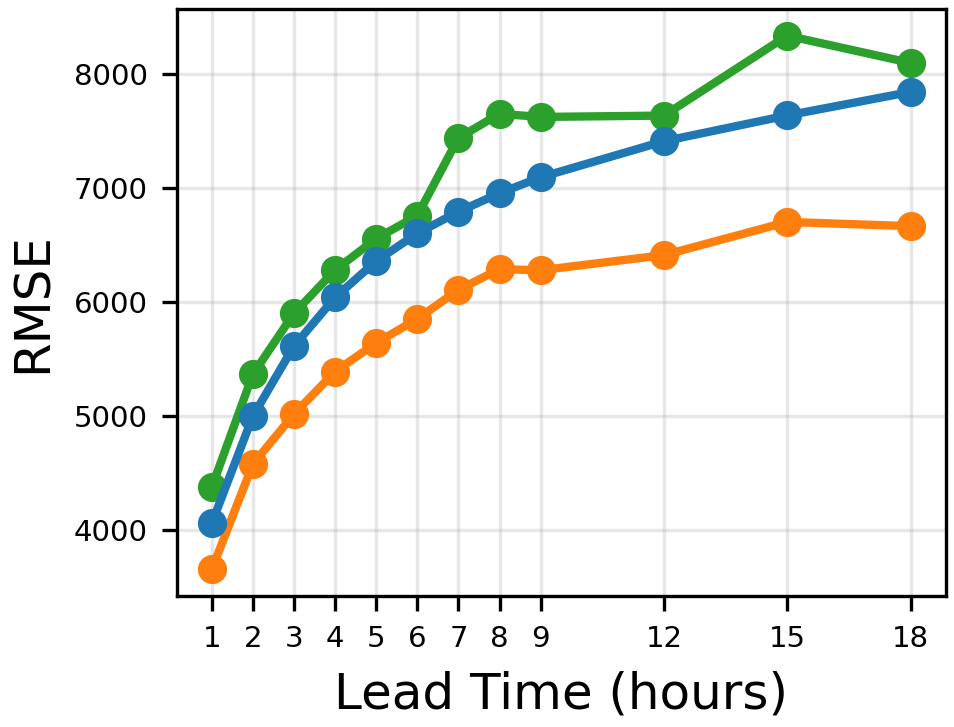}
\caption{Visibility (m)}
\end{subfigure}\hfill

\vspace{0.6em}

\begin{subfigure}[t]{0.23\textwidth}
\centering
\includegraphics[width=\textwidth]{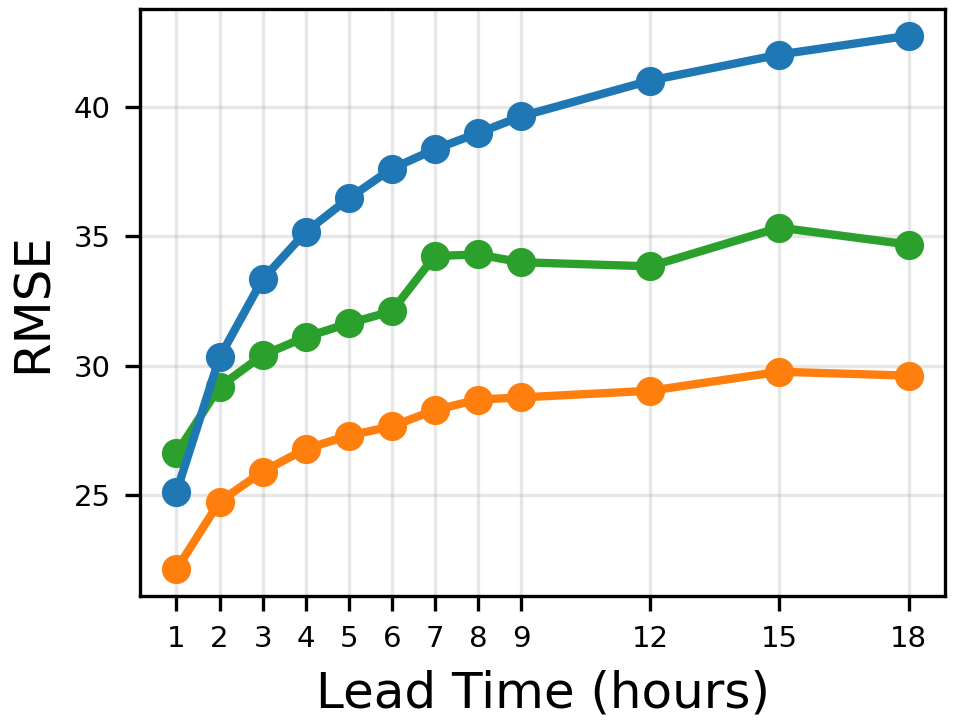}
\caption{Total cloud cover (\%)}
\end{subfigure}\hfill
\begin{subfigure}[t]{0.23\textwidth}
\centering
\includegraphics[width=\textwidth]{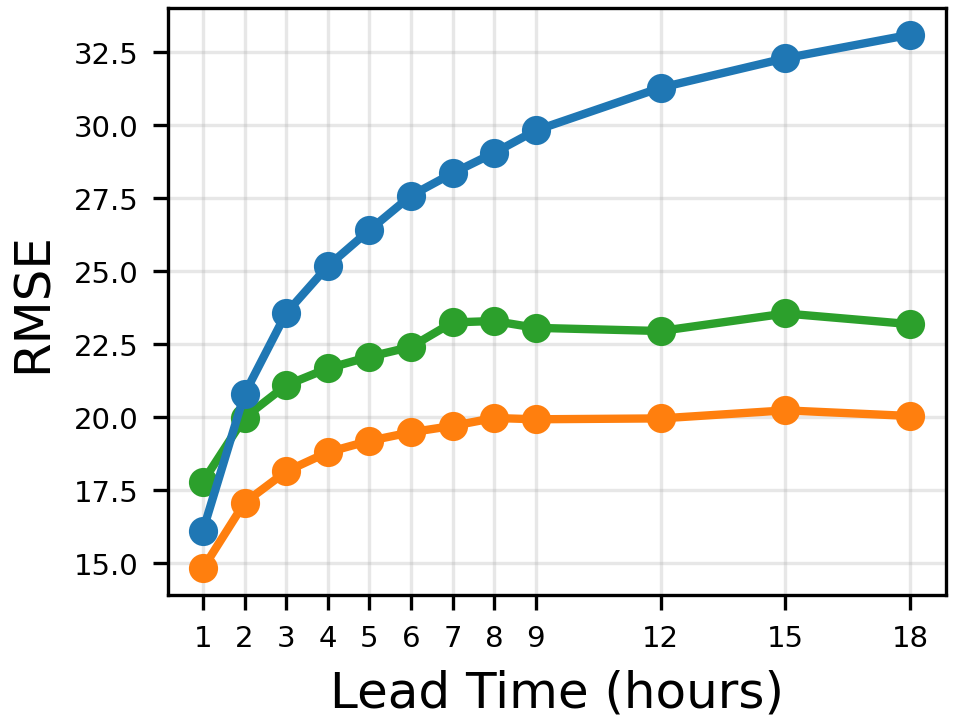}
\caption{Mid cloud cover (\%)}
\end{subfigure}\hfill
\begin{subfigure}[t]{0.23\textwidth}
\centering
\includegraphics[width=\textwidth]{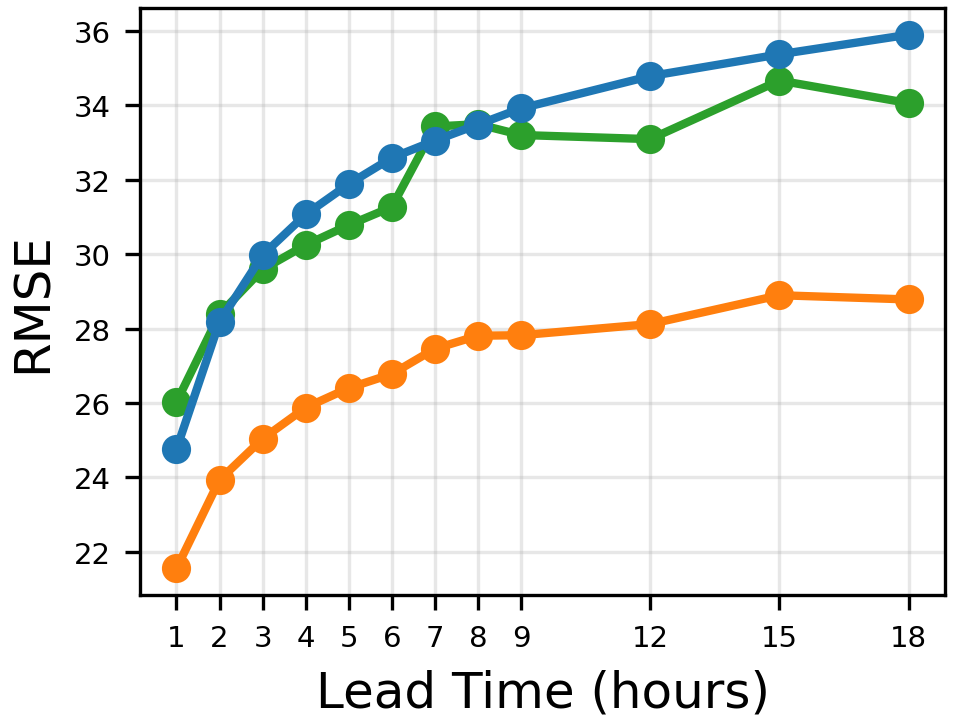}
\caption{Low cloud cover (\%)}
\end{subfigure}\hfill
\begin{subfigure}[t]{0.23\textwidth}
\centering
\includegraphics[width=\textwidth]{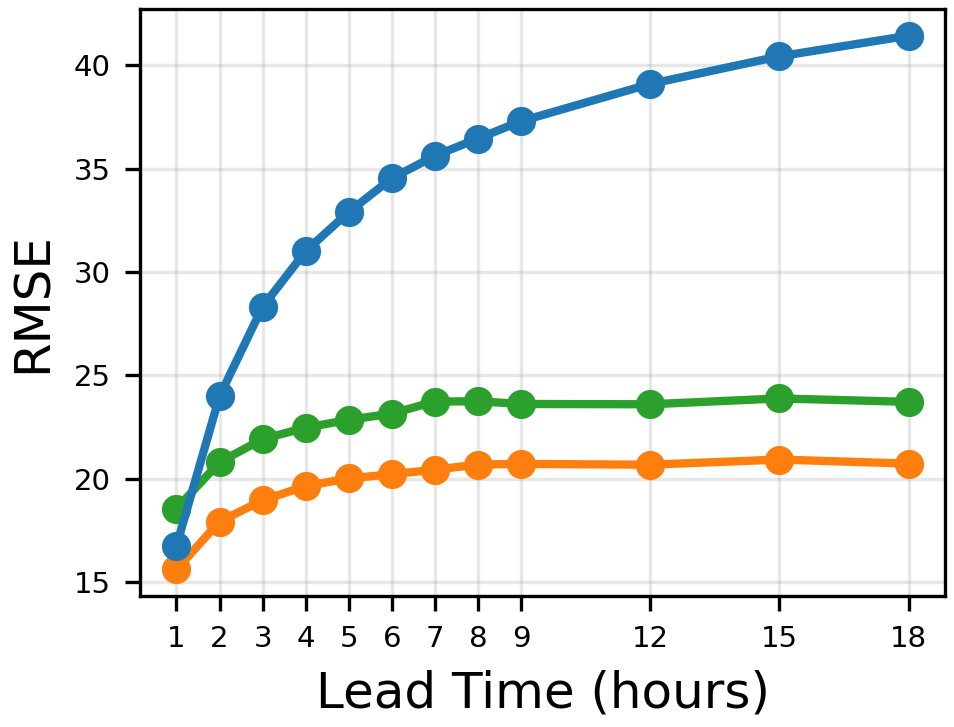}
\caption{High cloud cover (\%)}
\end{subfigure}
\caption{Selected RMSE verification fields against HRRR analyses.}
\label{fig:rmse_verification}
\end{figure*}

\begin{figure*}[tbp]
\centering

\begin{subfigure}[t]{0.24\textwidth}
\centering
\includegraphics[width=\textwidth]{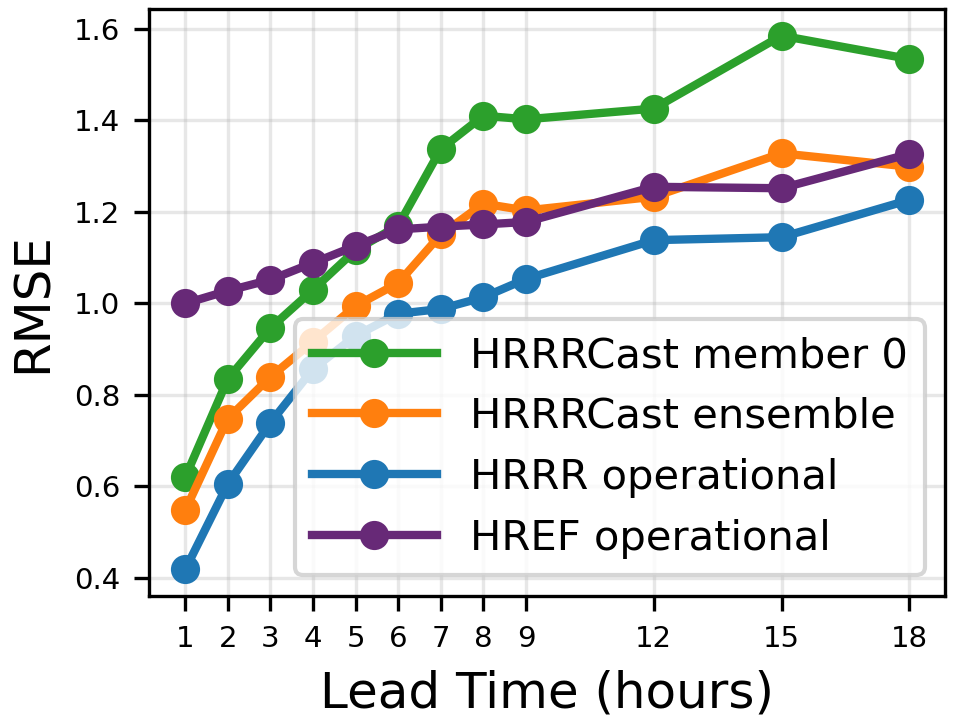}
\caption{2-m temperature (K)}
\end{subfigure}\hfill
\begin{subfigure}[t]{0.24\textwidth}
\centering
\includegraphics[width=\textwidth]{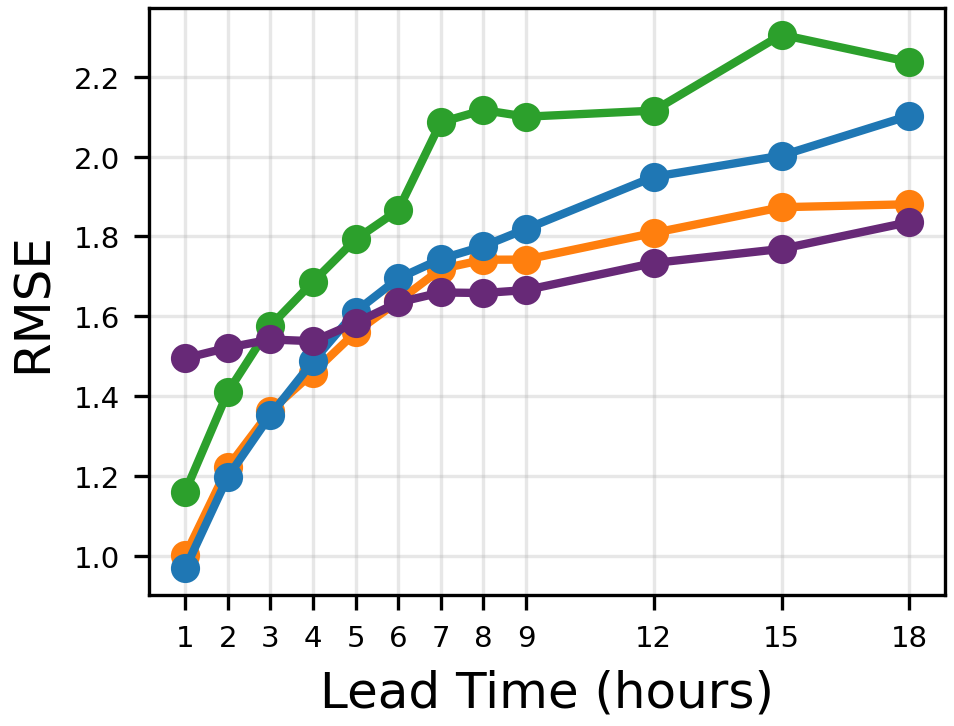}
\caption{2-m dewpoint (K)}
\end{subfigure}\hfill
\begin{subfigure}[t]{0.24\textwidth}
\centering
\includegraphics[width=\textwidth]{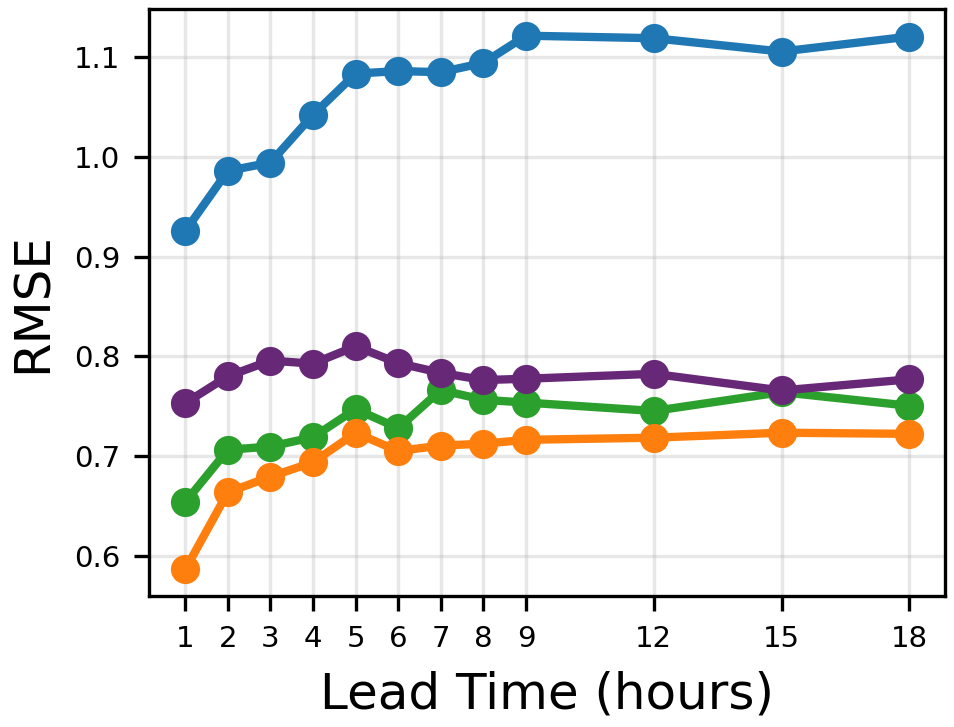}
\caption{Hourly precip. (mm)}
\end{subfigure}\hfill
\begin{subfigure}[t]{0.24\textwidth}
\centering
\includegraphics[width=\textwidth]{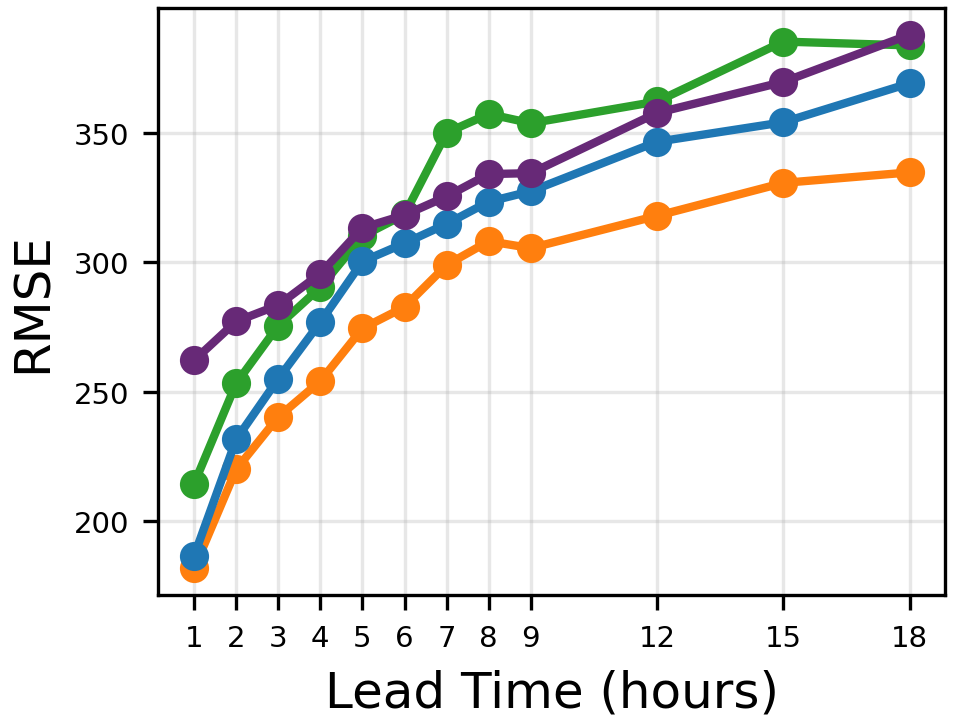}
\caption{CAPE (J kg$^{-1}$)}
\end{subfigure}

\vspace{0.6em}

\begin{subfigure}[t]{0.24\textwidth}
\centering
\includegraphics[width=\textwidth]{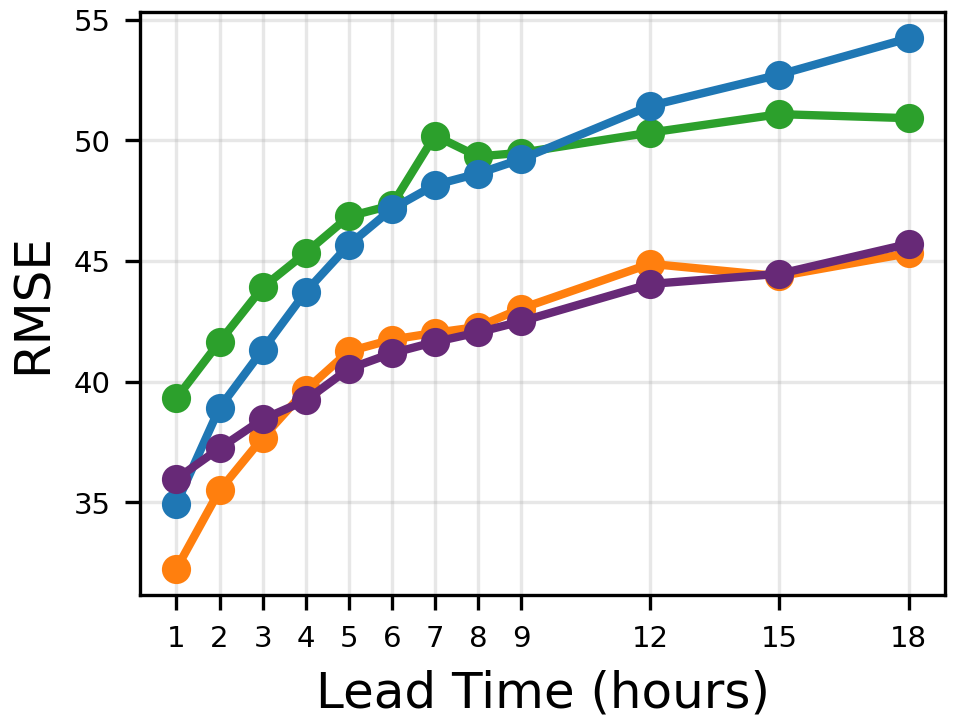}
\caption{CIN (J kg$^{-1}$)}
\end{subfigure}\hfill
\begin{subfigure}[t]{0.24\textwidth}
\centering
\includegraphics[width=\textwidth]{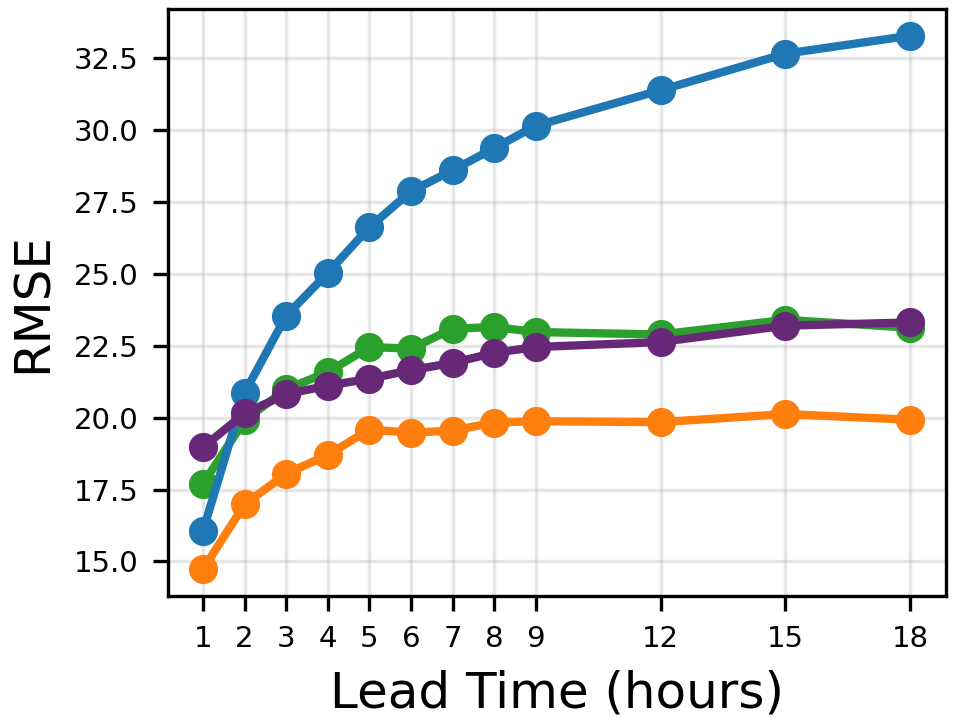}
\caption{Mid cloud cover (\%)}
\end{subfigure}\hfill
\begin{subfigure}[t]{0.24\textwidth}
\centering
\includegraphics[width=\textwidth]{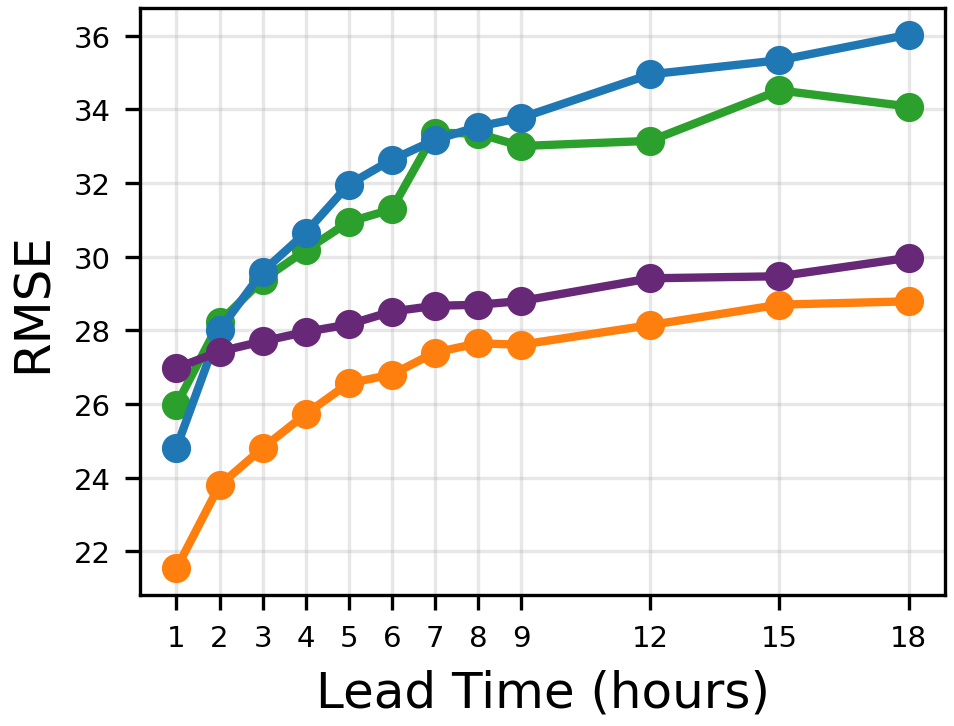}
\caption{Low cloud cover (\%)}
\end{subfigure}\hfill
\begin{subfigure}[t]{0.24\textwidth}
\centering
\includegraphics[width=\textwidth]{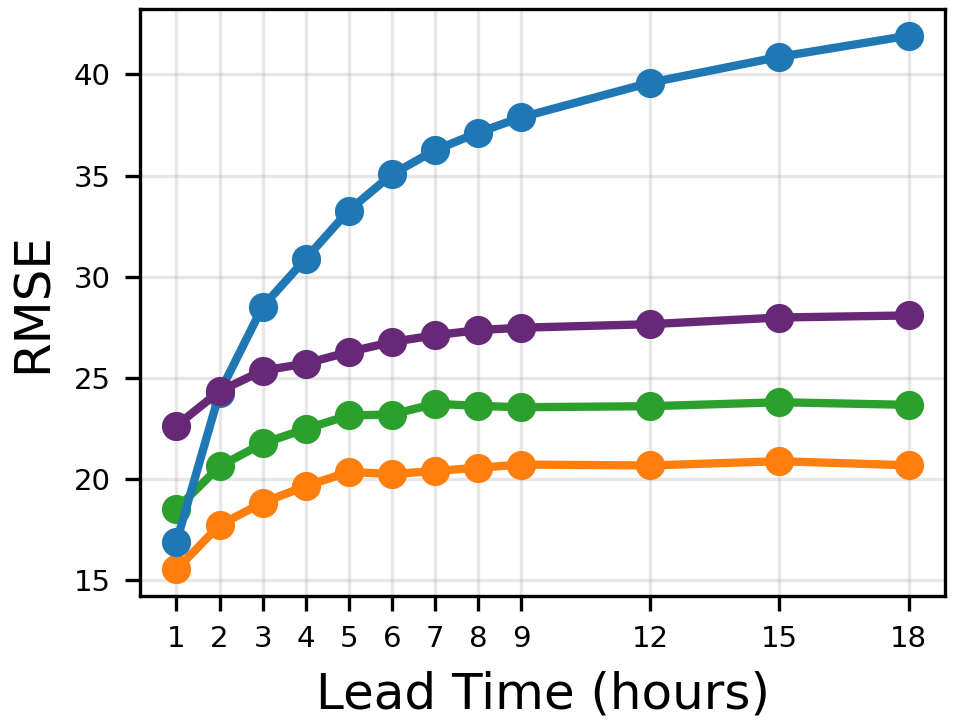}
\caption{High cloud cover (\%)}
\end{subfigure}
\caption{Selected RMSE verification fields for the subset of cycles when HREF was available
(00, 06, 12, and 18Z), comparing HRRRCast, operational HRRR, and HREF on a matched sample.}
\label{fig:rmse_href_subset}
\end{figure*}

Figure~\ref{fig:rmse_verification} shows representative RMSE curves for selected
2D diagnostics spanning low-level winds and convective metrics (80-m wind
components, CAPE, and CIN), precipitation and visibility, and cloud-cover fields at
multiple levels. The HRRRCast single member is competitive with HRRR across most fields, despite
being a single deterministic trajectory from a diffusion model, which is known to
sacrifice RMSE performance in favor of better probabilistic skill \citep{price2024}.
On the other hand, the HRRRCast ensemble is the best performer across nearly all fields;
this archive uses 9 members for most cycles, expanding to 18 members after July 1, 2026.
However, it is expected that a single member deterministic model trained with RMSE objective should 
outperform the HRRRCast ensemble in RMSE, as was the case in the previous version \citep{abdi39}.

Restricting the RMSE comparison to the synoptic-cycle subset for which HREF was available
(00, 06, 12, and 18Z; Fig.~\ref{fig:rmse_href_subset}) yields the same broad conclusion.
The HRRRCast ensemble remains the lowest-RMSE system for nearly all variables shown,
including dewpoint, precipitation, CAPE, CIN, and the mid-, low-, and high-cloud-cover fields, while HREF generally
improves upon operational HRRR and often also upon the single-member HRRRCast forecast. The main
exception is 2-m temperature, for which operational HRRR has the lowest RMSE over this matched
subset, with the HRRRCast ensemble remaining competitive and HREF lying in between the ensemble and
the single-member AI forecast. Overall, the matched HREF subset indicates that the ensemble benefit
in HRRRCast is not simply an artifact of the broader sample: even when evaluated only at cycles
where an operational convection-allowing ensemble exists, HRRRCast retains a consistent RMSE
advantage across most of the expanded diagnostic variable set.

\section{Calibration comparison against a CRPS-trained baseline}
\label{append:crps_comparison}

\begin{figure*}[tbp]
\centering
\begin{subfigure}[t]{0.49\textwidth}
\centering
\includegraphics[width=\textwidth]{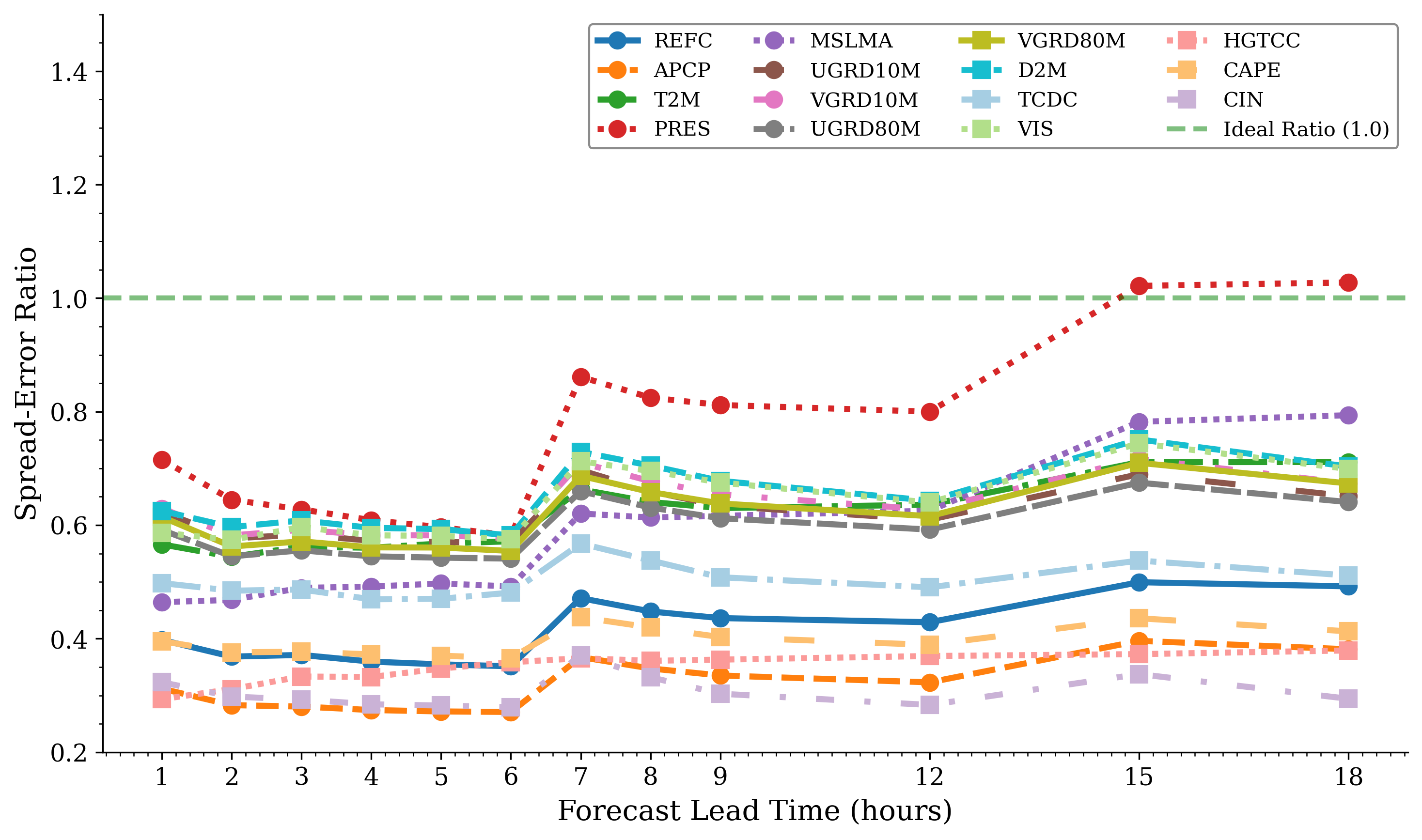}
\caption{Diffusion ensemble.}
\end{subfigure}\hfill
\begin{subfigure}[t]{0.49\textwidth}
\centering
\includegraphics[width=\textwidth]{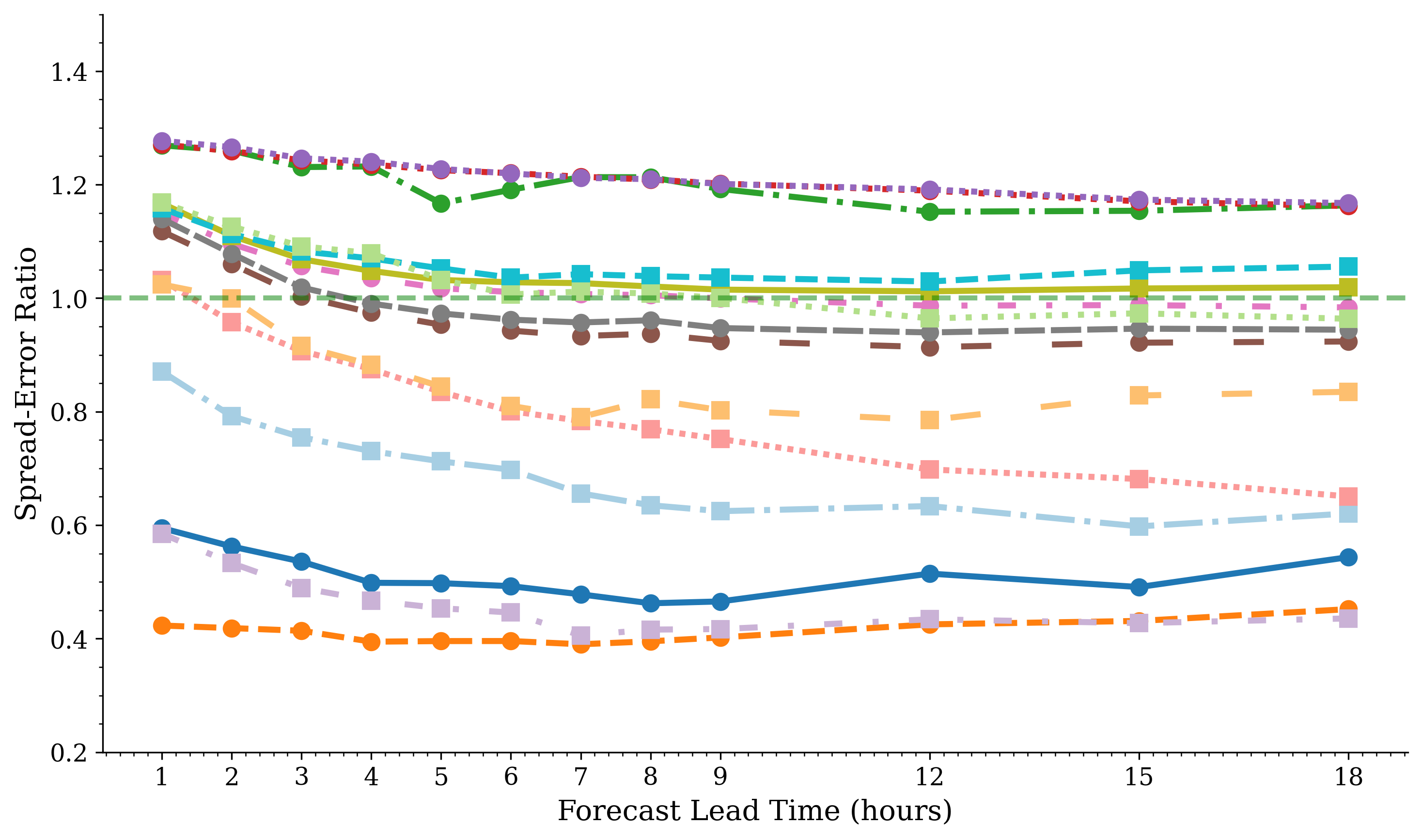}
\caption{CRPS-optimized ensemble.}
\end{subfigure}
\caption{Ten-member spread-error ratio as a function of forecast lead time for the diffusion and CRPS ensembles. The ideal value is 1.0. The diffusion model remains under-dispersive across many fields and lead times, while the CRPS model is substantially closer to calibrated spread and therefore better matches the spread-error criterion.}
\label{fig:ser_ablation}
\end{figure*}

To isolate the calibration trade-off, we compare the HRRRCast diffusion ensemble with a 
separate model trained with a Continuous Ranked Probability Score (CRPS) objective and 
without diffusion. The CRPS model follows the approach in \citet{alet2025}: it uses a 
single global noise vector of size 32, generates two ensemble members during training, and 
produces an almost-fair CRPS estimate. The spread-error ratio (SER), defined as the ratio of 
ensemble spread to RMSE, is near 1 for well-calibrated uncertainty and substantially below 1 
for under-dispersion. Figure~\ref{fig:ser_ablation} shows a consistent pattern: the 
CRPS-trained ensemble remains closer to the ideal value of 1 for most variables and across 
most lead times, whereas the diffusion ensemble is systematically under-dispersive. This 
behavior is expected given the optimization target: CRPS-based training explicitly rewards 
calibrated ensemble distributions, so the model learns to produce more realistic spread 
relative to forecast error.

This result does not imply that the CRPS ensemble is uniformly preferable for our application because
ensemble calibration is not the sole objective. The diffusion model remains attractive because its 
denoising formulation preserves sharper, more coherent storm structures and mesoscale gradients, 
which are especially important for reflectivity, precipitation, and hazardous-weather fields.
Interestingly, even the CRPS-trained ensemble remains under-dispersive for the most important variables,
namely REFC and APCP, indicating that rare, heavy-tailed convective regimes remain difficult to 
calibrate even with a CRPS objective. The CRPS-trained ablation does improve ensemble calibration overall, but
incorporating CRPS-like objectives into a diffusion model is non-trivial because diffusion
training operates on denoised samples rather than full forecast trajectories.

\section{Timeseries diagnostics}
\label{append:timeseries_diagnostics}

Here, we plot the daily model cycle for an approximately two-week
period (2026-08-11 to 2026-08-24), with forecasts initialized at 12Z each day
and verified through the subsequent 24-h forecast window.

\begin{figure*}[tbp]
\centering
\includegraphics[width=0.98\textwidth]{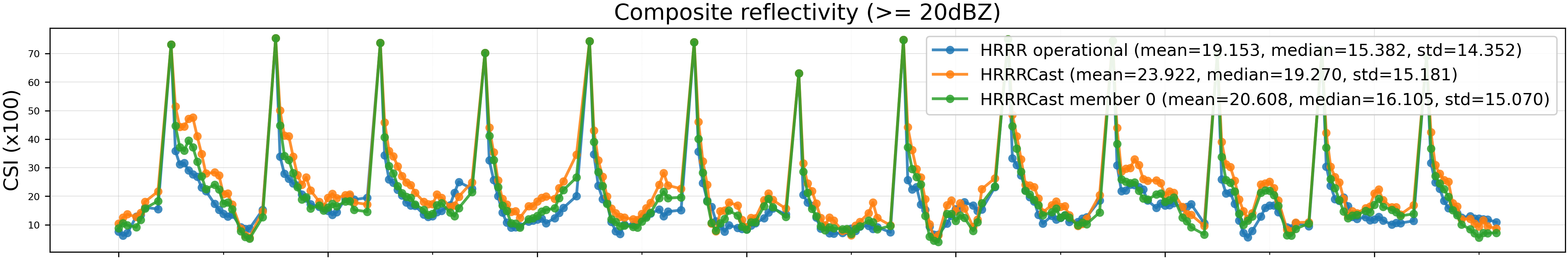}\\[0.35em]
\includegraphics[width=0.98\textwidth]{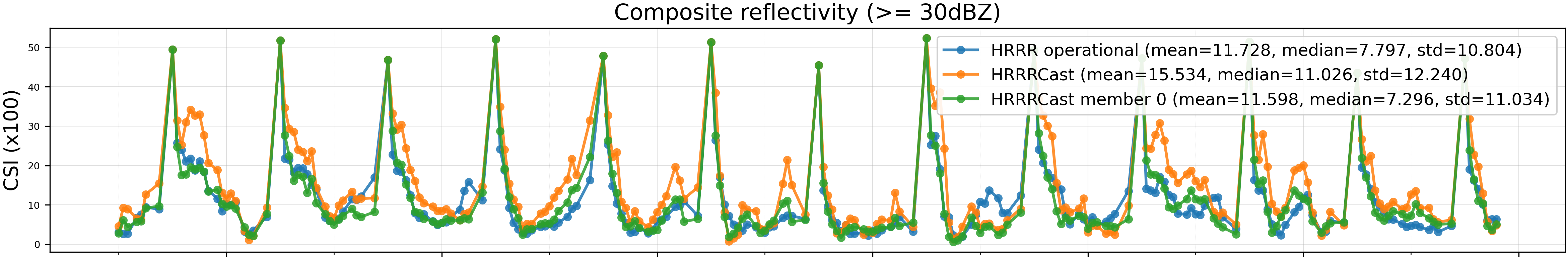}\\[0.35em]
\includegraphics[width=0.98\textwidth]{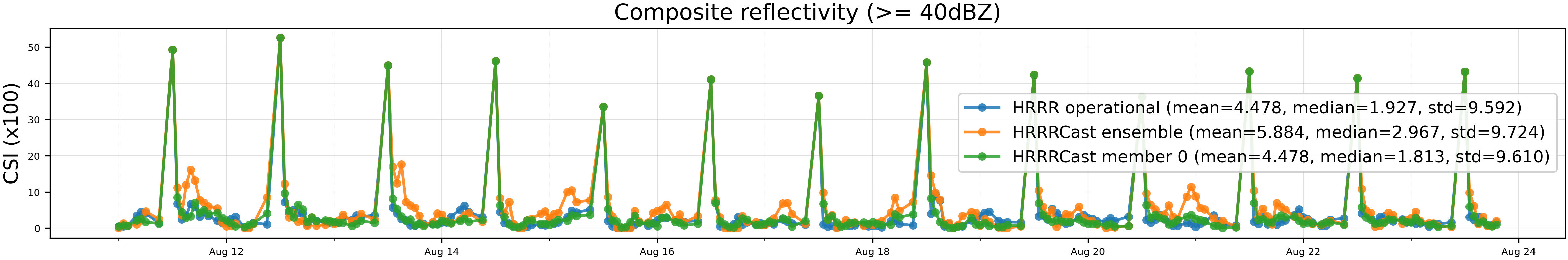}
\caption{Daily model cycle: composite-reflectivity CSI vs MRMS (multiple thresholds).}
\label{fig:appendix_mats_timeseries_2}
\end{figure*}

In composite-reflectivity CSI (Fig.~\ref{fig:appendix_mats_timeseries_2}), the
HRRRCast ensemble mean remains above HRRR operational at 20, 30, and 40 dBZ in
the aggregate and through many event cycles, while preserving similar event timing.
Across thresholds, the HRRRCast ensemble-mean advantage is expressed mainly in
amplitude: event peaks and post-peak shoulders are generally higher than HRRR,
especially at 30 dBZ. At 40 dBZ, deterministic HRRRCast member 0 stays close to
HRRR on many cycles, while the ensemble mean remains systematically higher,
indicating that most of the high-threshold gain is concentrated in the ensemble
signal rather than in a single deterministic trajectory.

\begin{figure*}[tbp]
\centering
\includegraphics[width=0.98\textwidth]{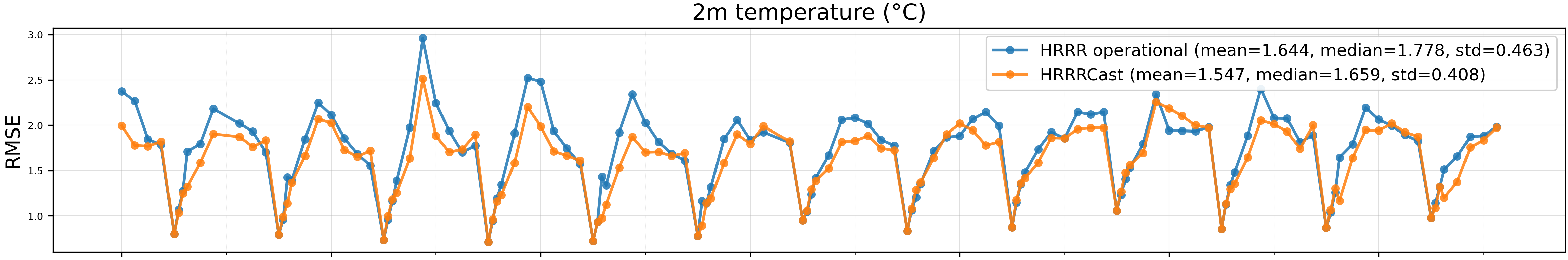}\\[0.35em]
\includegraphics[width=0.98\textwidth]{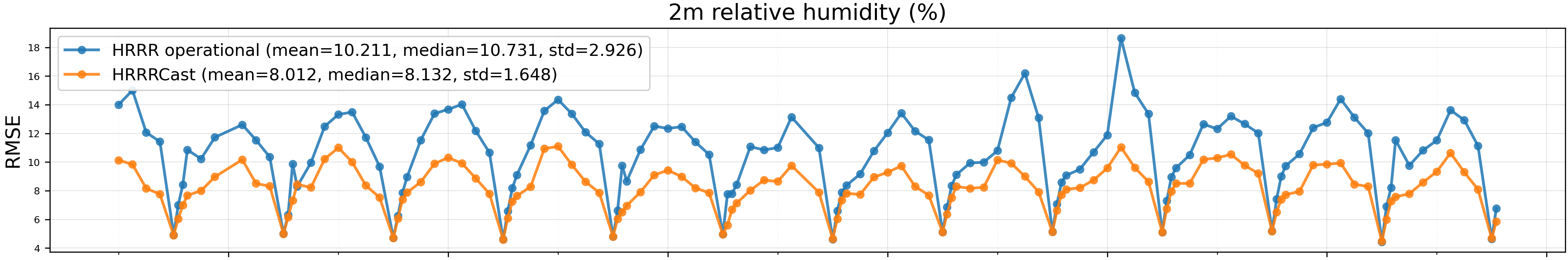}\\[0.35em]
\includegraphics[width=0.98\textwidth]{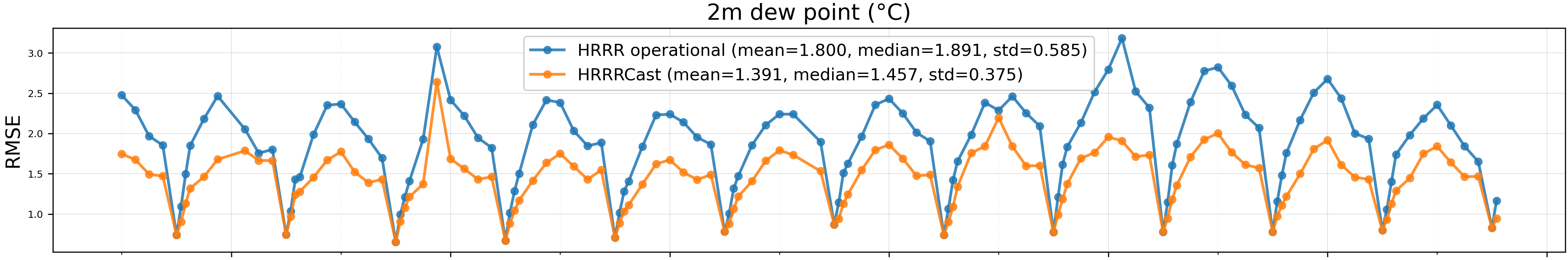}\\[0.35em]
\includegraphics[width=0.98\textwidth]{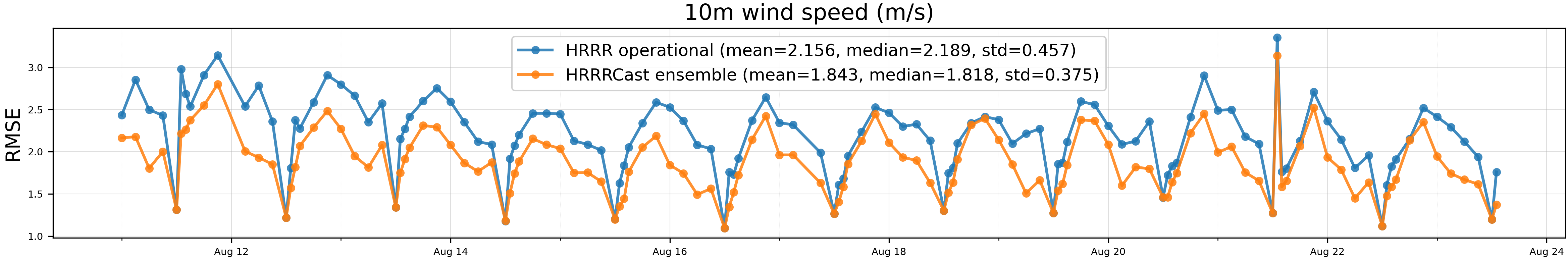}
\caption{Daily model cycle: RMSE for selected surface variables vs METAR.}
\label{fig:appendix_mats_timeseries_1}
\end{figure*}

Figures~\ref{fig:appendix_mats_timeseries_1}--\ref{fig:appendix_mats_timeseries_2}
show that HRRRCast broadly preserves the day-to-day and diurnal phase of HRRR
operational while changing error amplitude in a variable- and threshold-dependent
way. In the surface RMSE series (Fig.~\ref{fig:appendix_mats_timeseries_1}),
HRRRCast is consistently lower for 2-m relative humidity, 2-m dewpoint, and
10-m wind speed, and is also modestly lower for 2-m temperature in the aggregate.
The largest separations occur during recurring daily high-error periods, where
HRRRCast damps several of the strongest HRRR peaks.

\section{Illustrative forecast example}
\label{append:case_study}

To illustrate the forecast behavior of the newly added HRRRCast fields, we show the
forecast evolution for a nor'easter storm case initialized on 2026-09-26. This
example highlights the evolving precipitation pattern, instability structure,
cyclone evolution, and the associated 500-hPa flow pattern that steers the system.

This example is intended to show how the newly added HRRRCast fields evolve during a
realistic severe-weather episode, with the most prominent feature being the coexistence
of two distinct storm regimes: a land-falling cyclone along the northeastern coast and
a separate convective regime over the central Great Plains. The coastal system is most
evident in the mean-sea-level-pressure minimum and the associated corridor of heavy
precipitation and reflectivity that becomes increasingly focused along the coastline
from the 1-h to 6-h forecast. In contrast, the Great Plains signal is expressed more as
a broad region of elevated CAPE and cloud cover, with more diffuse precipitation and a
less compact pressure anomaly, consistent with a strongly convective, warm-sector
structure rather than a mature coastal cyclogenesis center. The 500-hPa
geopotential-height field reinforces this interpretation by showing a trough-and-ridge
pattern that supports both the East Coast cyclone pressure fall and the interior
instability burst.

At the 1-h lead, the fields already suggest two separate dynamical centers: a coastal
low-pressure system and a more inland convective plume over the Plains. By 3 h, the
northeastern cyclone becomes more organized, as indicated by the strengthened pressure
gradient, broader precipitation shield, and moistening of the near-surface
environment, while the Plains region remains characterized by elevated instability and
cloud growth. By 6 h, the land-falling northeastern storm is the dominant structure in
the precipitation and pressure fields, while the Great Plains feature remains as a
secondary maximum in CAPE and cloud cover, illustrating how multiple storm-scale
mechanisms can coexist within the same large-scale flow regime.


\FloatBarrier
\bibliographystyle{\bibliostyle}
\bibliography{references,mypublications}

\end{document}